\documentclass[fleqn,usenatbib]{mnras}

\usepackage{newtxtext,newtxmath}
\usepackage[T1]{fontenc}

\DeclareRobustCommand{\VAN}[3]{#2}
\let\VANthebibliography\thebibliography
\def\thebibliography{\DeclareRobustCommand{\VAN}[3]{##3}\VANthebibliography}

\usepackage{graphicx}	% Including figure files
\usepackage{amsmath}	% Advanced maths commands
\usepackage{caption}
\usepackage{subcaption}
\usepackage{longtable}
\title[Galaxy Morphologies and Stellar Properties]{MUSE-ALMA Haloes IX: Morphologies and Stellar Properties of Gas-rich Galaxies}

\author[Karki et al.]{Arjun Karki$^{1}$\thanks{E-mail: karkia@email.sc.edu}, 
Varsha P. Kulkarni$^{1}$,
Simon Weng$^{2,3,4,5}$,
C\'eline P\'eroux$^{2,6}$,
Ramona Augustin$^{7}$,
\newauthor
Matthew Hayes$^{8}$, 
Mohammadreza Ayromlou$^{9}$,
Glenn G. Kacprzak$^{10,5}$, 
J. Christopher Howk$^{11}$, 
\newauthor
Roland Szakacs$^{2}$, 
Anne Klitsch$^{12}$, 
Aleksandra Hamanowicz$^{7}$, 
Alejandra Fresco$^{13}$, 
\newauthor
Martin A. Zwaan$^{2}$,
Andrew D. Biggs$^{2}$, 
Andrew J. Fox$^{14}$, 
Susan Kassin$^{7}$, 
Harald Kuntschner$^{2}$
\\
$^{1}$ Department of Physics and Astronomy, University of South Carolina, Columbia, SC 29208, USA\\
$^{2}$ European Southern Observatory, Karl-Schwarzschildstrasse 2, D-85748 Garching bei M{\"u}nchen, Germany\\
$^{3}$ Sydney Institute for Astronomy, School of Physics, University of Sydney, NSW 2006, Australia\\
$^{4}$ ARC Centre of Excellence for All Sky Astrophysics in 3 Dimensions (ASTRO 3D)\\ 
$^{5}$ ATNF, CSIRO Astronomy and Space Science,  PO Box 76, Epping, NSW 1710, Australia \\
$^{6}$ Aix Marseille Universit\'e, CNRS, LAM (Laboratoire d'Astrophysique de Marseille) UMR 7326, 13388, Marseille, France \\
$^{7}$ Space Telescope Science Institute, 3700 San Martin Drive, Baltimore, MD21218, USA \\
$^{8}$ Stockholm University, Department of Astronomy and Oskar Klein Centre for Cosmoparticle Physics, AlbaNova University Centre, SE-10691, Stockholm, Sweden\\
$^{9}$Universit{\"a}t Heidelberg, Zentrum f{\"u}r Astronomie, Institut f{\"u}r theoretische Astrophysik, Albert-Ueberle-Str. 2, 69120 Heidelberg, Germany\\
$^{10}$ Centre for Astrophysics and Supercomputing, Swinburne University of Technology, Hawthorn, Victoria 3122, Australia\\
$^{11}$ Department of Physics, University of Notre Dame, Notre Dame, IN46556, USA\\
$^{12}$ DARK, Niels Bohr Institute, University of Copenhagen, Jagtvej 128, 2200 Copenhagen, Denmark\\
$^{13}$ Max-Planck-Institut f\"ur Extraterrestrische Physik (MPE), Giessenbachstrasse 1, D--85748 Garching, Germany\\
$^{14}$ AURA for ESA, Space Telescope Science Institute, 3700 San Martin Drive, Baltimore, MD 21218\\
}

\date{Accepted XXX. Received YYY; in original form ZZZ}

\pubyear{2022}

\begin{document}
\label{firstpage}
\pagerange{\pageref{firstpage}--\pageref{lastpage}}
\maketitle

\begin{abstract}
Understanding how galaxies interact with the circumgalactic medium (CGM) requires determining how galaxies’ morphological and stellar properties correlate with their CGM properties. We report an analysis of 66 well-imaged galaxies detected in \emph{HST} and VLT MUSE observations and determined to be within $\pm$500 km s$^{-1}$ of the redshifts of strong intervening quasar absorbers at $0.2 \lesssim z \lesssim 1.4$ with H \textsc{i} column densities $N_{\rm H I}$ $>$ $10^{18}$ $\rm cm^{-2}$. We present the geometrical properties (S\'ersic indices, effective radii, axis ratios, and position angles) of these galaxies determined using \textsc{GALFIT}. Using these properties along with star formation rates (SFRs, estimated using the H$\alpha$ or [O \textsc{II}] luminosity) and stellar masses ($M_{*}$ estimated from spectral energy distribution fits), we examine correlations among various stellar and CGM properties. Our main findings are as follows: (1) SFR correlates well with $M_{*}$, and most absorption-selected galaxies are consistent with the star formation main sequence (SFMS) of the global population. (2) More massive absorber counterparts are more centrally concentrated and are larger in size. (3) Galaxy sizes and normalized impact parameters correlate negatively with $N_{\rm HI}$, consistent with higher $N_{\rm H I}$ absorption arising in smaller galaxies, and closer to galaxy centers. (4) Absorption and emission metallicities correlate  with $M_{*}$ and sSFR, implying metal-poor absorbers arise in galaxies with low past star formation and faster current gas consumption rates. (5) SFR surface densities of absorption-selected galaxies are higher than predicted by the Kennicutt-Schmidt relation for local galaxies, suggesting a higher star formation efficiency in the absorption-selected galaxies.

\end{abstract}


\section{Introduction}

The circumgalactic medium (CGM) has become increasingly recognized as an important component of the baryonic Universe. It serves as a transition region between the galaxy disk and the intergalactic medium (IGM) \citep{2017ARA&A..55..389T}. Metal-poor IGM gas is believed to flow into the galaxy, passing through the CGM. This gas is converted into stars and progressively enriched chemically. The outflows driven by the supernovae [or active galactic nuclei (AGN)] transfer the enriched gas back into the IGM, also passing through the CGM. This cosmic baryon cycle regulates star formation in the galaxy \citep{2020ARA&A..58..363P}. Given the central role of the CGM in this cycle, it is expected to play a major role in the evolution of the galaxy. Understanding how the CGM interacts with galaxies requires analyzing how the stellar properties of the galaxies are dictated by and, in return, influence the CGM properties. 

The stellar properties of the galaxies are described by measurements of various properties that depend directly on their stellar populations -- for example, photometric magnitudes, colors, star formation rates (SFRs), and stellar masses. The morphologies of the galaxies are closely coupled to these properties. They can be expressed in terms of various quantitative measures of the surface brightness distribution such as the effective radius ($ R_{e}$), axis ratio (\emph{b/a}), and S\'ersic index ($n$). 
Broadly speaking, galaxies fall into two main types in the color-magnitude diagrams -- the ``blue cloud'' consisting of the late-type, more actively star-forming galaxies, and the ``red sequence'' of early-type, more passive galaxies that have their star formation quenched \citep[e.g.][]{2003MNRAS.Kauffmann, 2011arXiv1102.0550B}.

The ``size'' of a galaxy detected in optical light depends on its stellar mass and the rate of star formation, which are governed by the dark matter halo and the galaxy's formation history. Massive galaxies are larger and have higher SFRs. \citep{Mowla_2019}. The slope of the size versus stellar mass relation is shallower for late-type galaxies than for early-type galaxies \citep[e.g.][]{2003MNRAS..Shen, 2014ApJ...788...28V}. This difference may be due to dry minor mergers, which can lead to a size growth caused by adding an outer envelope without adding as much mass. Galaxies that undergo repeated dry minor mergers tend to have a larger size \citep[e.g.][]{10.1093..Hilz, Carollo_2013}. The star formation caused by gas-rich mergers can also lead to the formation of larger disks. The different evolutionary paths taken by early-type and late-type galaxies thus result in different size-mass relationships. 

%Studying galaxy formation and evolution in the circum-galactic medium or CGM requires in-depth knowledge about physical and chemical processes that drive the gas to flow into and out from the galaxies in the CGM. The CGM is a region where the gas from the intergalactic medium (IGM) flows into the galaxy and accumulates. Those accreted gases become the source of fuel for star formation for the galaxies once they cool down. At the same time, the gas outflows from the galaxy, ejecting materials such as metals and dust to the surrounding (\cite{2017ARA&A..55..389T}). 

The direct observation of emissions from the gas inflows and outflows passing through the CGM is difficult due to the very low gas density. 
%the gas in-flowing into and out-flowing from the galaxies. 
Absorption spectroscopy of background sources such as quasars or gamma-ray bursts (GRBs) provides an alternative and powerful technique to study these gas flows. 
%Such spectroscopy technique consists of absorption features that are detected in the spectra of the background of bright sources such as quasar or gamma-ray bursts (GRBs). 
%The densest among these quasar absorption systems are the Damped Lyman Alpha (DLA) and Sub-Damped Lyman (sub-DLA) systems. 
Damped Ly$\alpha$ (DLA) and sub-Damped Ly$\alpha$ (sub-DLA) systems provide a huge reservoir of neutral hydrogen required for star formation  \citep[e.g.][]{2003MNRAS.345..480P,2005ARA&A..43..861W,2009ApJ...696.1543P,2012A&A...547L...1N,2022ApJ...929..150K}. DLAs ($N_{\rm H I}$ $\geq$ 2 $\times$ $10^{20}$ $\rm cm^{-2}$) and sub-DLAs ($10^{19}$ $\leq$  $N_{\rm H I}$ < 2 $\times$ $10^{20}$ $\rm cm^{-2}$) permit measurements of a variety of metal ions, and are therefore among the best-known tracers of element abundances in distant galaxies \citep[e.g.][]{2005ApJ...618...68K,2012ApJ...755...89R, 2015ApJ...806...25S,2016MNRAS.455.4100F}.

Although the absorption technique provides an effective tool to probe gas along the sight line to the background object, it cannot provide information 
about the galaxy in which the absorption arises \citep[e.g.][]{1986A&A...155L...8B, 1991A&A...243..344B}. %(\cite{2019MNRAS..Celine}. 
Also, detecting galaxies associated with absorbers in the vicinity of the quasar using imaging and spectroscopy was not always effective in past studies, since the galaxies selected for spectroscopic study were  sometimes found to be offset in redshift from the absorber. The technique of integral field spectroscopy (IFS) provides an efficient method for detecting galaxies associated with the absorbers, and thus connecting stellar properties with gas properties \citep[][]{2019MNRAS.485.1595P,2022MNRAS.516.5618P}. Several surveys (e.g., MusE GAs FLOw and Wind \citep[MEGAFLOW;][]{2016ApJ...833...39S}, MUSE Ultra Deep Field \citep[MUDF;][]{2019MNRAS.490.1451F} and MUSE Analysis of Gas around
Galaxies \citep[MAGG;][]{2020MNRAS.499.5022D,2020MNRAS.491.2057L,2023MNRAS.518..305L}) have used the power of IFS to search for sources traced by Mg \textsc{ii} and H \textsc{i} absorption and studied the gas properties of CGM. While the Bimodal Absorption System Imaging Campaign \citep[BASIC survey;][] {2022arXiv220413229B}, has studied H \textsc{i} selected partial Lyman limit systems (pLLSs) and Lyman limit systems (LLSs) to search for absorber-associated galaxies, the Cosmic Ultraviolet Baryon Survey \citep[CUBS;][]{2020MNRAS.497..498C} has studied the galactic environments of Lyman limit systems
(LLSs) at $z_{\rm abs} < 1$ using IFS observations. Another survey, MUSE Quasar-field Blind Emitters Survey \citep[MUSEQuBES;][]{2020MNRAS.496.1013M} searched for Ly$\alpha$ emitters (LAEs) at the redshift of the absorbers using guaranteed time observations with the Multi-Unit Spectroscopic Explorer \citep[MUSE;][]{2010SPIE.7735E..08B}  on the Very Large Telescope (VLT). 
%IFS can be used as an unbiased method to detect associated galaxies. 

We have surveyed a large number of absorption-selected galaxies with VLT  MUSE and the Atacama Large Millimetre/submillimetre Array \citep[ALMA;][]{2009IEEEP..97.1463W} as part of our MUSE-ALMA Haloes Survey,  and have recently imaged  
these galaxies with the Hubble Space Telescope (HST). The \emph{HST} images provide high-resolution broadband continuum imaging of the galaxies, while the MUSE data provide IFS of the galaxies, thus providing information about the spatial distribution of gas kinematics, SFRs and emission-line metallicity. The kinematics measured from the IFS also allows estimates of the dynamical masses of the galaxies \citep[e.g.,][]{2012MNRAS..Celine, 2013Sci..Bouche}. Thus, combining \emph{HST} images and MUSE IFS provides a powerful approach to studying the morphologies and stellar content of absorption-selected galaxies and relating them to the CGM gas properties. Together, the stellar and gas properties can be used to put improved constraints on the evolution of galaxies and their CGM. 
%Such study can improve our understanding of the formation and evolution trends of the galaxies in the CGM.

The MUSE-ALMA Haloes (MAH) survey targets the fields of 32 H \textsc{i} rich absorbers at redshift 0.2 $\le$ $\rm $z$_{\rm abs}$ $\le$ 1.4 detected in sight lines to 19 quasars. These 32 absorbers were detected in HST FOS, COS, or STIS UV spectra. Most of our quasars also have optical high-resolution spectra from VLT/UVES, X-Shooter, or Keck/HIRES with spectral resolutions ranging from 4000-18 000 (X-shooter), 45 000-48 000 (HIRES) and up to 80 000 (UVES). The H \textsc{i} column densities determined from these UV spectra (with resolution R $=$ 20 000$-$30 000) are found to be $N_{\rm HI}$ > $10^{18}$ $\rm cm^{-2}$. Information about quasar spectra and an overview of the survey are provided in \citet{2022MNRAS.516.5618P}. In this paper, we focus on 66 galaxies observed in the HST images that lie within a radial velocity range of $\pm ${ }$ $500 km $\rm s^{-1}$ of the redshifts of 25 H \textsc{i} rich absorbers (the remaining 7 absorbers having no associated galaxies within $\pm ${ }$ $500 km $\rm s^{-1}$). The paper is organized as follows: section 2 presents the sample selection and observations. Section 3 details the results derived from the observation. Section 4 summarizes our findings. We adopt the following cosmology parameters: $\rm H_{\rm o} = 70$  $\rm km $ $\rm s^{-1}$ $\rm Mpc^{-1}$, $\Omega_{M} = 0.3$ and $\Omega_{\Lambda} = 0.7$ throughout the paper.

\begin{table*}
\begin{center}
\caption{{\bf Summary of HST observations.} Listed are the quasar name, quasar and absorber redshifts, the full-width half maximum of the PSF of \emph{HST} image in each  filter, the field of view, the \emph{HST} camera and filter used, the number of exposures $\&$ exposure time of each exposure and the 3-$\sigma$ limiting magnitude in each filter.  A complete table of quasar fields and their properties is available as machine-readable online material.
}
\begin{tabular}{lcccclcc}
\hline\hline
Quasar & $z_{\rm qso}$ &$z_{\rm abs}$ & FWHM  & FOV & Instrument  & $\rm n_{\rm exp}$ $\times$ $\rm t_{\rm exp}$   & Magnitude \\
       &    &      &  (")  &   (" $\times$ ")  &  &   [s]  &  limit \\
\hline                                                        
Q0058$+$0019 &  1.959  & 0.6127  &  0.188 &  80 $\times$ 80  & WFPC2/WF3 F702W &   4 $\times$ 1100  & 27.02\\

Q0138$-$0005   & 1.340 & 0.7821 & 0.073& 160 $\times $80  &WFC3/UVIS F475W &  4 $\times$ 567 & 28.70 \\

... & ...& ... & 0.083 & 160 $\times$ 80 &WFC3/UVIS F625W &  4 $\times$ 591 & 28.90\\

... & ... & ... & 0.130 & 123 $\times$ 123 &WFC3/IR F105W &  4 $\times$ 603  & 27.93 \\

J0152$-$2001  & 2.060 & 0.3897 &  0.083 & 160 $\times$ 80  &WFC3/UVIS F336W & 2 $\times$ 587 + 2 $\times$ 595 &  29.39 \\

... & ...& 0.7802 & 0.074 & 160 $\times$ 80 &WFC3/UVIS F475W &  2 $\times$ 587 + 2 $\times$ 595 & 29.29 \\

... & ... & ... & 0.182 & 80 $\times$ 80 &WFPC2/WF3 F702W &  2 $\times$ 1100 + 3 $\times$ 1300 + 1 $\times$ 1400 & 27.68\\

Q0152$+$0023  & 0.589 & 0.4818 & 0.083&  160 $\times$ 80 &   WFC3/UVIS F336W &  2 $\times$ 535 + 2 $\times$ 518 & 28.54\\

... & ...& ... &  0.074 & 160 $\times$ 80 &WFC3/UVIS F475W &  1 $\times$ 635 + 3 $\times$ 618  & 28.39\\

... & ... & ... & 0.082 & 160 $\times$ 80 &WFC3/UVIS F814W &  1 $\times$ 635 + 3 $\times$ 618 & 27.92\\

... & ...& ... & ...  & ... & ...  & ... & ... \\ 

\hline\hline 				       			 	 
\label{table1}
\end{tabular}			       			 	 

\end{center}			       			 	 
\end{table*}

\begin{figure*}

    \includegraphics[width=0.91\textwidth]{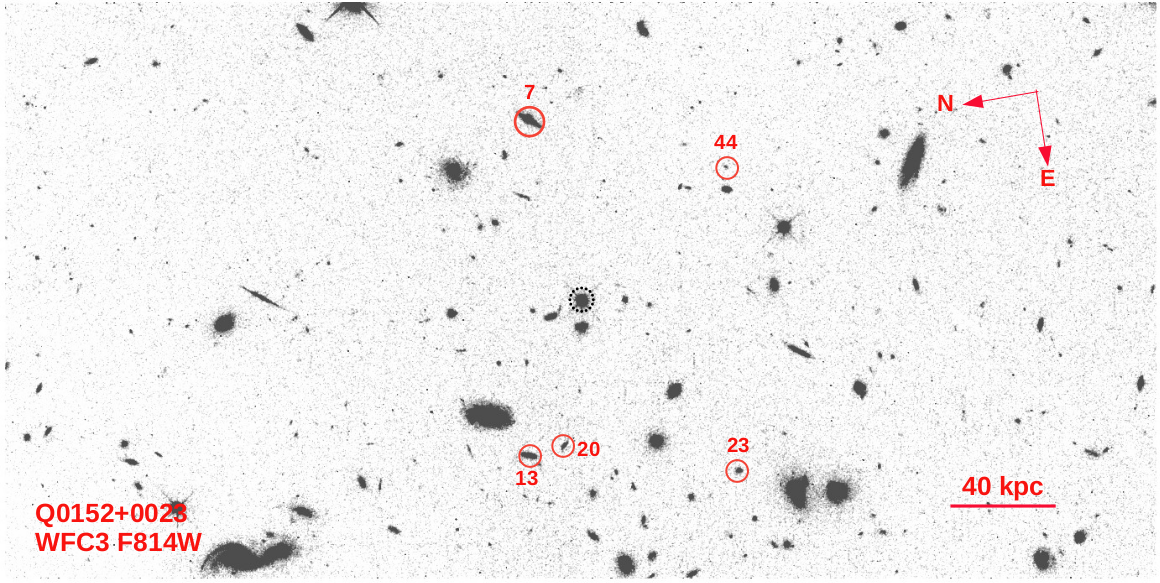}   
    \caption{Median stacked image of Q0152+0023 in the F814W UVIS filter-band. The image is created by combining all the individual exposure images, which helps to remove the cosmic ray effects and bad pixels and improve the signal-to-noise ratio. This enables us to study the morphological properties of the galaxy populations robustly. The solid red circles represent the position of the associated galaxies in the field of view, while a dotted black circle denotes the QSO's position. The astropy PHOTUTILS package was used to detect the associated galaxies in the given quasar field. The associated galaxies are located within $\pm$ 500  km $\rm s^{-1}$ of the absorber's redshift. The redshift of those galaxies came from MUSE data. The scale corresponds to 40 kpc at $z_{\rm abs}$ = 0.4818. The object identification number (ID) of these galaxies came from the MUSE-ALMA Haloes master table listed in \citet{2022MNRAS.516.5618P}.
    }
    \label{fig1}
\end{figure*}

\section{HST images and Observations}

Our primary \emph{HST} imaging dataset comes from the broad-band imaging observations performed in GO Program ID 15939 (PI: P\'eroux) with the Wide Field Camera-3 (WFC3). These data were complemented by archival Wide Field and Planetary Camera-2 (WFPC2) or WFC3 images obtained in programs 5098, 5143, 5351, 6557, 7329, 7451, 9173, and 14594 (PIs Burbidge, Macchetto, Bergeron, Steidel, Malkan, Smette, Bechtold, Bielby, respectively). The observations consisted of multiple dithered exposures in a variety of filters. Further details about these observations and the observation strategy used for our own program (PID 15939) can be found in \citet{2022MNRAS.516.5618P}.

\begin{figure*}
    \includegraphics[width=0.95\textwidth]{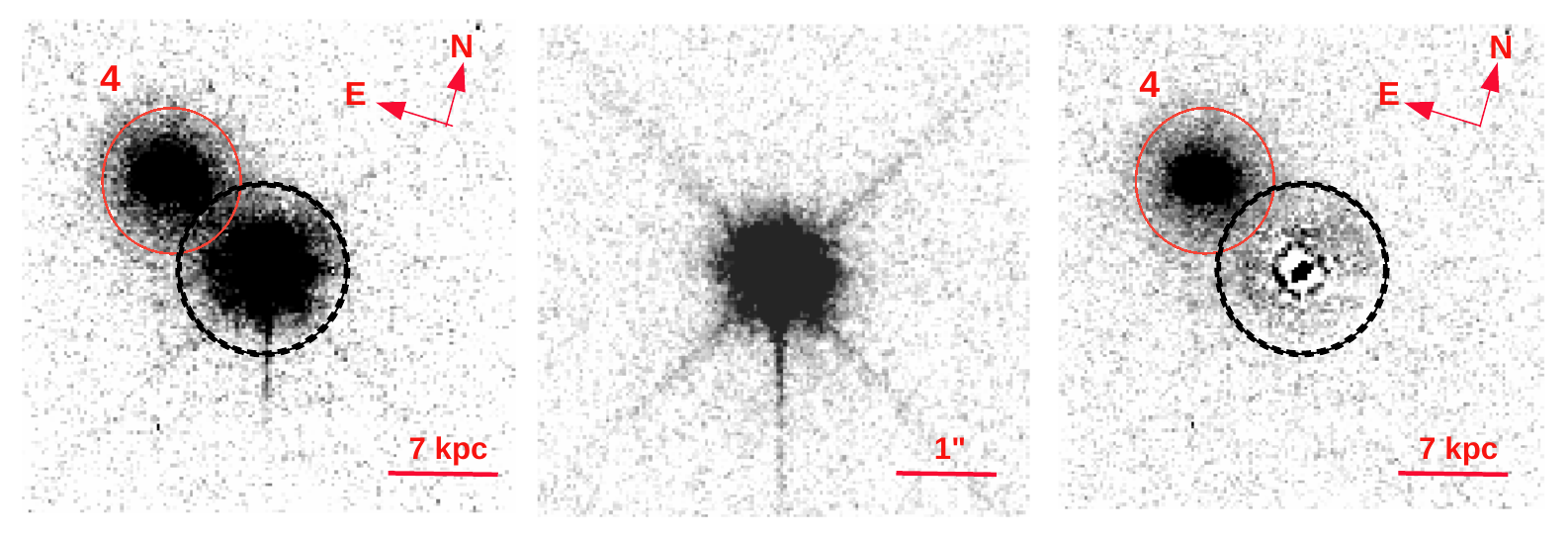}
    \caption{Left panel: Median stacked zoomed-in image of Q1515+0410 in the F814W UVIS filter-band. The image shows the bright background quasar exhibiting strong diffraction spikes where a galaxy is in front of it. The dotted black and solid red circles denote the location of the quasar and the galaxy in the image. The quasar's light contaminates the galaxy's image. Middle panel: A Point Spread Function (PSF) constructed to remove quasar's contamination. The PSF is designed using observations of all other quasar fields from our sample in the same filter. Right panel: PSF subtracted image of the same zoomed-in image. The contamination of the  quasar on the galaxy is removed after performing a careful PSF-subtraction. Detailed information about PSF construction and PSF subtraction is provided in \citet{2022MNRAS.516.5618P}.
    }
    
    \label{fig2}
\end{figure*}

The data were reduced using the \textsc{calwf3} or \textsc{calwf2} pipelines. For each filter, the sky-subtracted, aligned images from the individual exposures were median-combined to produce the final images. \autoref{fig1} and \autoref{fig2} show examples of final full-frame images and zoomed-in sections near the quasar. The quasar point spread function (PSF) has been subtracted in the zoomed-in frame to search for galaxies at small angular separations from the quasar. The PSF in each filter and instrument for each given quasar field was constructed using observations of all remaining quasar fields from our sample in the same band and instrument. For each such image used in making the PSF, we masked the objects other than the central quasar and performed sky subtraction. All such processed images were aligned spatially and coadded after the flux levels in the outer wings of the PSF were scaled to match with each other. The resulting PSF thus constructed was subtracted from the quasar field of interest after matching the flux levels of the two images. The PSF-subtracted images thus produced were used to search for galaxies near the quasars. This approach is similar to that used in previous works \citep[e.g.][]{2000ApJ...536...36K, 2001ApJ...551...37K, 2010AJ....139..296C, 2011AJ....141..206S, 2018MNRAS.478.3120A}. Object detections and astrometric and photometric measurements were performed using the \textsc{astropy} package \textsc{photutils}. The data reduction, PSF construction, PSF subtraction, and photometric measurements are further detailed in \citet{2022MNRAS.516.5618P}.

Using the MUSE observations along with the \emph{HST} imaging, a total of 3658 sources were detected in all of our fields. The MUSE Line Emission Tracker (MUSELET) tool of the MPDAF package was used to detect sources with emission lines, and the \textsc{profound} R Package was used to identify continuum sources in the MUSE fields. A final master table was produced by matching the sources detected in the HST images with the MUSE results using the \textsc{topcat} tool. Spectroscopic redshifts were determined for 703 objects out of the 3658 sources detected in all fields using the \textsc{marz} tool on the VLT/MUSE spectra. The remaining objects do not have spectroscopic redshifts, because they were either detected outside of the MUSE field of view or too faint to estimate a redshift. We refer the reader to \citet{2022MNRAS.516.5618P} for a detailed methodology regarding the redshift measurements and for the master table of all targets. 
  
  \autoref{table1} provides a summary of \emph{HST} imaging observations for 18 quasar fields. 
 %Using these images, we detected 3222 sources in the 18 \emph{HST} fields, while 1097 sources were detected in the MUSE data cubes. The wider field of view and high spatial resolution of the \emph{HST} images compared to the MUSE cubes allowed us to detect more objects in the \emph{HST} images. 
 The high spatial resolution of the broad-band  \emph{HST} images allowed us to detect several sources nearby or far away from the quasar sightlines, which were undetected in the MUSE cubes. The \emph{HST} images also enabled us to detect and resolve several sources near the quasars whose redshifts are not well known. While the MUSE data allowed us to study the emission properties of galaxies, the study of geometrical properties of the galaxies and structural features such as tidal tails are difficult to discern in MUSE images due to insufficient spatial resolution. The high-resolution \emph{HST} images also allowed us to study the morphologies of the galaxies and structural details.

The MUSE data show 79 associated galaxies within $\pm$ 500 km s$^{-1}$ of the absorber redshifts. Out of the total of 32 absorbers, 19 (59$\%$) have two or more associated galaxies, 7 (22$\%$) have one associated galaxy and the remaining 6 (19$\%$) have no associated galaxy detected within $\pm$ 500 km s$^{-1}$ of the  absorption redshift. A detailed explanation of how these associated galaxies were selected is provided in \cite{2023MNRAS.519..931W}. Out of these 79 galaxies, 9 galaxies were detected in emission only in MUSE fields (but not in \emph{HST} images), one galaxy was detected at the edge of the \emph{HST} image, and three galaxies were detected in the archival \emph{HST} images but were too faint to perform reliable morphological measurements. Therefore, we analyzed the remaining 66 associated galaxies in the 18 \emph{HST} fields.

\begin{table*}
%\begin{center}
%\caption{{\bf } The listed parameters are the object IDs, galaxies' redshift, the k-corrected absolute magnitude, the effective radius, the sersic index, the axis ratio, and the position angle of the galaxies.}
\caption{{\bf Results of our GALFIT morphological modeling, impact parameters (b), and surface brightness of the gas-rich galaxies.} The listed parameters are the quasar name, object IDs from \citet{2022MNRAS.516.5618P}, impact parameters of the quasar sightlines from the galaxy centers, k-corrected absolute magnitude in the UVIS F814W filter, S\'ersic index, effective radius,  axis ratio, position angle (PA) of the major axis, absolute surface brightness averaged within the effective radii of the galaxies and goodness of fit. A complete table of associated galaxies and their properties is available as machine-readable online material.
}

\begin{tabular}{llccccccccc}
\hline\hline
Object & ID &  b & Absolute mag & \emph{n}  &  \emph{$R_{e}$} & \emph{b/a} &  PA  &  $<\mu_{\rm eff}>$& $\chi^2_{\nu}$ \\
    &   &  [kpc] & [F814W]  &       &  [kpc]    &     &  [deg.]    & [mag $\rm arcsec^{-2}$]   &  \\
\hline  

Q0138$-$0005	&	14	&	79.7	&	$	-21.39	\pm	0.02	$	&	$	0.84	\pm	0.01	$	&	$	1.92	\pm	0.02	$	&	$	0.70	\pm	0.01	$	&	$	106.38	\pm	0.74	$	&	$	18.59	\pm	0.03	$	&	1.093	&	\\
Q0152$-$2001	&	13	&	83.8	&	$	-20.60	\pm	0.07	$	&	$	1.88	\pm	0.08	$	&	$	6.12	\pm	0.34	$	&	$	0.57	\pm	0.01	$	&	$	136.06	\pm	1.66	$	&	$	21.90	\pm	0.14	$	&	1.127	&	\\
Q0152$-$2001	&	4	&	170	&	$	-22.29	\pm	0.01	$	&	$	1.53	\pm	0.02	$	&	$	5.01	\pm	0.06	$	&	$	0.30	\pm	0.00	$	&	$	140.92	\pm	0.22	$	&	$	19.77	\pm	0.03	$	&	1.365	&	\\
Q0152$-$2001	&	5	&	60.7	&	$	-21.46	\pm	0.04	$	&	$	1.24	\pm	0.03	$	&	$	6.55	\pm	0.16	$	&	$	0.32	\pm	0.00	$	&	$	154.40	\pm	0.37	$	&	$	21.19	\pm	0.07	$	&	1.136	&	\\

... & ...& ... & ...  & ... & ...  & ... & ... & ... & ... \\

\hline\hline 				       			 	 
\label{Tab2}
\end{tabular}			       			 	 

%\end{center}			       			 	 
\end{table*}

\begin{figure*}
  
    \includegraphics[width=\textwidth]{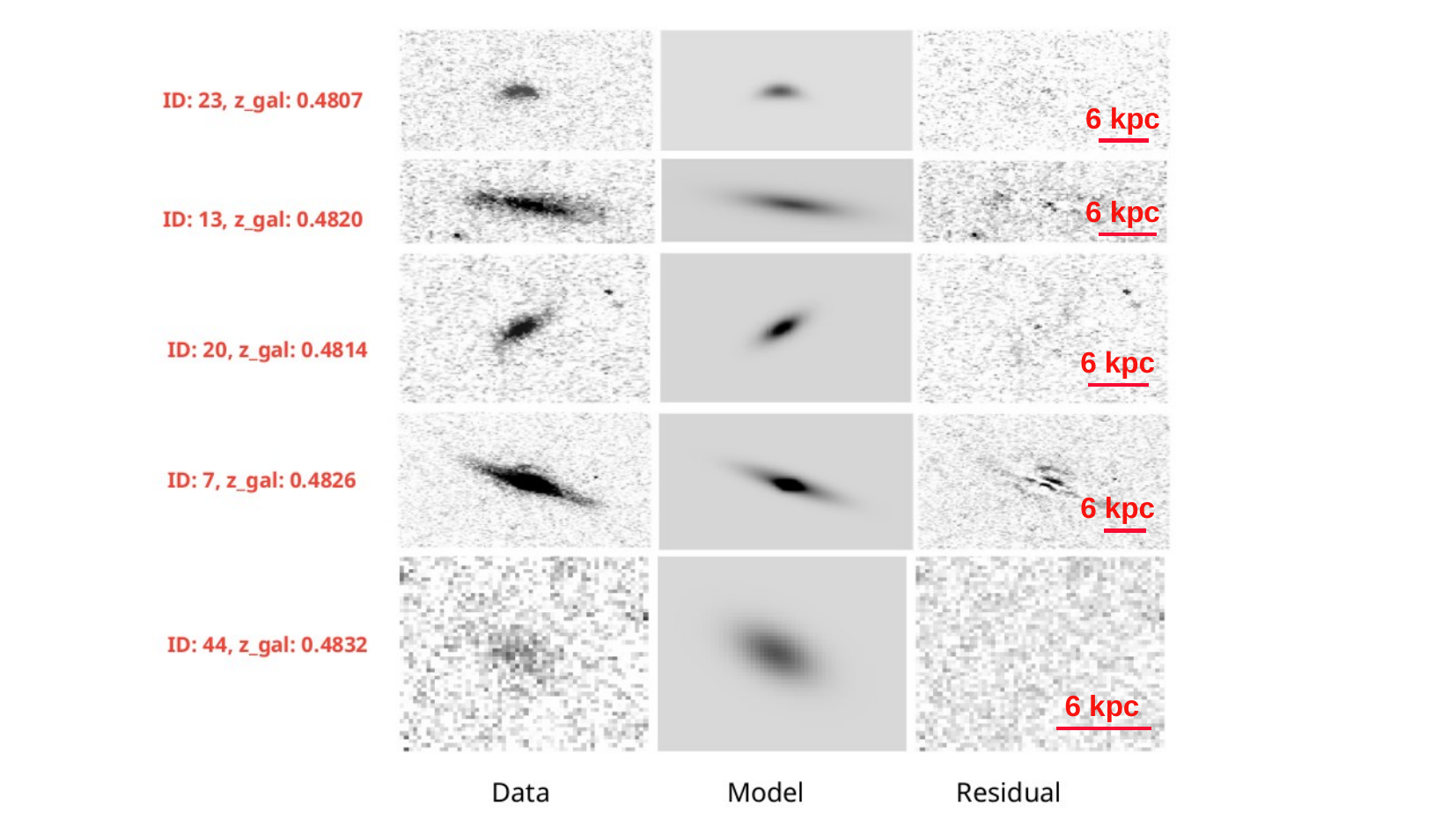}
    \caption{Results of the S\'ersic profile fitting using \textsc{galfit} for the associated galaxies presented in the \autoref{fig1}. The left panels show galaxy images in UVIS/F814W band, the best-fitted S\'ersic model (middle panels), and residual images (right panels). The orientation of the images is the same as \autoref{fig1}.
    }
    \label{fig3}
\end{figure*}

\begin{figure*}
  
    \includegraphics[width=0.85\textwidth]{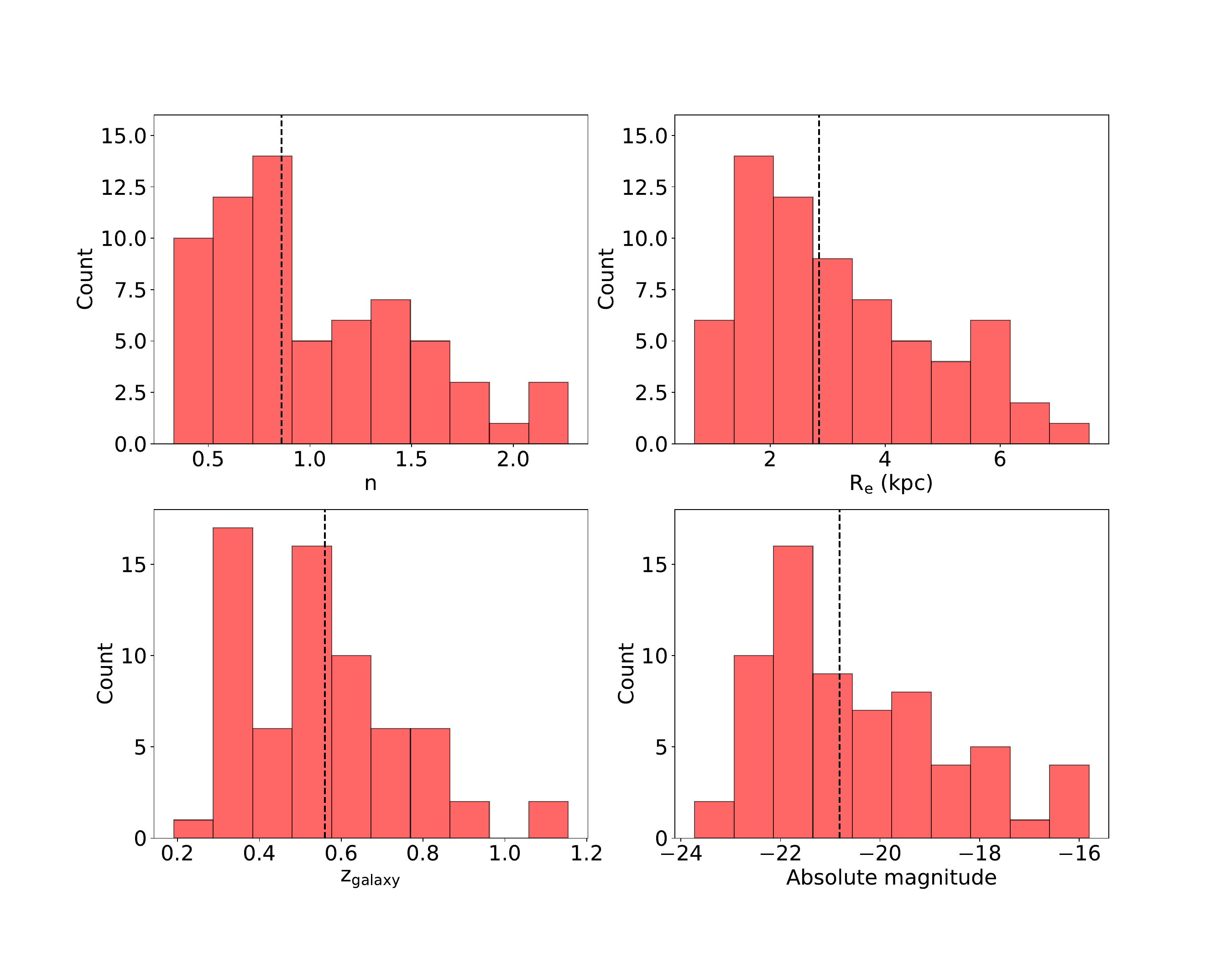}
    \caption{The frequency distribution of the S\'ersic index (upper-left), effective radius (upper-right), galaxy's redshift (lower-left), and the k-corrected absolute magnitude (lower-right) in UVIS/F814W filter of the 66 associated galaxies. In each panel, the vertical dashed black line represents the median value. We used IDL-based software \textsc{kcorrect} to calculate $\rm k$-corrections to the absolute magnitudes.
    }
    \label{fig4}
\end{figure*}

\section{Analysis and Results}
\subsection{Morphological properties of associated galaxies}

The \textsc{galapagos} software  \citep{2011ASPC..442..155H} was used to study the morphologies of the 66 galaxies associated with the absorbers in our MUSE-ALMA halo sample. \textsc{Galapagos} uses \textsc{sextractor} \citep{1996A&AS..117..393B} for object detection and \textsc{galfit}  \citep{2002AJ....124..266P} for two-dimensional image decomposition, and can be run in batch mode to analyze multiple objects. This \textsc{galapagos} analysis was performed on the reddest images obtained for each quasar field because these images in the red or near-infrared filters (WFPC2 F702W, F814W, WFC3 F105W, WFC3 F140W, or NICMOS F160W) provide far more sensitive detection thresholds and enable better deblending of extended sources compared to the bluer filters. The redder bands better sample the cooler and older stars, which in turn more accurately follow the gravitational potential. 

S\'ersic profiles were fitted to each of the sample galaxies. In each case, a cutout region was created with the galaxy centered in the image. Depending on the size of the galaxy, the cutout region ranged from 3" $\times$ 3" to 7" $\times$ 7" in size, and included a sufficient number of pixels at the sky level surrounding the source. Most of the cutout images show only the galaxy without the presence of nearby sources. In cases where nearby sources were present, they were masked using the segmentation map made using \textsc{sextractor}. 
%The PSF (constructed using our observations as described earlier) was adopted. 
%We used \textsc{galapagos} software to run \textsc{galfit} in batch mode. 
%The x and y coordinates of the source are determined using \textsc{sextractor} in Galapagos. 
For each object, the best-fitting values of the morphological parameters were  determined so as  to minimize the residual between the data and the fitted S\'ersic model. In cases where there were significant residuals or the parameter values returned by \textsc{galapagos} had large errors, the morphological parameters were determined by adding more S\'ersic components and individually running \textsc{galfit} iteratively until the parameter converged. As an example,  \autoref{fig3} shows the outcomes of the S\'ersic profile fitting for the five associated galaxies detected in the UVIS/F814W image of the quasar field presented in the \autoref{fig1}. 
%The orientation of these images is similar to that of the image of the quasar field shown in the \autoref{fig1}. 
Each of these five galaxies were well-fitted using a single S\'ersic component.

    %\includegraphics[width = \columnwidth]{Figures/SBVsAbswithZ_crop.pdf}
   
   % \caption{Second image}

 \begin{figure*}
 
    \includegraphics[width = 0.9\textwidth]{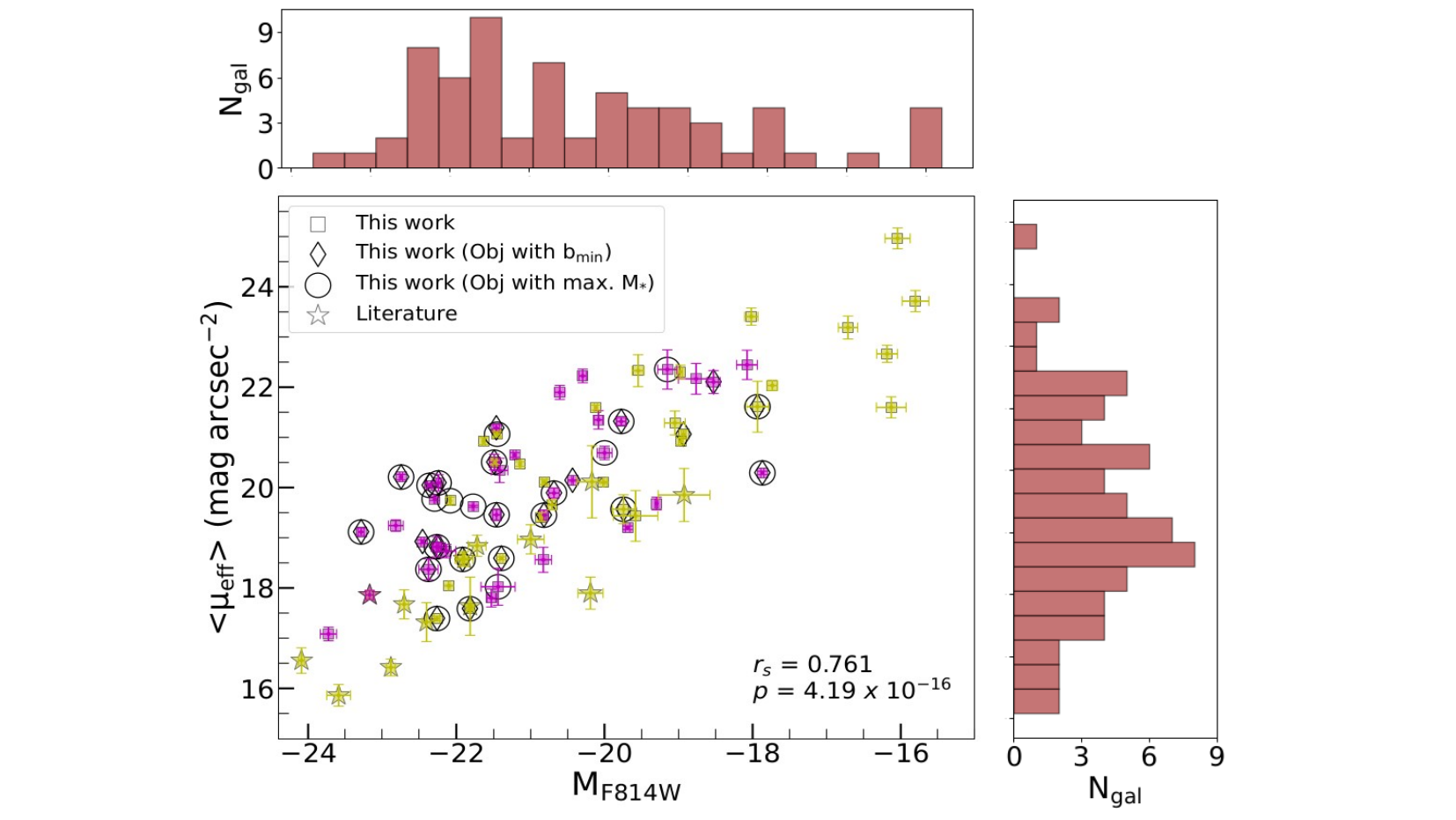}

     \caption{The absolute surface brightness averaged within the effective radius of the galaxies associated with H \textsc{i} absorbers plotted against the absolute magnitude in the UVIS/F814W band. Sample and literature galaxies are divided into two bins in terms of  H \textsc{i} column density using the median value of column density of H \textsc{i} ($N_{\rm HI, med}$ = 6.03 $\times$ $\rm 10^{19}$ $\rm cm^{-2}$). Magenta and yellow symbols denote galaxies with $N_{\rm HI}$ < $N_{\rm HI, med}$ and $N_{\rm HI}$ $\geq$ $N_{\rm HI, med}$, respectively.  Diamonds denote galaxies nearest to the quasar sight lines (in impact parameter) from our \emph{HST} measurements, while open circles  denote the most massive galaxies in each quasar field at the redshift of the absorber.
     The literature values are taken from \citet{2011MNRAS.413.2481F},  \citet{2018MNRAS.478.3120A}, \citet{2021MNRAS.506..546R}. For literature galaxies, we calculated the surface brightness within the $R_{e}$ using the photometric magnitudes listed in the literature papers. The stacked histograms show the distribution of the galaxies in both axes.
    }
  %The green and black dashed lines are the scaling relations between B-band absolute magnitude and surface brightness in the g and z- bands of the early-type galaxies (ETG) of the Virgo Cluster. The scaling relations came from \cite{2006ApJS..164..334F}.

        \label{fig5}
\end{figure*}

\autoref{Tab2} lists the morphological parameters ($R_{e}$, $\emph{n}$, $\emph{b/a}$ and PA of the major axis (in degrees east of north)) along with impact parameter, k-corrected absolute magnitude, absolute effective surface brightness and reduced chi-squared values determined from our analysis for each of the galaxies using \emph{HST} images. The galaxy redshifts are based on the emission line measurements from our MUSE observations, as described in \cite{2022MNRAS.516.5618P}. The absolute magnitude for each galaxy was determined using the apparent magnitude and luminosity distance for the assumed cosmological parameters and $\rm k$-corrected.

%For each galaxy, the luminosity distance was calculated using Ned cosmology calculator \cite{2006PASP.NedCalc}. 
Since our sample consists of galaxies at different redshifts observed in various bandpasses, applying a $\rm k$ correction is essential to make a meaningful comparison among them \citep{2007AJ..Blanton}. The IDL-based software  \textsc{kcorrect} was used to calculate $\rm k$-corrections to the absolute magnitudes. \autoref{fig4} shows the distribution of the morphological properties provided by  \textsc{galfit} and the galaxies redshift of the associated galaxies. For our sample galaxies, the S\'ersic index ranges from 0.33 to 2.27 with a median value of 0.86 $\pm$ 0.07. This suggests that most of the sample galaxies are disks or dwarf spheroidals. Such disk galaxies exhibit an exponential light profile \citep{10.1111/j.1365-2966.2012.20355.x}. The effective radii of the sample galaxies range from 0.68 kpc to 7.55 kpc with a median value of 2.85 $\pm$ 0.15 kpc, and their absolute magnitudes (in F814W filter) range from -15.80 to -23.73 with a median value of -20.81 $\pm$ 0.07. The redshifts of the 
galaxies range from $z$ = 0.19 to $z$ = 1.15, with a median value $z$ = 0.56.

\begin{table*}
\begin{center}
\caption{{\bf Stellar Properties of the gas-rich galaxies and other galaxies from the literature.} The listed parameters are %delta RA, delta DEC, impact parameter, 
the quasar name, galaxy ID, galaxy's redshift, specific star formation rate, SFR surface density, stellar mass surface density and sSFR surface density. The entries marked ``-999''   correspond to measurements that are unavailable. A complete table of associated galaxies and their properties is available as machine-readable online material.
}
\begin{tabular}{llcccccccc}
\hline\hline
Object & ID  & $\rm z_{gal}$  & Log sSFR  & Log $\Sigma_{\rm SFR}$ & Log $\Sigma_{\rm M*}$ & Log $\Sigma_{\rm sSFR}$ \\

&  &     &  [$\rm yr^{-1}$]    &   [M\textsubscript{\(\odot\) } $\rm yr^{-1}$ $\rm kpc^{\rm -2}$] &[M\textsubscript{\(\odot\)} $\rm kpc^{-2}$]    & [$\rm yr^{-1}$ $\rm kpc^{\rm -2}$]\\
\hline                                                        

Q0138-0005	&	14	&	0.7821	&	$	-8.96	\pm	0.21	$	&	$	-0.53	\pm	0.15	$	&	$	8.44	\pm	0.15	$	&	$	-10.33	\pm	0.21	$	\\
Q0152-2001	&	13	&	0.3814	&	$	-11.46	\pm	0.62	$	&	$	-3.33	\pm	0.58	$	&	$	8.13	\pm	0.22	$	&	$	-13.83	\pm	0.62	$	\\
Q0152-2001	&	4	&	0.3814	&	$	-999			$	&	$	-999			$	&	$	9.33	\pm	0.08	$	&	$	-999			$	\\
Q0152-2001	&	5	&	0.3826	&	$	-10.92	\pm	0.17	$	&	$	 -2.10	\pm	0.10	$	&	$	8.82	\pm	0.14	$	&	$	-13.35	\pm	0.17	$	\\

... & ...& ... & ...  & ... & ...  & ...   \\

\hline\hline 				       			 	 
\label{table3}
\end{tabular}			       			 	 
\begin{minipage}{180mm}
\emph{Note:} The SFRs and \emph{M*} for our sample galaxies are from \citet{2023MNRAS.519..931W} and Augustin et al. (in prep.), respectively. The SFR and \emph{M*} for the literature sample are from  
$^a$ \cite{2018MNRAS.478.3120A},
$^b$ \cite{2021MNRAS.506..546R},
$^c$ \cite{2014MNRAS.445..225C},
$^d$ \cite{2019MNRAS.485.1961Z},
$^e$ \cite{2012ApJ...760...49L} and
$^f$ \cite{2018ApJ...857L..12M}.

\end{minipage}
\end{center}			       			 	 
\end{table*}

\subsection{Stellar Properties of associated galaxies}

The SFRs of the associated galaxies were measured using the H$\alpha$ emission line for the sample galaxies with z $\leq$ 0.4, and with [O \textsc{ii}] emission lines for the remaining galaxies (where the H$\alpha$ emission lines fall outside the MUSE wavelength coverage). Dust-corrected SFRs were calculated for 13 galaxies at redshift z $<$ 0.4 with detections of H$\alpha$ and H$\beta$ using the measured H$\beta$ and H$\alpha$ emission-line fluxes. For 15 galaxies, only a 3-$\sigma$ upper limit could be placed on the SFR. A complete description of the SFR estimates and dust correction on SFRs is provided by \cite{2023MNRAS.519..931W}. The stellar masses ($M_{*}$), estimated by using the \emph{HST} broad-band magnitudes and performing spectral energy distribution (SED) fits using the photometric redshift code \textsc{le phare} \citep{1999MNRAS.310..Arnouts,2006A&A...457..Ilbert}, were found to span a  wide range 7.8 $<$ log $M_{*}$ $<$ 12.4.  Further detail about the stellar mass determination is provided in Augustin et al. (in prep).

The SFR surface density and average stellar mass density within the effective radius were calculated as
\begin{equation}
   \qquad \qquad \qquad \qquad  \Sigma_{\rm SFR} = \frac{\rm SFR}{\rm 2 \pi R^2_{e}}
\end{equation}

\begin{equation}
   \qquad \qquad \qquad \qquad  \Sigma_{\rm M_{*}}  = \frac{ M*}{2 \pi \rm R^2_{e}}
\end{equation}

The absolute effective surface brightness,  <$\mu_{\rm eff}$> (mag $\rm arcsec^{-2}$), averaged within the effective radius for our galaxies were calculated based on the $\rm k$-corrected absolute magnitude and the effective radius using the following expression \citep{2005PASA...22..118G}: 

\begin{equation}
   \qquad \qquad \rm <\mu_{\rm eff}>  = \rm M  + 5  \log_{10} \rm (R_{\rm e})   + 38.57 
\end{equation}

\noindent \autoref{table3} lists the derived stellar properties of the galaxy populations.

\begin{table*}
\begin{center}
\caption{{\bf Absorption properties of the gas-rich galaxies and other absorbers from the literature.} Listed are the name of the quasar field, absorber redshift, impact parameter of the nearest galaxy, normalized impact parameter (b/$\rm R_{e}$), H \textsc{i} column density,  rest equivalent widths of Mg II $\lambda$2796 and Fe II $\lambda$2600 and the absorption metallicities. The absorption metallicities listed here are the observed values (based on generally Zn) in the quasar sightlines. 
The last column lists the references for the impact parameter, absorber rest-frame equivalent widths, absorption  metallicity, and 
the H \textsc{i} column density. See the text for more details. The entries marked ``-999''   correspond to measurements that are unavailable. A complete table of absorption properties is available as machine-readable online material.
} 

\begin{tabular}{llllcccccl}
\hline\hline
Object  &  $\rm z_{abs.}$  &   b & b/$\rm R_{e}$  & log $N_{\rm H I}$  & $\rm W_{\rm r}$ Mg II 2796  & $\rm W_{\rm r}$ Fe II 2600  & $\rm [X/H]_{\rm abs}$ &  X & References \\
&  & [kpc] &  & [$\rm cm^{-2}$]&   [\AA] & [\AA]   &    &    & \\
\hline 

Q0138$-$0005	&	0.7821	&	79.70	&	$	41.50	\pm	0.22	$	&	$	19.81	\pm	0.08	$	&	$	1.21	\pm	0.10	$	&	$	1.10	\pm	0.11	$	&	$	0.28	\pm	0.16	$	&	Zn &[1, 4, 21, 30]	\\
Q0152$-$2001	&	0.7802	&	54.30	&	$	9.28	\pm	0.27	$	&	$	18.87	\pm	0.12	$	&	$	0.36	\pm	0.04	$	&	$	$<$ 0.30		$	&	$		-999		$	& $ -999 $	& [1, 30]	\\
Q0152$-$2001	&	0.3830	&	60.70	&	$	9.26	\pm	0.22	$	&	$	$<$ 18.80	$	&	$	0.58	\pm	0.05	$	&	$		-999		$	&	$	-999		$	& $ -999 $	& [1, 5, 30]	\\
Q0152+0023	&	0.4818	&	120.90	&	$	62.64	\pm	2.31	$	&	$	19.78	\pm	0.08	$	&	$	1.34	\pm	0.06	$	&	$	0.88	\pm	0.06	$	&	$		$ -999 $		$	&	$ -999 $ &[1, 30]	\\

... & ... & ... & ... & ... & ... & ... & ... & ... & ...  \\

\hline\hline
\label{table4}
\end{tabular}			       			 	 
\begin{minipage}{180mm}
References: 
$[1]$ This work 
$[2]$ \cite{2020MNRAS.492.2347H},  
$[3]$ \cite{1998A&A...333..841B}, 
$[4]$ \cite{2008MNRAS.386.2209P}, 
$[5]$ \cite{2018MNRAS.480.5046R},
$[6]$ \cite{2007ApJS..171...29P},
$[7]$ \cite{2011MNRAS.413.2481F},
$[8]$ \cite{2019NewA...66....9B},
$[9]$ \cite{2017MNRAS.469.2959K},
$[10]$ \cite{2015MNRAS.447.2738H},
$[11]$ \cite{2017MNRAS.465.1613Z},
$[12]$ \cite{2012MNRAS.419....2B},
$[13]$ \cite{2016MNRAS.463.3021B},
$[14]$ \cite{2008ApJ...681..881W},
$[15]$ \cite{2014ApJ...780..116K},
$[16]$ \cite{2014MNRAS.445..225C},
$[17]$ \cite{2018MNRAS.478.3120A},
$[18]$ \cite{2021MNRAS.506..546R},
$[19]$ \cite{2011MNRAS.410.2237P},
$[20]$ \cite{2019MNRAS.485.1595P},
$[21]$ \cite{2008MNRAS.386.2209P},
$[22]$ \cite{2018A&A...618A.129R},
$[23]$ \cite{1998A&A...333..841B},
$[24]$ \cite{2016MNRAS.463..980R},
$[25]$ \cite{2016MNRAS.457..903P},
$[26]$ \cite{1977ApJ...218...39W},
$[27]$ \cite{2011MNRAS.416.3118K},
$[28]$ \cite{2017MNRAS.464L..56B},
$[29]$ \cite{2001ApJ...560...92C},
$[30]$ \cite{2006ApJ...636..610R},
$[31]$ \cite{2007MNRAS.376..557M},
$[32]$ \cite{2006A&A...457...71L},
$[33]$ \cite{2018ApJ...857L..12M},
$[34]$ \cite{2012ApJ...760...49L},
$[35]$ \cite{2019MNRAS.485.1961Z}
\end{minipage}
\end{center}			       			 	 
\end{table*}

\subsection{Absorption Properties of associated galaxies}

 The MUSE-ALMA Haloes survey analyzed the absorption properties of 32 H \textsc{i} rich absorbers with $N_{\rm HI}$ > $10^{18}$ $\rm cm^{-2}$ using 19 quasar fields. 
 %These 32 absorbers include 5 DLAs, 13 sub-DLAs, and 14 LLSs. 
 %\textbf{The observation of those 18 quasar spectra was done using either HST/FOS, COS, or STIS instrument. The quasar spectra of most of the targets have ground-based optical high-resolution from VLT/UVES, X-Shooter, or Keck/HIRES observations. Detail information about quasar spectra is provided in \cite{2022MNRAS.516.5618P}.} 
 Using the \emph{HST} images of the 18 quasar fields, we detected 66 galaxies within $\pm$ 500 km $\rm s^{-1}$ of the absorber redshift for 25 absorbers.  \autoref{table4} lists the impact parameters of the quasar sightlines from the galaxy centers and the absorption properties along these sightlines for the sample galaxies and the galaxies from the literature. The impact parameters of these 66 associated galaxies are measured from the astrometry of the \emph{HST} images. The measurements of the column density of H \textsc{i}, absorption metallicities, and the equivalent widths of the Fe II $\lambda$ 2600 and Mg II $\lambda$ 2796 absorption lines  are from various references in the literature. 
 
 The absorption metallicities listed for the DLAs and sub-DLAs in the sample (available for 4 out of the 25 absorbers) are based on the Zn abundance without corrections for ionization and (in most cases) dust depletion. The effect of dust depletion is expected to be modest, since Zn is a volatile element that is far less depleted in the interstellar medium (ISM) of the Milky Way compared to refractory elements  such as Fe \citep[e.g.][]{Jenkins_2009, 2011A&A...530A..33V}. Indeed, Zn is often used as the metallicity indicator in DLA/sub-DLAs for these reasons. Ionization corrections are known to be small for DLAs \citep[]{2016ApJ...823...66M,2020ARA&A..58..363P}. For sub-DLAs (which tend to be more ionized than DLAs), the ionization corrections for Zn abundance are found to be $\lesssim$0.2 dex  \citep[e.g.,][]{2009MNRAS.397.2037M}. We, therefore, select only the absorbers with Zn 
 abundance measurements for comparison of the absorption metallicities with the stellar properties of the galaxies in the following sections. Finally, we note that emission-line metallicities are also available for our galaxies, 
 and were  measured using the $R_{3}$ calibration from \citet{2017MNRAS.465.1384C}. A complete description of the galaxy metallicity estimates is provided in \cite{2023MNRAS.519..931W}.

%XXXX Please check absolute magnitudes and $\mu_{eff}$ values are correct. 

%\begin{figure*}
%    \includegraphics[width=\textwidth]{Figures/sfrSDVsSB_NHI.pdf}

  %  \caption{The star formation rate and stellar mass surface densities averaged within effective radius plotted against the average effective surface brightness. Left panel: Plot of $\Sigma_{\rm SFR}$ versus surface brightness. Right panel: Plot of $\Sigma_{\rm M*}$ versus surface brightness. \textbf{Galaxies with $N_{\rm HI}$ < $N_{\rm HI,med}$ are colored with magenta, while the yellow color represents galaxies with $N_{\rm HI}$ $\geq$ $N_{\rm HI,med}$. Diamonds denote galaxies nearest to the quasar sight lines from our \emph{HST} measurements, while open circles stars denote the most massive galaxies of each quasar at the redshift of the absorber.} The stars are the literature values taken from \citet{2011MNRAS.413.2481F}, \citet{2021MNRAS.506..546R} and  \citet{2018MNRAS.478.3120A}. %\textbf{The dashed black and blue lines are the linear fit between the surface brightness and absolute magnitude for the two redshift bins ranging from 0.1 < z < 0.6 and 0.6 $\leq$ z < 3, respectively.}
 %  }
 %   \label{fig7}
%\end{figure*}

\begin{figure*}
    \includegraphics[width=0.75\textwidth]{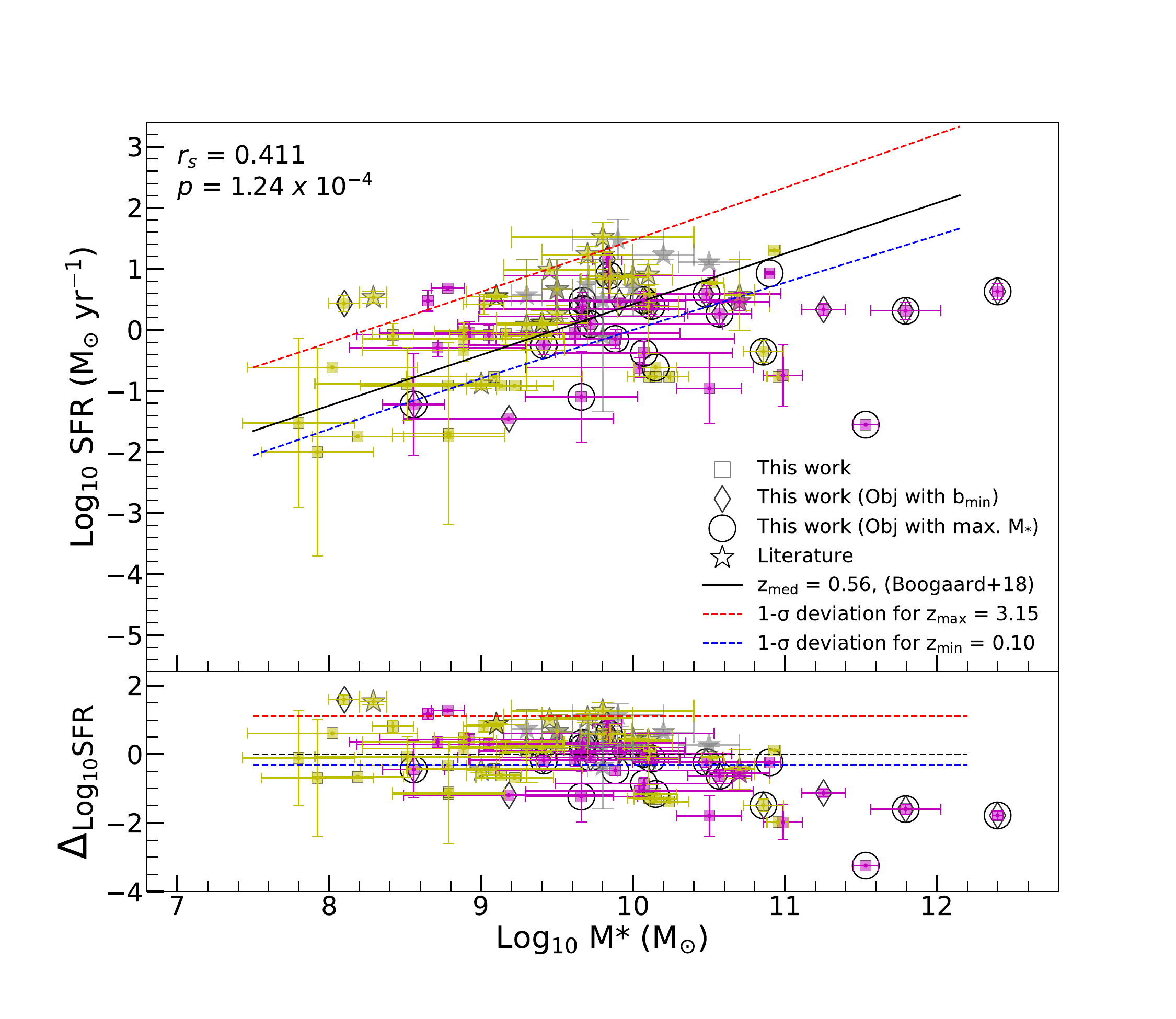}

    \caption{Upper panel: SFR vs. stellar mass for galaxies associated with gas-rich absorbers. The gray stars denote galaxies from \citet{2019MNRAS.485.1961Z} whose H \textsc{i} column density is more uncertain [not based on the actual Ly$\alpha$ absorption profile, but estimated from the Mg \textsc{ii} $\lambda$2796 equivalent width using the relation from \citet{2009MNRAS.393..808M}]. Yellow and magenta stars denote galaxies from \citet{2011MNRAS.413.2481F},
    \citet{2012MNRAS.424L...1K}, \citet{2014MNRAS.445..225C}, \citet{2018MNRAS.478.3120A} and \citet{2021MNRAS.506..546R}. All other symbols are as in \autoref{fig5}. The solid black line shows the star formation main sequence from \citet{2018A&A...619A..27B} at the median redshift of the sample galaxies. The blue and red dashed lines show the 1-$\sigma$ deviations from the SFR-\emph{M*} relations at $z_{\rm min}$ and $z_{\rm max}$ of the full sample of absorber-associated galaxies (including galaxies from both our MAH sample and the literature). Lower panel: plot of the difference in the observed SFR and the expected SFR from the SFMS at $z_{\rm med}$ taken from \citet{2018A&A...619A..27B} at the observed stellar mass vs. the stellar mass. The blue and red dashed lines show the difference relative to the SFMS at $z_{\rm med}$ for the lower and upper 1-$\sigma$ SFMS at $z_{\rm min}$ and $z_{\rm max}$. 
    }
    \label{fig6}
    
\end{figure*}

\begin{figure*}

     \includegraphics[width=\textwidth]{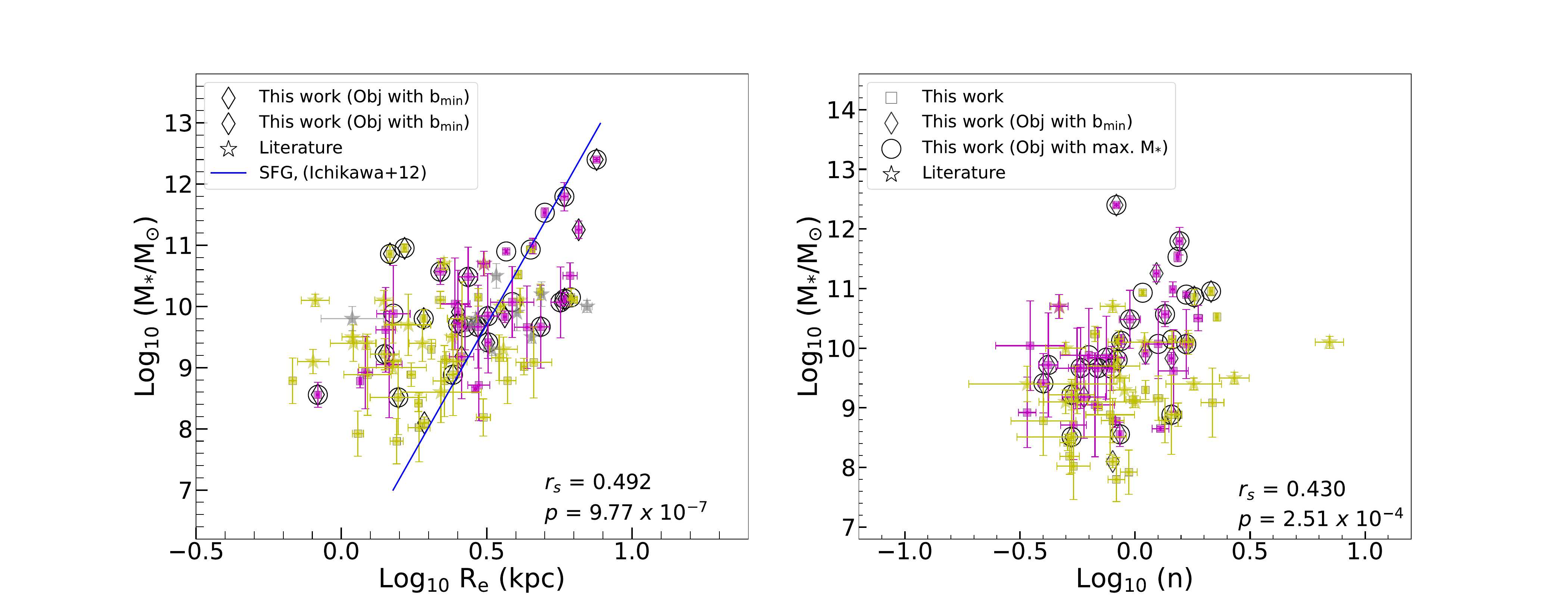} 
    
    \caption{%Morphological parameters of the galaxies in our sample plotted against the absolute magnitude. Upper right: plot of absolute magnitude versus the log of SFR. Upper right: plot of absolute magnitude versus the logarithm of stellar mass. The solid black line is the scaling relation between mass-luminosity of the nearby blue spheroid (BSph) in r-band absolute magnitude taken from \cite{2018MNRAS.475..788M} where the two dashed lines are the $\pm 1\rm \sigma$ deviations. Lower left: plot of absolute magnitude versus the log of \emph{n}. The blue dashed line represents the scaling relationship between the luminosity-concentration of early-type galaxies of the Virgo Cluster in B-band absolute magnitude taken from \cite{2006ApJS..164..334F}. Lower right: plot of absolute magnitude versus the log of \emph{$R_{e}$}. Color coding and symbols are similar to those in \autoref{fig5}. 
    Mass-size and mass-concentration relationships of the associated galaxies. Left panel: plot of stellar mass against the effective radius. The solid blue
    line is the scaling relationship between the mass and size of the star-forming galaxies (SFG) from \citet{2012MNRAS.422.1014I} for the redshift of 0.50 < z < 0.75. Right panel: plot of stellar mass against the S\'ersic index. The plots suggest that massive galaxies are bigger in size and more concentrated. 
    %Galaxies with $N_{\rm HI}$ < $N_{\rm HI,med}$ are colored with magenta, while the yellow color represents galaxies with $N_{\rm HI}$ $\geq$ $N_{\rm HI,med}$. Diamonds denote galaxies nearest to the quasar sight lines from our \emph{HST} measurements, while open circles stars denote the most massive galaxies of each quasar at the redshift of the absorber. The stars are the literature values taken from \citet{2011MNRAS.413.2481F},\citet{2012MNRAS.424L...1K}, \citet{2014MNRAS.445..225C}, \citet{2018MNRAS.478.3120A} and \citet{2021MNRAS.506..546R}. 
    The gray stars in the left panel are the literature sample from \citet{2019MNRAS.485.1961Z} whose H \textsc{i} column density is estimated from the Mg \textsc{ii} $\lambda$2796 equivalent widths using the relation from \citet{2009MNRAS.393..808M}. All other symbols are as in \autoref{fig5}. 
    }

    \label{fig7}
\end{figure*}

%\begin{figure}
%\includegraphics[width=\columnwidth]{Figures/KormendyRelation.pdf}
 %   \caption{The rest-frame surface brightness averaged within effective radii plotted against the logarithm of effective radius for the galaxies with $\emph{n}$ greater than 1.0. \textbf{The galaxies are subdivided into two bins using the median value of the stellar mass (log \emph{M*$_{\rm med}$} = 9.67). Galaxies with log \emph{M*} < log \emph{M*$_{\rm med}$} are colored with blue squares and the galaxies with log \emph{M*} $\geq$ log \emph{M*$_{\rm med}$} are represented with green squares. The diamond symbols are the sample galaxies that lie closest to the quasar sightline, while open circles stars denote the most massive galaxies of each quasar at the redshift of the absorber. The stars are the literature galaxies.} The dashed red line is the linear scaling relation from \cite{Longhetti2007} for z $=$ 0, while the dashed blue line is similar but for our galaxy sample's median value of $\rm z_{gal}$. Larger galaxies appear to be fainter than smaller ones.
%    }
%\label{fig10}
%\end{figure}

\section{Discussion}
Using the high-resolution \emph{HST} images and the integral field spectroscopy provided by MUSE, we have studied and analyzed the morphological and stellar properties of 66 associated galaxies. The two powerful software: \textsc{galapagos} and \textsc{le phare}, provide ample information about the structural and stellar properties of those gas-rich galaxies. The following sections discuss the scientific results, compare the associated galaxies with the general galaxy population, and explore the connection between their stellar and absorption properties.

\subsection{Literature Sample}

To examine how the properties of the galaxies in our sample compare to other absorption-selected galaxies, we use a comparison sample compiled from the literature. Specifically, this literature sample 
consists of 61 galaxies detected at the redshifts of known gas-rich absorbers from 
the IFS surveys of the CGM (see references listed in \autoref{table4}). These literature galaxies range in redshift 
from 0.10 to 3.15, and in impact parameter from $\sim$3 kpc to 88 kpc. The H \textsc{i} column density of some of the literature galaxies \citep[denoted as grey stars in the figures] {2019MNRAS.485.1961Z} are estimated from the Mg \textsc{ii} $\lambda$2796 equivalent widths using the relation from \citet{2009MNRAS.393..808M}. Three absorbers from this literature sample have multiple galaxies at the absorber redshifts \citep{2018MNRAS.478.3120A}. In these cases, if combined measurements of stellar properties were available, we adopted those values and treated those multiple sources as a single source. The stellar masses for the galaxies in the literature sample were derived from SED fitting in most cases, and 
using the tight correlation between $M_{*}$ and a dynamical estimator, i.e. a function of galaxy velocity dispersion  and rotational velocity \citep{2019MNRAS.490.4368S} in a few cases. The SFRs were measured using H$\alpha$, Ly$\alpha$, or [O \textsc{ii}] emission line fluxes. The emission metallicities for the literature sample are from \cite{2014MNRAS.445..225C}, \cite{2018MNRAS.478.3120A}, and \citet{2018A&A...618A.129R}.

\subsection{Correlations between stellar and absorption properties}

To assess the correlations between the various properties of the galaxies associated with the absorbers, we use the  Spearman rank-order correlation method to  evaluate the correlation coefficient ($ r_{s}$) and the probability ($p$) that the observed value of $ r_{s}$ could occur purely by chance. In cases where there is a mixture of detections and limits (i.e., in the presence of censored data), we use the survival analysis method to calculate the Spearman correlation coefficient (as implemented in the Image Reduction and Analysis Facility  \citep[IRAF;][]{1986SPIE..627..733T} task \textsc{spearman}). The survival analysis method uses the Kaplan-Meier estimate of the survival curve to assign ranks to the observations that include censored points. 
%A censor indicator is assigned to the data points while running the task, depending on whether the data point is a detection, upper limit, or lower limit.
Censored points are assigned half (for upper limits) or twice (for lower limits) the rank that they would have had were they uncensored.

Given that the literature sample is based on observations obtained with different 
selection methods, it is useful to ask how much the correlations are affected by the inclusion or exclusion of the literature sample. With this in mind, we computed the $ r_{s}$ and $p$ values between the various properties for both our own sample, and a larger sample after including the literature galaxies. 
\autoref{table5} lists the results of these correlation calculations. In the following subsections, we discuss some of the key implications of our analysis to address two major questions: How do absorber-selected galaxies compare to the general galaxy population? How do stellar and absorption properties relate?

%\begin{figure}
   % \includegraphics[width=\columnwidth]{Figures/MvsSSMD.pdf}
   % \caption{The stellar mass of associated galaxies plotted against the stellar mass surface density averaged over the effective radius. The blue dotted line is the scaling relations for the star-forming galaxies (SFG) at the redshift ranging from 0.25 $\leq$ z $<$ 3. The scaling relation came from \cite{2012MNRAS.422.1014I}.  Due to the star formation process or merging with other galaxies, they grow in mass and size in such a way that  they follow the scaling relation as shown in the figure. Color coding and symbols are similar to those in \autoref{fig5}.  }
    %\label{fig10}

%\end{figure}

 \begin{figure*}
    \includegraphics[width=0.8\textwidth]{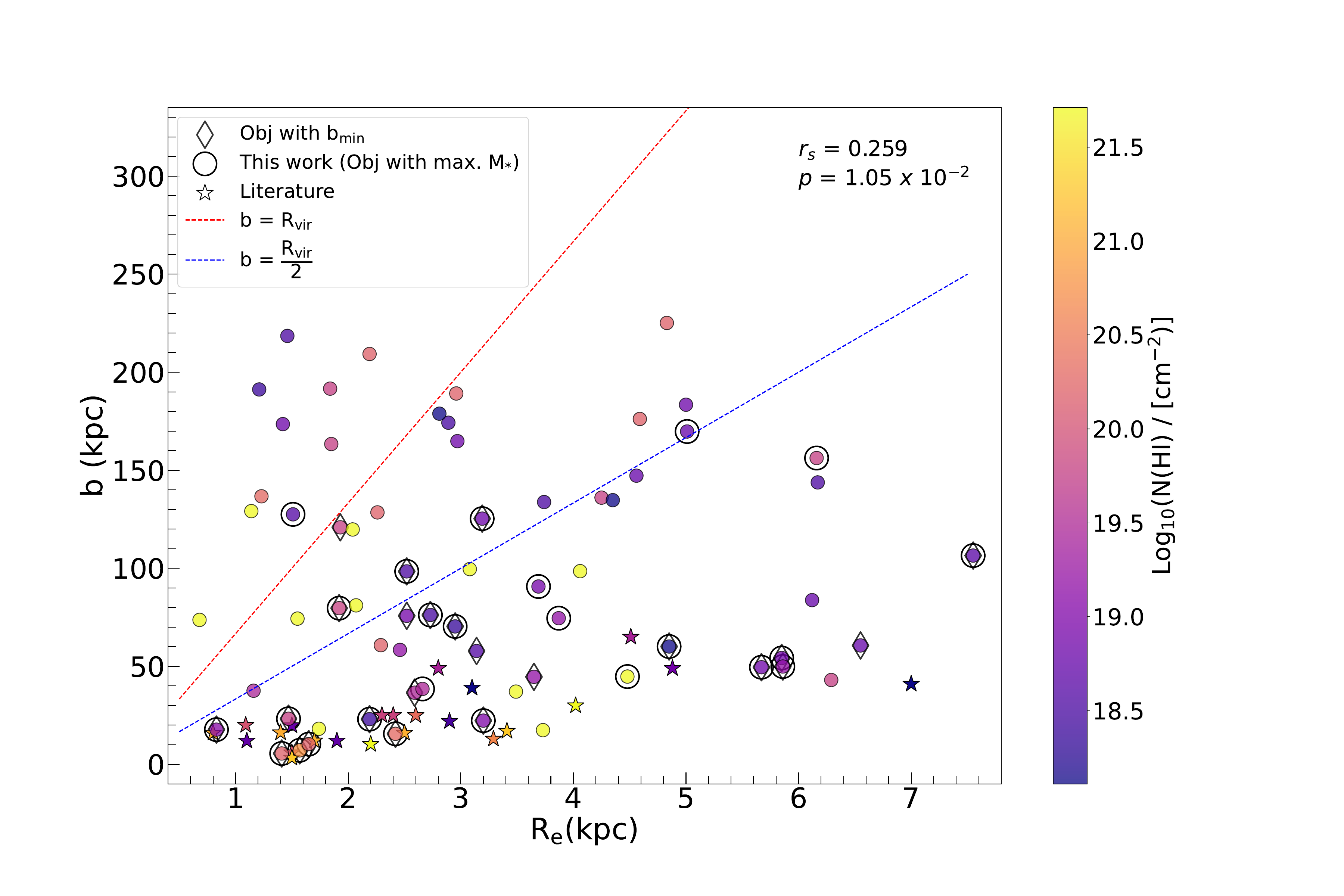}

    \caption{The effective radii plotted against the impact parameter color-coded by column density of neutral hydrogen of the galaxies sample. The diamond symbols are the sample galaxies that lie closest to the quasar sightline, while open circles stars denote the most massive galaxies of each quasar at the redshift of the absorber. The stars are the literature sample galaxies. The red and blue dashed lines correspond to $\rm b=R_{vir}$, and $\rm b=R_{vir}/2$, respectively, taking the approximate relation between the effective radius and the virial radius. From the plot, we see most of the sample galaxies are found at or within the $\rm R_{vir}$ and more than half of the sample galaxies are found at or within the half of the $\rm R_{vir}$. It is also interesting to see that all the galaxies that are closest to the quasar's sightlines are within the virial radius. Most of the gas-rich galaxies are located in the region near the galactic center and fewer gas-rich absorbers are found to be tracing CGM at a larger distance.
    }    
    \label{fig8}
\end{figure*}

\subsection{Do gas-rich galaxies differ from the general population?}

We now compare the absorber-associated galaxies from our sample and the literature with the properties of the overall galaxy population. While making these comparisons, we examine whether there is a difference between galaxies associated with high and low H \textsc{i} column densities, and also between small vs. large impact parameters. The galaxies with the lowest impact parameters may be thought of as the host galaxies of the absorbers  \citep[e.g.][]{2016ApJ...833...39S, 2023MNRAS.519..931W}. In most cases (18 out of 24), the galaxies with the lowest impact parameters are also the most massive galaxies.

%\textbf{The section discusses the comparison between the various morphological and stellar properties of our sample galaxies together with the literature galaxies. Using the median value of column density of H \textsc{i}, we divide the sample and literature galaxies into two bins in terms of the column density of H \textsc{i}. We use magenta color to denote the galaxies with $N_{\rm HI}$ < 6.03 $\times$ $\rm 10^{19}$ $\rm cm^{-2}$, while we use yellow color to represent galaxies with $N_{\rm HI}$ $\geq$ 6.03 $\times$ $\rm 10^{19}$ $\rm cm^{-2}$. Galaxy with the smallest impact parameter is termed as the host galaxy \citep[e.g.][]{2016ApJ...833...39S,2022arXiv221201395W} and is helpful in studying gas distribution in the CGM. Out of 66 associated galaxies, we have found 24 galaxies as the host galaxies, while we have detected 25 galaxies as the most massive galaxies for each quasar field at the redshift of the absorber. We also found that most of the host galaxies (18/24) are also massive galaxies. We denote the host galaxies with diamonds and open circles to represent the most massive galaxies.  } 

\subsubsection{Dependence of $\mu_{\rm eff}$ on luminosity}

%shows the absolute (rest-frame) effective  surface brightness plotted versus the absolute magnitude for our sample galaxies.  %The upper panel shows our sample galaxies color-coded by redshift, divided into three redshift bins with an equal number of galaxies per bin. No substantial difference is observed in the slope of the relation or the scatter with respect to the mean trend between the highest-redshift and intermediate-redshift bins. However, 
%The trend appears flatter for the lowest luminosity galaxies ($M_{}$$\lesssim$-18), especially in the lowest redshift bin. The prevalence of low redshifts for these dwarf galaxies may potentially be a selection effect since dwarf galaxies are more difficult to detect at higher redshifts. We caution, however, that the sample is still not large enough to allow robust comparisons between different subsamples. 

\autoref{fig5} shows the absolute effective rest-frame surface brightness against absolute magnitude relation, where the galaxies are color-coded
by the H \textsc{i} column density of the absorber associated with the galaxy. A clear positive correlation ($ r_{s}$ $=$ 0.76 and \emph{p} $=$ 4.19 $\times$ $10^{-16}$) is observed, consistent with results from past studies \citep[e.g.][]{McConnachie_2012, Karachentsev_2013, 10.1093...Seo}. However, we find no substantial difference between the trends for the lowest and intermediate column density bins. The trend appears to be flatter for the lowest luminosity galaxies, especially for galaxies associated with the highest H \textsc{i} column density absorbers. This finding is consistent with past  suggestions that the highest H \textsc{i} column density absorbers may be associated with dwarf galaxies  \citep{2010NewA...15..735K}. We also analyzed a plot similar to \autoref{fig5} by sub-sampling our sample galaxies into two redshift bins using the median redshift of the sample galaxies. No significant differences are observed between z and $N_{\rm HI}$ sub-samples.

\subsubsection{Dependence of SFR on stellar mass}

\autoref{fig6} shows a plot of the SFR vs. stellar mass, revealing a strong correlation with $ r_{s}$ = 0.41 and $p$ = 1.24  $\times$ $10^{-4}$. Also shown for comparison is the SFR-$M_{*}$ relation, i.e. the star formation main sequence (SFMS) for galaxies at  $\rm z$ $=$ 0.56, the median redshift of our full sample of absorber-associated galaxies \citep[based on the SFMS from ][]{2018A&A...619A..27B}. Since the SFMS evolves with redshift and our full sample covers a wide redshift range, we show the 1-$\sigma$ deviations from the SFMS relations at $\rm z$ $=$ 0.10 and $\rm z$ $=$ 3.15 (the minimum and maximum redshifts of our full sample). 

Most galaxies associated with strong intervening quasar absorbers appear to be consistent with the SFMS within the uncertainties. We note, however, that a  small fraction of high-mass galaxies lie below the SFMS. The lower panel of \autoref{fig6} shows the deviation from the SFMS vs. $M_{*}$. The deviation seems to be highest  for galaxies with the highest stellar mass. Similar conclusions were also reached by \cite{2022ApJ...929..150K}. We note, however, that dust corrections for SFRs were not possible for most of these galaxies (12 out of the 14 galaxies with 
$M_{*} > 10^{10} \rm M_{\odot}$ showing $> 2 \sigma$ deviation from the SFMS), and the true SFRs for these galaxies could be higher.

\begin{figure}

    \includegraphics[width=\columnwidth]{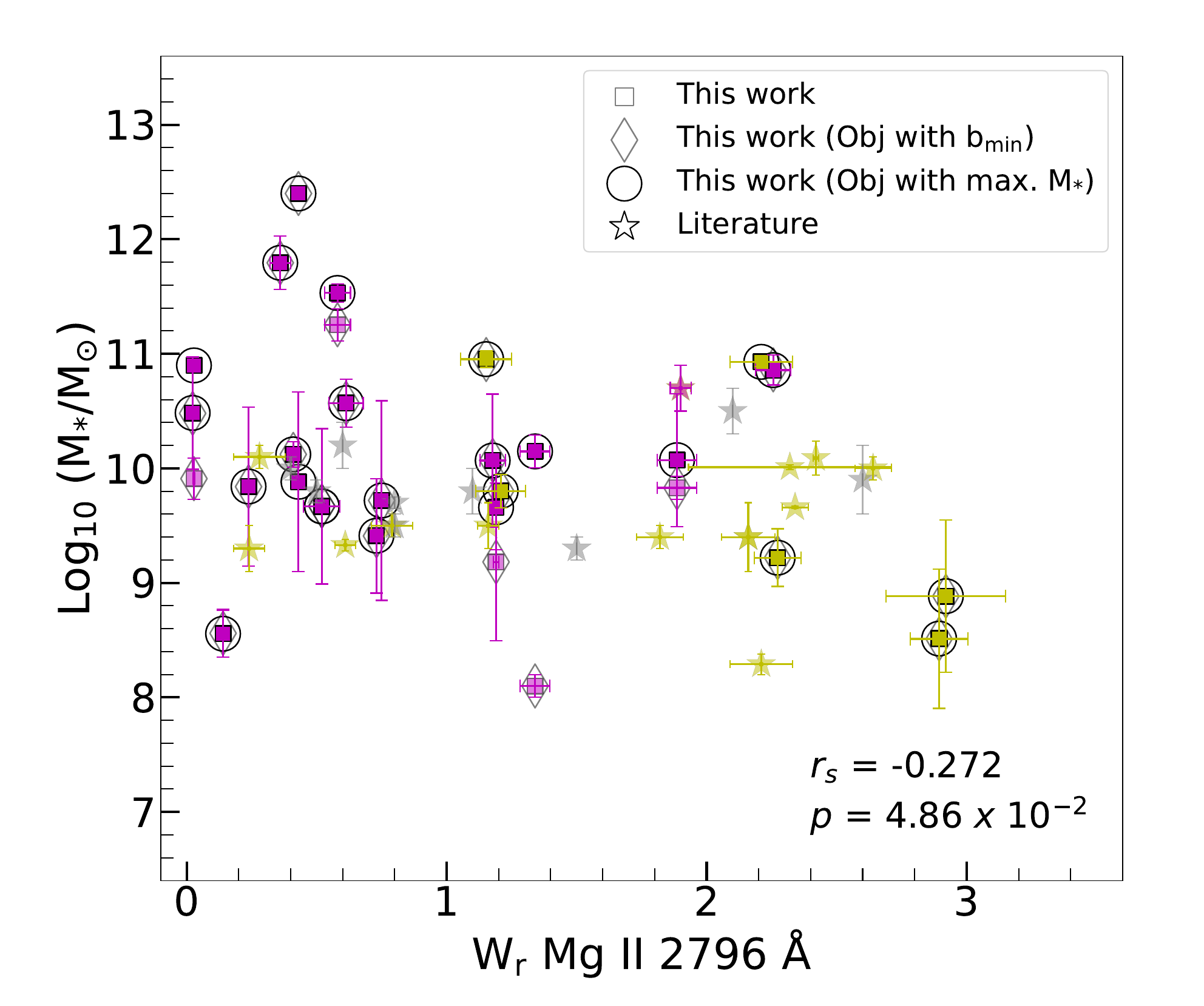}

    \caption{The rest-frame equivalent width of Mg II $\lambda$ 2796  plotted against the stellar mass. All symbols are as in \autoref{fig5}. 
    The stellar mass of the galaxies is anti-correlated with the column density of H \textsc{i} gas \citep{2018MNRAS.478.3120A} while the $N_{\rm HI}$ positively correlates with the equivalent width of the metal lines \citep{2006ApJ...636..610R}. The massive galaxies will have lower metal line strengths, while the less massive galaxies are found to possess strong metal line strengths.
    }

    \label{fig9}
\end{figure}

\subsubsection{Dependence of Stellar mass on S\'ersic index and size}

%The top panels of \autoref{fig6} show plots of the SFR and stellar mass vs. the absolute magnitude. 

\autoref{fig7} shows plots of the stellar mass vs. the effective radius and S\'ersic index. Also shown for comparison is the mass-size scaling relationship for galaxies at  0.50 $<$ $z$ $<$ 0.75 adopted from \citet[e.g.][]{2012MNRAS.422.1014I}. $M_{*}$ appears to be strongly correlated with the effective radius ($ r_{s}$ $=$ 0.49, $p$ = 9.77 $\times$ $10^{-7}$). A similar result was found in previous studies \citep[e.g.][]{2012MNRAS.422.1014I}. A positive correlation is  observed between the S\'ersic index and stellar mass with $ r_{s}$ $=$ 0.43 and $p$ = 2.51 $\times$ $10^{-4}$. 
%This latter result agrees with the previous studies \citep[e.g.][]{2021MNRAS..Lima}. 
The more massive galaxies tend to be more centrally concentrated and, therefore, earlier-type. %Nevertheless, %possess higher SFR compared to late-type galaxies. 
%Hence, the former type of galaxies due to having more star forming rate are luminous than the latter type of galaxies. 
Similar results were also observed in previous studies \citep[e.g.][]{2014ApJ...788...28V, Mowla_2019,2021MNRAS..Lima}. However, we caution that our sample is relatively small, and the correlation between $M_{*}$ and $n$ is  sensitive to the presence of a few galaxies with the largest masses. It seems clear that galaxies with S\'ersic index below 1.2 are primarily below $\rm 10^{10} M_{\odot}$ in stellar mass, while those with higher S\'ersic indices span the full mass range, consistent with observations of a wide mass range among local early-type galaxies (e.g., dwarf spheroidal and giant elliptical galaxies)

To summarize, we have compared the various morphological and stellar properties of our galaxies (which were selected for strong H \textsc{i} absorption) with the properties of the global galaxy population. We find that the absorption-selected galaxies exhibit similar properties as shown by the general population.

\begin{table*}
\begin{center}
\caption{{\bf Results of correlation tests between various properties of our sample galaxies and literature galaxies.} Column 1 lists the parameter pairs for which the correlation is computed. Columns 2, 3, and 4 list the number of paired parameters, the Spearman rank order correlation coefficient ($ r_{s}$), and the probability ($\emph{p}$) that the observed value of $ r_{s}$ could arise by chance for both sample and literature galaxies. Columns 5, 6, and 7 list similar values for our sample galaxies only. The last column lists the figure and the panel number showing the corresponding data, where the first index denotes the figure number, and the second and third indices denote the column number (from left to right) and row number (top to bottom) of the panel in the figure.
}

\begin{tabular}{lccccccc}
\hline\hline
Parameters & $N_{\rm total}$  &  $ r_{\rm s, total}$  &   $p_{\rm total}$  & $N_{\rm sample}$ & $r_{\rm s, sample}$  &  $ p_{\rm sample}$ & Figure panel \\
&  &    &    \\
\hline  

%Significant Correlations:     \\

$\rm M_{\rm F814W}$, $<\mu_{\rm eff}>$  & 79 &  0.761 &  4.19 $\times$ $10^{-16}$ & 66    &  0.734     &  2.17 $\times$ $10^{-12}$    &     5, 1, 1  \\

%log SFR, $ M_{\rm }$  &  65  &  -0.669  &  2.88 $\times$ $10^{-9}$  & 8, 1, 1 \\

log $ M_{\rm *}$, log SFR  &   88   &  0.411 &  1.24 $\times$ $10^{-4}$ &  60   &  0.404 &  2.19 $\times$ $10^{-3}$ &  6, 1, 1  \\

log $ M_{\rm *}$, $R_{\rm e}$      &    97      &   0.492  &  9.77 $\times$ $10^{-7}$  &   60    &   0.578   &  1.33 $\times$ $10^{-6}$  &  7, 1, 1  \\

log $ M_{\rm *}$, $ n$      &     75      &   0.430  &   2.51 $\times$ $10^{-4}$ &  60      &   0.474  &  1.31 $\times$ $10^{-4}$ &   7, 2, 1  \\

%$ M_{\rm }$, $ R_{\rm e}$  & 80   &  -0.276 &  1.39 $\times$ $10^{-2}$ & 8, 2, 2  \\

%$ M_{\rm }$, $ n$ &  75   &  -0.459 &  3.39 $\times$ $10^{-5}$ & 8, 1, 2  \\

%$\Sigma_{\rm M*}$, $<\mu_{\rm eff}>$  &   \textbf{79}  &  \textbf{-0.538}  &   \textbf{1.07} $\times$ $10^{-6}$ & \textbf{60}  &  \textbf{-0.556}  &   \textbf{4.00} $\times$ $10^{-6}$ &  7, 2, 1  \\

%$\Sigma_{\rm SFR}$, $<\mu_{\rm eff}>$ &   \textbf{79}  &   \textbf{-0.638}   &    \textbf{2.09}  $\times$ $10^{-9}$  &  \textbf{60}  &   \textbf{-0.614}   &    \textbf{1.45}  $\times$ $10^{-7}$  &  7, 1, 1   \\

%\emph{i}, $<\mu_{\rm eff}>$ &  \textbf{79} &  \textbf{0.264}  &    \textbf{2.23}  $\times$ $10^{-2}$  & \textbf{66} &  \textbf{0.347}  &    \textbf{4.26}  $\times$ $10^{-3}$  &  6, 1, 1   \\

%$R_{e}$, $<\mu_{\rm eff}>$ &   28  &    \textbf{0.509}  &   \textbf{5.63}   $\times$ $10^{-3}$ &   \textbf{27}  &    \textbf{0.453}  &   \textbf{1.77}   $\times$ $10^{-2}$  & 10, 1, 1   \\

$R_{e}$, \emph{b} &   97  &    0.259  &  1.05  $\times$ $10^{-2}$  & 66  &    0.096  &  4.44  $\times$ $10^{-1}$  &  8, 1, 1   \\

log $M_{*}$, $W_{r}$ (Mg II 2796 \AA) &  51  &    -0.272  &  4.86   $\times$ $10^{-2}$  &  60  &    -0.397  &  1.68   $\times$ $10^{-3}$  &  9, 1, 1   \\

%\emph{$R_{e}$}, $W_{r}$ (Mg II 2796 \AA)&   35  &    -0.562  &   1.50   $\times$ $10^{-3}$  & 13, 1, 2   \\

%\emph{$R_{e}$}, $W_{r}$ (Fe II 2600 \AA)&   19  &    -0.485  &   3.95   $\times$ $10^{-2}$  & 13, 2, 2   \\

%log $N_{\rm HI}$, $<\mu_{\rm eff}>$ &   37  &    -0.405  &   1.50   $\times$ $10^{-2}$  & 18, 1, 1   \\

log $ M_{\rm *}$, $\rm [X/H]_{emi}$ &  70 &    0.434  &  5.68  $\times$ $10^{-3}$  &   35 &    0.330  &  8.10  $\times$ $10^{-2}$   & 10, 2, 1   \\

log $\rm sSFR$, $\rm [X/H]_{emi}$ &   70   &    -0.401  &  2.07  $\times$ $10^{-2}$ &   35 &    -0.246  &  2.37  $\times$ $10^{-1}$  & 10, 2, 3   \\

%Insignificant Correlations:     \\

log $\rm SFR$, $\rm [X/H]_{emi}$ &   70 &    -0.012  &  9.42  $\times$ $10^{-1}$   &   35 &    0.170  &  3.88  $\times$ $10^{-1}$   &   10, 2, 2  \\

log $N_{\rm HI}$, \emph{$b/R_{e}$} &  45  &    -0.556  &   2.00  $\times$ $10^{-4}$ &   31  &    -0.393  &   3.14  $\times$ $10^{-2}$  & 12, 1, 1   \\

log $N_{\rm HI}$, \emph{$R_{e}$} &   45  &    -0.517  &   6.00   $\times$ $10^{-4}$ &   31  &    -0.326  &   7.45  $\times$ $10^{-2}$   & 12, 2, 1   \\

log $N_{\rm HI}$, log $\Sigma_{\rm sSFR}$ &    41  &  0.418  &   8.30   $\times$ $10^{-3}$ &   22  &   0.121  &   5.23   $\times$ $10^{-1}$  & 13, 1, 2   \\

%log $N_{\rm HI}$, log $\Sigma_{\rm M*}$ &   \textbf{49}  &    \textbf{-0.212}  &   \textbf{1.77}   $\times$ $10^{-1}$ &   \textbf{30}  &    \textbf{-0.169}  &   \textbf{3.62}   $\times$ $10^{-1}$  & 16, 1, 1   \\

%log $M_{*}$, $W_{r}$ (Fe II 2600 \AA)&   \textbf{39} &    \textbf{-0.183}  &     \textbf{2.84} $\times$ $10^{-1}$  &   \textbf{17} &    \textbf{-0.177}  &     \textbf{4.79} $\times$ $10^{-1}$  & 12, 2, 1   \\

%$<\mu_{\rm eff}>$, $\rm [X/H]_{abs}$&   18 &    -0.228  &     3.48  $\times$ $10^{-1}$  & 18, 2, 1   \\

%log $\rm sSFR$, $\rm [X/H]$&   14 &    -0.396  &    1.54  $\times$ $10^{-1}$  & 14, 2, 2   \\

log $N_{\rm HI}$, log $\Sigma_{\rm SFR}$ &   42  &    0.287  &   6.65   $\times$ $10^{-2}$ &   23  &    -0.215  &   3.14   $\times$ $10^{-1}$  & 13, 1, 1  \\

$b/R_{e}$, $\rm [X/H]_{abs}$&   20 &    0.433   &    5.12  $\times$ $10^{-2}$  &  5 & & & 11, 2, 1   \\

\emph{b}, $\rm [X/H]_{abs}$&   28 &    0.423  &    2.80  $\times$ $10^{-2}$  &  5 &  & & 11, 1, 1   \\

log $ M_{\rm *}$, $\rm [X/H]_{abs}$&  25 &    0.436  &    2.58  $\times$ $10^{-2}$  &   5 & & & 10, 1, 1   \\

log SFR, $\rm [X/H]_{abs}$&  18   &    -0.447  &    6.52  $\times$ $10^{-2}$  &   4  & & & 10, 1, 2   \\

log sSFR, $\rm [X/H]_{abs}$&    17   &     -0.597  &   1.11  $\times$ $10^{-2}$  &   4 & & & 10, 1, 3   \\

log sSFR, log $ M_{\rm *}$  &  80  &  -0.465  &  1.38 $\times$  $10^{-5}$ &   55 & -0.580 &   3.53  $\times$  $10^{-6}$  \\

\hline\hline 				       			 	 
\label{table5}
\end{tabular}
\begin{minipage}{180mm}

\textit{Notes.}  $\rm [X/H]_{abs}$ is the absorption metallicity and $\rm [X/H]_{emi}$ is the emission metallicity. Some rows do not report $ r_{\rm s, sample}$  and   $p_{\rm sample}$ values since the number of paired parameters is small, which is insufficient to perform correlation tests. 

%\footnotetext{\textit{a}: $\rm [X/H]_{abs}$ is the absorption metallicity.}

%\footnotetext{\textit{b}: $\rm [X/H]$ is the estimated metallicity at the center of the galaxy after correcting for the metallicity gradient.}

\end{minipage}

\end{center}			       			 	 
\end{table*}

\subsection{How do stellar and absorption properties relate?}

We now examine the stellar properties of the galaxies within the velocity range of $\pm$ 500 km $\rm s^{-1}$ of the absorption redshift with the absorption properties. We only include the absorbers with reliable metallicities for comparisons of absorption metallicities and stellar properties of the galaxies.  To perform these comparisons, we selected galaxies with the smallest impact parameters from the quasar sightlines and the most massive galaxies in each quasar field detected within $\pm$ 500 km $\rm s^{-1}$ of the absorption redshift. We note that it is not possible to consider all the galaxies in such cases or even use the average values of the stellar properties of all the galaxies since doing so would either underestimate or overestimate the true values, given that the impact parameters of the individual galaxies are different.

%Selecting all the galaxies found within the absorber's redshift would lead to multiple values of stellar properties such as SFR, \emph{M*}, and b/$\rm R_{e}$ for each quasar field. This would make it more challenging to compare the stellar properties with the absorption properties. 
%We note that it is not possible to consider all the galaxies in such cases, or  even use the average values of the stellar properties of all the galaxies, since doing so would either underestimate or overestimate the true values, given that the impact parameters of the individual galaxies are different.

%\autoref{table4} lists down some of the absorption properties of the sample galaxies and the literature sample. 

\subsubsection{Impact parameter and galaxy's effective radius}

\begin{figure*}
    \includegraphics[width=0.85\textwidth]{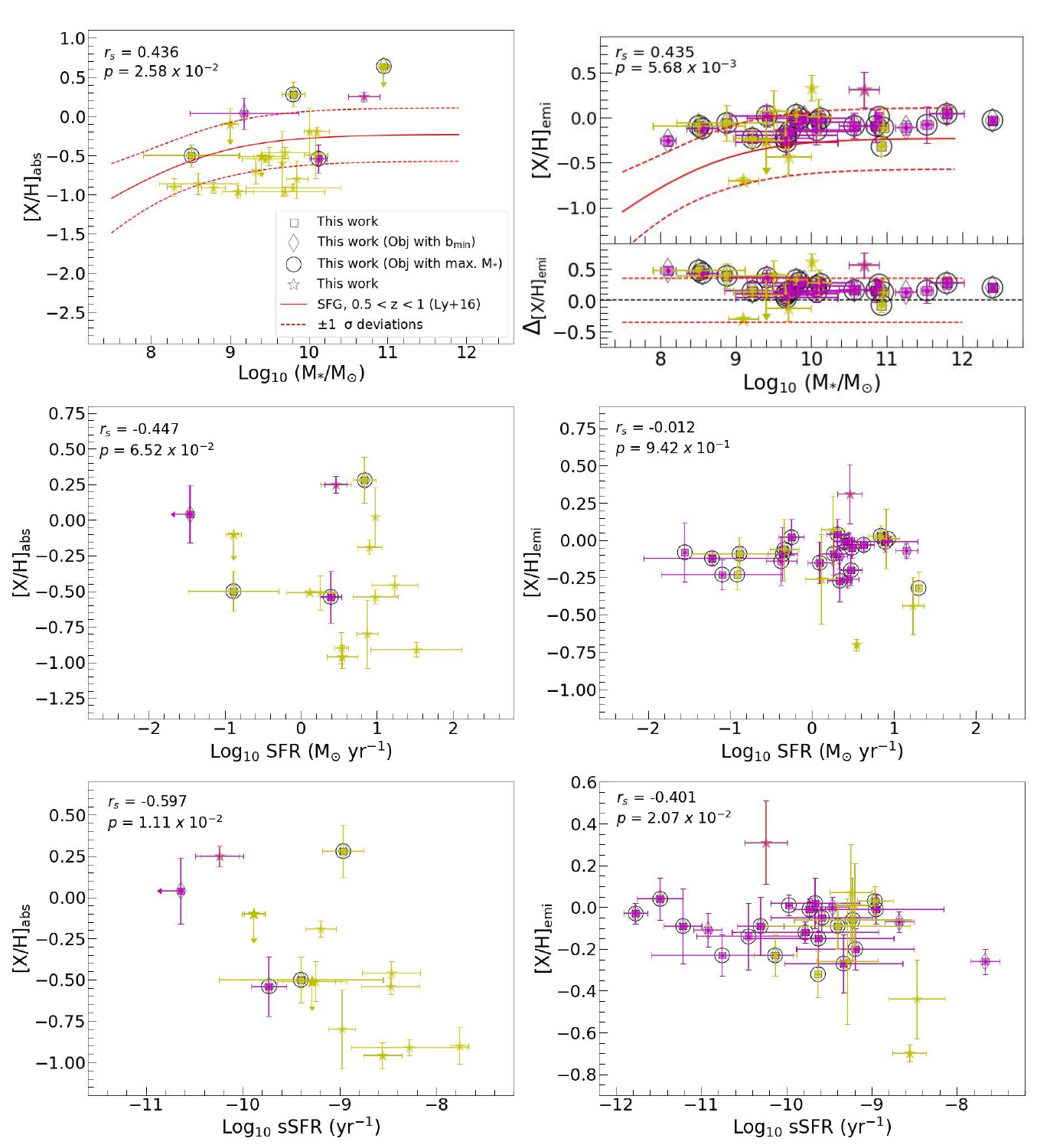}
    \caption{Absorption and emission metallicities plotted against stellar properties of gas-rich galaxies. Left panel: Absorption metallicities plotted against the stellar mass, star formation rate, and specific star formation rate (from top to bottom). All symbols are as in \autoref{fig5}. The absorption metallicities are based on Zn for most cases, while for one absorber, we adopt dust-free absorption metallicity. Right Panel: Emission metallicities plotted against the stellar mass, star formation rate, and specific star formation rate (from top to bottom). The  solid and dashed black lines show, respectively, the mass-metallicity relation for 0.5 < $z$ < 1.0 from \citet{2016ApJ...828...67L} and the $\pm$1-$\rm \sigma$ uncertainties in this relation. The subpanel at the bottom of the upper right panel shows the difference in the observed emission metallicities and the expected emission metallicities from the MZR \citep{2016ApJ...828...67L} plotted vs. the stellar mass.
    }
    
    \label{fig10}
\end{figure*}

    %Galaxies with $N_{\rm HI}$ < $N_{\rm HI,med}$ are colored with magenta, while the yellow color represents galaxies with $N_{\rm HI}$ $\geq$ $N_{\rm HI,med}$. Diamonds are the galaxies nearest to the quasar sight lines from our \emph{HST} measurements, while open circles stars denote the most massive galaxies of each quasar at the redshift of the absorber. Stars denote the literature values taken from \citet{2011MNRAS.413.2481F}, \citet{2014MNRAS.445..225C}, \citet{2018MNRAS.478.3120A} and \citet{2021MNRAS.506..546R}.

\autoref{fig8} shows the impact parameter versus the effective radius for our galaxies and other absorption-selected galaxies from the literature. For reference, the red and blue dashed lines correspond to $\rm b=R_{vir}$, and $\rm b=R_{vir}/2$, respectively, assuming the approximate relation between the effective radius and the Virial radius \citep{2013ApJ...764L..31K}. Most of the  sample galaxies ($\sim$85 $\%$) lie below $\rm b=R_{vir}$, and more than half of the sample galaxies ($\sim$59 $\%$) lie below $\rm b=R_{vir}/2$. All of the galaxies at the smallest impact parameters (i.e., the most probable host galaxies) are located below $\rm b=R_{vir}$ while almost all  massive galaxies ($\sim$96 $\%$) are present below the red dash line, as shown in the \autoref{fig8}.
Given the approximate relation between the effective radius and the Virial radius \citep{2013ApJ...764L..31K}, $\rm b/R_{e}\sim$70 corresponds to $\rm b\sim R_{vir}$. It is thus interesting to note that most of the galaxies have impact parameters below $\rm R_{vir}$ and that the galaxies with impact parameters larger than $\rm \sim0.3 ${ }$\rm R_{\rm Vir}$ are almost all below the sub-DLA  limit in H \textsc{i} column density \citep{2014A&A...566A..24N}. While most DLAs and sub-DLAs are associated with galaxies at impact parameters less than 0.2 $\rm R_{vir}$, a small fraction ($\sim$26 $\%$) of sample galaxies have impact parameters in the range of 0.5 $\rm R_{vir}$ to $\rm R_{vir}$. All of the galaxies from the  literature lie below $\rm b=R_{vir}/2$. This suggests that while these gas-rich absorbers usually trace regions close to galaxy centers, they occasionally trace the CGM at large distances extending out to the Virial radius.

\subsubsection{Dependence of Mg II $\lambda$ 2796 rest-frame equivalent width on stellar mass}

Studies of the interdependence of the Mg II $\lambda$ 2796 absorption strength and the galaxy's stellar properties have reached different conclusions. \citet{2011ApJ...743...10B} reported that the Mg II equivalent widths [estimated from low-resolution spectra of 
zCOSMOS galaxies, obtained with the Visible Multi-Object Spectrograph (VIMOS) on the Very Large Telescope (VLT), by fitting a single Gaussian profile across the Mg II $\lambda$ 2796 and $\lambda$ 2803 lines]  increase with stellar mass for blue galaxies at lower impact parameter (b $<$ 50 kpc). However, such dependence is absent in the red galaxies and even in blue galaxies with larger impact parameters (b $>$ 65 kpc). \citet{2013ApJ...776..115N} suggested  that more massive galaxies have larger Mg II equivalent width in the MAG\textsc{ii}CAT sample, based on a positive correlation between the K-band luminosity (assumed to be a proxy of $M_{*}$) and the Mg II equivalent width.

For impact parameters smaller than 50 kpc, \citet{2014ApJ...795...31L} showed that the Mg II $\lambda$ 2796 equivalent width shows a positive correlation with \emph{M*} for star-forming galaxies but not for  passive galaxies. Taking the larger dispersion values in median Mg II $\lambda$ 2796 equivalent widths,  \citet{2018ApJ...853...95R} found that the median Mg II equivalent widths increase with the stellar mass for blue galaxies at a transverse distance 30 kpc $<$ $\rm R_{\perp}$ $<$ 50 kpc, however the median Mg II $\lambda$ 2796 equivalent widths values for red galaxies are found to be smaller compared to those for high-mass blue galaxies. Including both detections and upper limits in Mg II $\lambda$ 2796 equivalent widths measurements, \citet{2020MNRAS.499.5022D} reported a weak positive correlation between Mg II $\lambda$ 2796 equivalent widths and \emph{M*} in the MAGG sample.  

For our MAH sample and literature galaxies, we find a weak negative correlation ($ r_{s}$ $=$ -0.27, $p$ = 4.86 $\times$ $10^{-2}$) between the Mg II $\lambda$ 2796 equivalent widths and stellar mass. Part of the reason for the difference between this trend and the weak or positive relations found for Mg II absorbers seems to be the difference in the selection techniques. Our sample absorbers are selected by high $N_{\rm H I}$, and therefore have higher Mg II $\lambda$ 2796 equivalent widths. 
For example, $\sim 70\%$ of the absorbers in our sample have Mg II $\lambda$ 2796 rest equivalent widths $>$ 0.5 {\AA}, while only $< 15\%$ of the absorbers in the sample of \citet{2021MNRAS.508.4573D} fall in this category. Moreover, $\sim 60\%$ of the galaxies in our sample are at impact parameters less than 100 kpc, while only $< 10\%$ of the galaxies in the sample of \citet{2021MNRAS.508.4573D} fall in this category. We also note that our $M_{*}$ values have substantial uncertainties, making a robust detection of the trend difficult. \autoref{fig9} shows plots of Mg II $\lambda$ 2796 absorption strengths versus the stellar mass. The H \textsc{i} column density is known to be positively correlated with the Mg II and Fe II absorption strengths \citep[e.g.][]{2006ApJ...636..610R}. 
Furthermore, the H \textsc{i} column density is negatively correlated with the stellar mass  and with metallicity, suggesting that the lower $N_{\rm H I}$ absorbers are associated with more massive galaxies that have had high past star formation and gas consumption \citep{2010NewA...15..735K, 2018MNRAS.478.3120A}. 
The negative correlation seen in the \autoref{fig9} is thus expected.

\begin{figure*}
    \includegraphics[width=\textwidth]{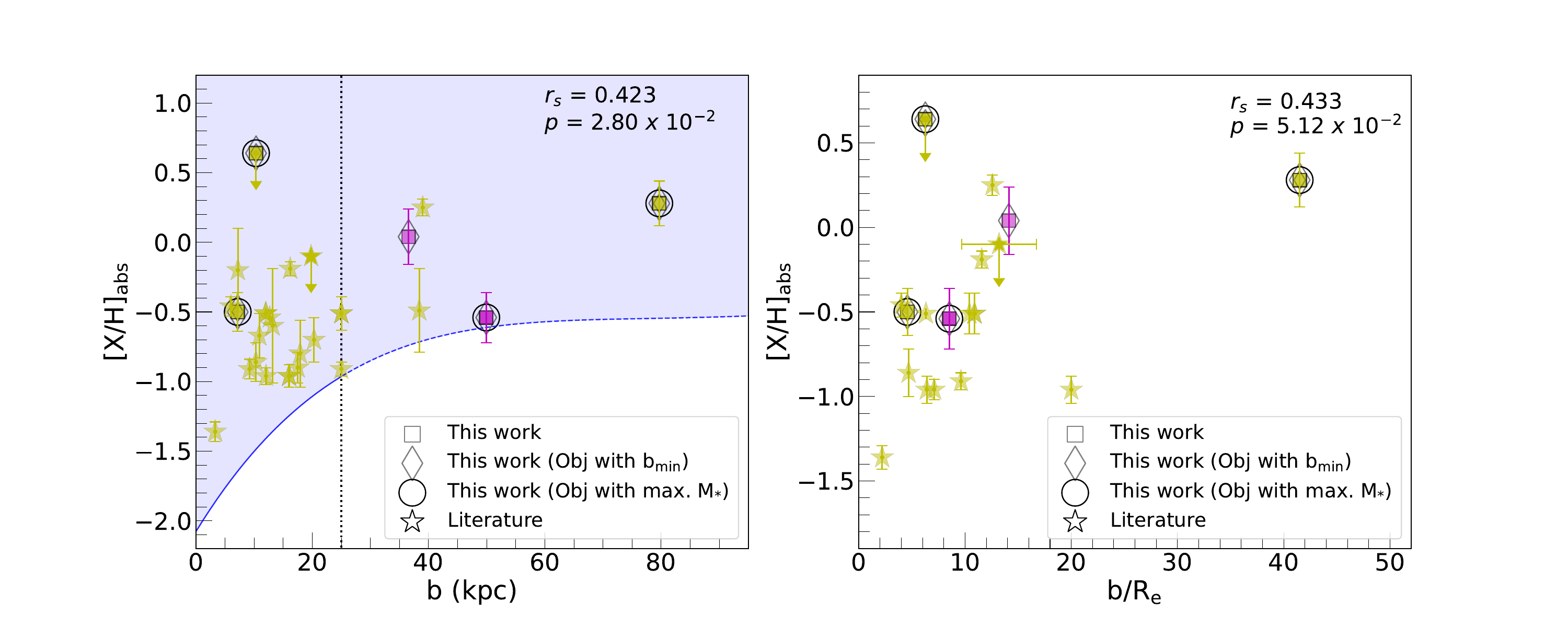}
    \caption{Absorption metallicities plotted versus the impact parameter (left) and the normalized impact parameter (right). 
    All symbols are as in \autoref{fig5}. 
    The light blue region below the solid blue line in the left panel figure denotes the simulated distribution of impact parameters plotted against the absorption metallicities for using the model described in \citet{2008ApJ...683..321F} for the DLA galaxies at $z$ $=$ 3 and is taken from  \citet{2012MNRAS.424L...1K}. Extrapolating the model's prediction, the region is further extended up to the impact parameters of 100 kpc to include the sample galaxies detected at the larger impact parameters and is denoted by a light blue region below the dashed blue line. The trends suggest that metal-poor strong H \textsc{i} absorbers are found in smaller haloes and, therefore, at smaller impact parameters  \citep{2012MNRAS.424L...1K}.
    }
    
    \label{fig11}
    
\end{figure*}

\subsubsection{Dependence of metallicity on stellar properties}

The left panels of \autoref{fig10} show plots of absorption metallicities vs. stellar properties ($\emph{M}_{*}$, SFR and sSFR) of the galaxies. For most cases, we use  absorption metallicities based on Zn, which is less depleted on dust grains. For one absorber, we use the dust-free absorption metallicity inferred from multiple elements using the method of \cite{Jenkins_2009} based on depletion trends observed in the local interstellar medium of the Milky Way. The right panels of \autoref{fig10} show similar plots, but for the emission metallicities [from \citet{2023MNRAS.519..931W} for our sample and from \cite{2014MNRAS.445..225C}, \cite{2018MNRAS.478.3120A}, \citet{2018A&A...618A.129R} for the literature sample]. Also shown, for reference, in the left and right upper panels of \autoref{fig10} are the mass-metallicity relation (MZR) for star-forming galaxies at 0.5 < $z$ < 1.0 and the 1-$\rm \sigma$ uncertainties in this relation \citep[][]{2016ApJ...828...67L}. The bottom sub-panel in the upper right panel shows the difference in the observed emission metallicities and the expected emission metallicities based on the MZR from \citet{2016ApJ...828...67L} at the observed stellar mass. It is clear that both the absorption-based metallicity (away from the galaxy center, at the impact parameter of the corresponding quasar sight line) and the emission-based metallicity (typically at the galaxy center) are positively correlated with the stellar mass, and are generally consistent with the MZR for star-forming galaxies.

%For some of the literature galaxies, we used the metallicities that were estimated at the center of each galaxy. Such metallicities were calculated by adding a term, $\Gamma b$, to the absorption metallicity in order to correct for the effect of the metallicity gradient $\Gamma$, and adopting $\Gamma = 0.023 $ as described in \cite{2014MNRAS.445..225C}. }

%We note, however, that the impact parameters of our galaxies cover a wider range (6 to 225 kpc) compared to the range (6 to 39 kpc) of impact parameters of the galaxies analyzed in \cite{2014MNRAS.445..225C}; so adopting the value of $\Gamma$ from \cite{2014MNRAS.445..225C} may not be completely appropriate. 

\begin{figure*}
    \includegraphics[width=\textwidth]{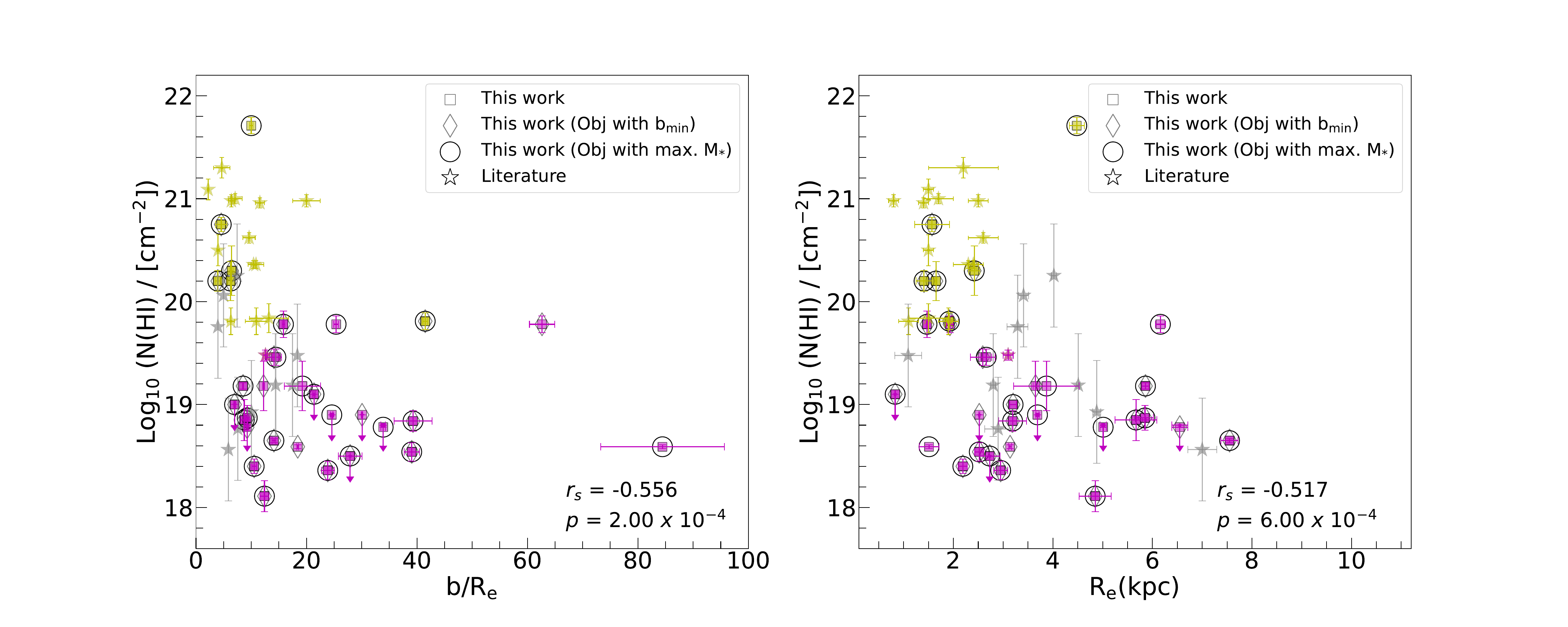}
    \caption{The normalized impact parameter (b/$ \rm R_{e}$) and the effective radius  plotted against the column density of H \textsc{i} gas. Left Panel: Plot of b/$ \rm R_{e}$ against the log of $N_{\rm HI}$. The plot infers that the H \textsc{i} column density is higher for the galaxies found at low-impact parameters to the quasar's sightline. At the same time, there is low H \textsc{i} column density in the regions located at larger distances. Right Panel: Plot of $ \rm R_{e}$ against the log of $N_{\rm HI}$. The gas-rich galaxies are smaller in size than the gas-poor galaxies. All symbols are as in \autoref{fig5}. 
    The stars are the literature values taken from \citet{2011MNRAS.413.2481F},
    \citet{2012MNRAS.424L...1K}, \citet{2014MNRAS.445..225C}, \citet{2018MNRAS.478.3120A} and \citet{2021MNRAS.506..546R}. The gray stars are the literature galaxies from \citet{2019MNRAS.485.1961Z} whose H \textsc{i} column density is estimated from the Mg \textsc{ii} $\lambda$2796 equivalent widths using the relation from \citet{2009MNRAS.393..808M}.
    }
    \label{fig12}
\end{figure*}

The middle panels of \autoref{fig10} show plots of the absorption and emission metallicity vs. the SFR. No correlation is observed between metallicity (both absorption and emission metallicity) and SFR. The bottom panels of \autoref{fig10} show plots of the absorption and emission metallicity vs. the specific star formation rate (sSFR). Both absorption and emission metallicity are negatively correlated with the sSFR. This suggests that the low-metallicity absorbers are associated with galaxies that are forming stars (and consuming  gas) more vigorously. However, they are less enriched due to low past star formation activity. While the plot suggests that the negative correlation is mostly driven by the presence of literature galaxies, it is essential to expand the sample to verify this correlation.

%; but some of the galaxies appear to be more metal-rich than predicted from the MZR, while some of the absorbers appear to be less metal-rich than the MZR prediction. These inconsistencies may arise partly from the use of $\Gamma$ derived by \cite{2014MNRAS.445..225C} for a narrower range of impact parameters.

\begin{figure*}
    \centering
    \includegraphics[width=\textwidth]{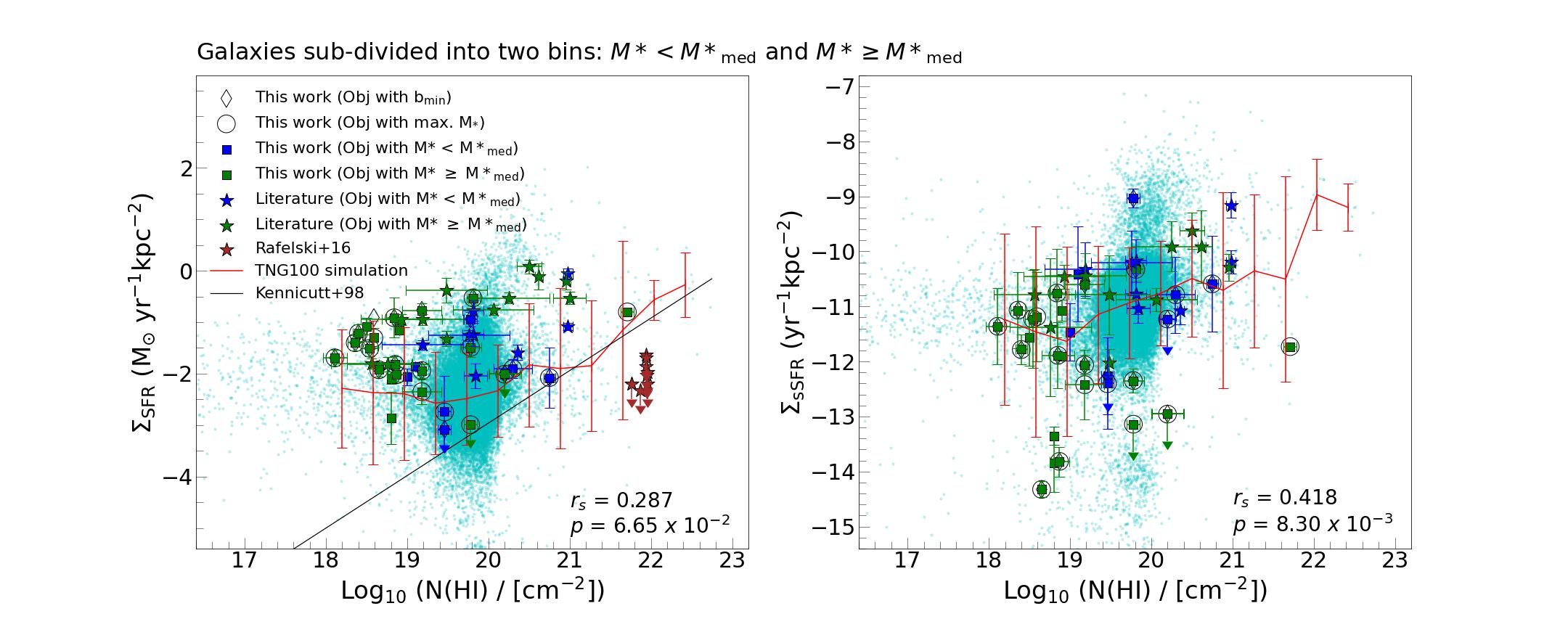}
    \centering
    \includegraphics[width=\textwidth]{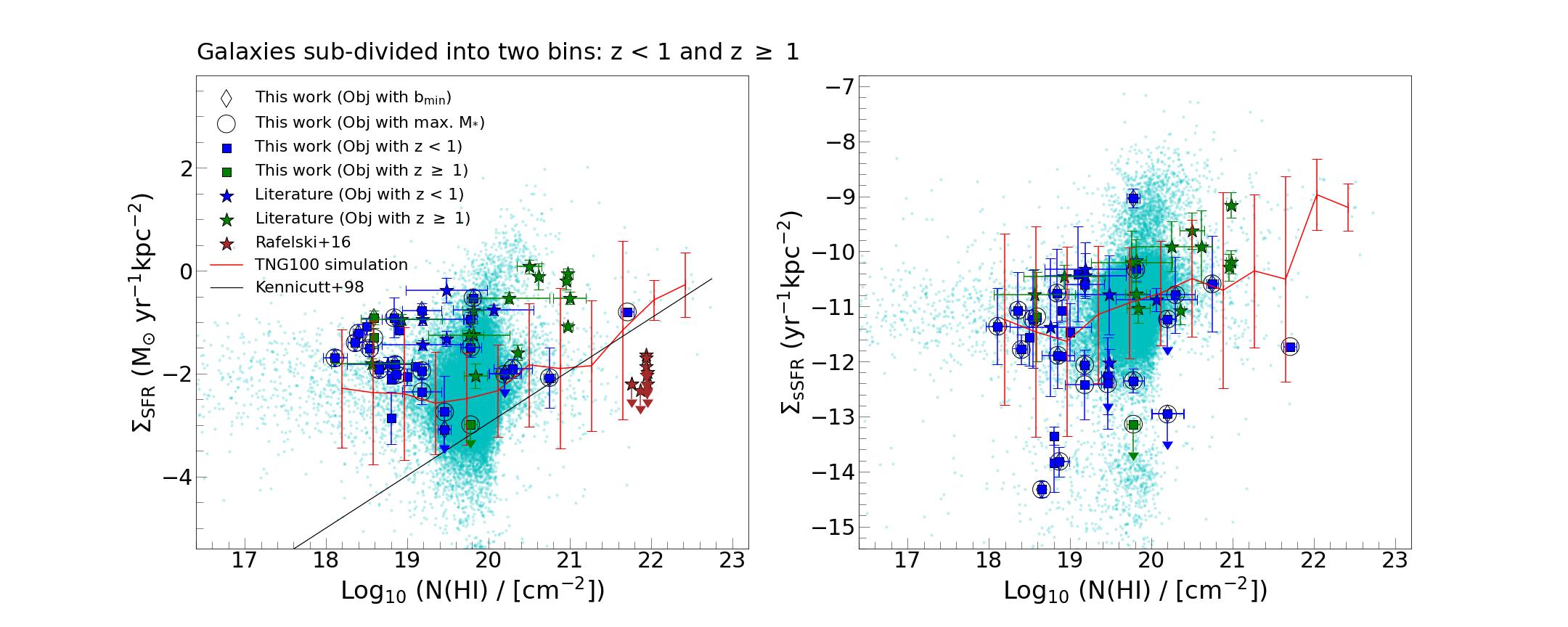}
    \caption{The surface densities of SFR (left panels) and sSFR (right panels) measured within the effective radius plotted versus the H \textsc{i} column density. The gas-rich absorbers tend to have higher 
    surface densities of SFR and sSFR. In the top panels, the galaxies are subdivided into two bins using the median value of the stellar mass. Galaxies with log \emph{M*} < log \emph{M*$_{\rm med}$} are colored with blue squares and the galaxies with log \emph{M*} $\geq$ log \emph{M*$_{\rm med}$} are represented with green squares. The bottom panels show similar plots where the galaxies are subdivided into two bins in terms of galaxies' redshift. Blue-colored galaxies have $\rm z_{gal}$ < 1 while green-colored galaxies have $\rm z_{gal}$ $\geq$ 1. Diamonds denote galaxies nearest to the quasar sight lines from our \emph{HST} measurements, while open circles stars denote the most massive galaxies of each quasar at the redshift of the absorber. One of the sample galaxies does not have \emph{M*} measurement and is denoted by an unfilled diamond. The stars are the literature values taken from \citet{2011MNRAS.413.2481F},
    \citet{2012MNRAS.424L...1K}, 
    \citet{2014MNRAS.445..225C}, 
    \citet{Rafelski_2016},
    \citet{2018MNRAS.478.3120A},
    \citet{2018ApJ...857L..12M}
    and \citet{2021MNRAS.506..546R}. In the left panel, the solid black line is the Kennicutt-Schmidt relationship for nearby galaxies taken from \citet{1998ApJ...498..541K}, while in both panels, the solid red line is the median value of the surface densities obtained from the TNG100 simulation. The solid red vertical lines denote 1-$\sigma$ uncertainties in the median value. The cyan dots are the simulated galaxies at $z$ = 0.5. No correlation is observed between SFR surface density and H \textsc{i} column density but a positive correlation exists between sSFR surface density and H \textsc{i} column density. Absorption-selected galaxies have a higher star formation efficiency  than predicted by the K-S law for local galaxies. See the text for more details. 
    }
    \label{fig13}
\end{figure*}

\subsubsection{Dependence of metallicity on impact parameter and galaxy's size}

\autoref{fig11} shows plots of  the metallicities of the absorption-selected galaxies vs. the impact parameter and the normalized impact parameter for galaxies from our sample and the literature. To place these in context, we use a model for the metallicity-size relation for DLAs suggested by  \citet{2008ApJ...683..321F}. In this simple model, galaxies are assigned a size, metallicity, and metallicity gradient based on their luminosities. The metallicity distributions of the quasar DLAs and GRB DLAs are predicted using the luminosity function of UV-selected galaxies (Lyman Break Galaxies), a metallicity vs. luminosity relation, and the radial distribution of H \textsc{i} gas along with luminosity. Using such a model prediction,  \citet{2012MNRAS.424L...1K} generated 4000 simulated data points to compare to the measured data points. The light blue region under the solid blue line in the right panel of \autoref{fig11} shows these 4000 simulated data points. For further details, see \citep[e.g.][]{2012MNRAS.424L...1K}. We extended the region to impact parameters of 100 kpc to include our sample galaxies. To do so, we extrapolated the metallicity-impact parameter relationship predicted by those simulation points. The light blue region under the dashed blue line in the \autoref{fig11} shows the extrapolated region.

Both panels of \autoref{fig11} suggest that the  metallicity is higher for galaxies sampled at higher impact parameters. This result is in agreement with previous studies \citep[e.g.][]{2012MNRAS.424L...1K, 2017MNRAS.469.2959K}, and would be  consistent with the model expectation \citep{2008ApJ...683..321F}. This suggests that the metal-poor systems are found in smaller haloes and, therefore, detectable only in galaxies at lower impact parameters \citep[][]{2012MNRAS.424L...1K}.

\subsubsection{Dependence of H \textsc{i} column density on impact parameter and galaxy  size}

\autoref{fig12} shows the impact parameter normalized by the effective radius and the effective radius plotted versus the H \textsc{i} column density. A  negative correlation ($ r_{s}$ $=$ -0.56 and \emph{p} $=$ 2.00 $\times$ $10^{-4}$) is observed between the normalized impact parameter and the H \textsc{i} column density. The effective radius of the absorbing galaxies also shows a strong negative correlation ($ r_{s}$ $=$ -0.52 and \emph{p} $=$ 6.00 $\times$ $10^{-4}$) with H \textsc{i} column density. This shows that galaxies associated with higher H \textsc{i} absorbers are smaller in size, and therefore likely to show strong absorption at small impact parameters. 
Indeed, all galaxies in the current sample associated with DLAs have effective radii smaller than 3 kpc, while all galaxies with $\rm R_{e}>$3 kpc are associated with lower H \textsc{i} column density absorbers. This finding is consistent with past suggestions that DLAs are associated 
with dwarf galaxies \citep[e.g.][]{1986ApJ...311..610Y, 2010NewA...15..735K}.

\subsubsection{Relation between H \textsc{i} column density and  surface density of star formation}

\autoref{fig13} shows plots of the surface densities of the SFR and sSFR averaged within the effective radius versus the column density of H~\textsc{i}. In the upper two panels, the galaxies are binned into two groups by stellar mass (above and below the median stellar mass log \emph{M*$_{\rm med}$} = 9.67). 
In the lower two panels, the galaxies are binned into two groups by redshift ($z<1$ and $z \geq 1$). For comparison with local galaxies, we also show the relationship observed between the surface densities of SFR and H~\textsc{i}  \citep[][]{1998ApJ...498..541K} for low-redshift spirals (what we refer to as the ``atomic Kennicutt-Schmidt (K-S)  relation''). This relation is based on the best fit (log $\Sigma_{\rm SFR}$ = 1.02 log $\Sigma_{\rm HI}$ - 2.89) between the tabulated values of $\rm \Sigma_{SFR}$ and $\Sigma_{\rm H I}$ listed in \citet[][]{1998ApJ...498..541K}. We use this relation (which translates to log $\Sigma_{\rm SFR}$ = 1.02 log $N_{\rm HI}$ - 23.36) rather than the standard KS relation between $\rm \Sigma_{\rm SFR}$ and $\Sigma_{\rm gas}$ since the latter also includes molecular gas, and molecular gas measurements are not available for most of the absorber-selected galaxies plotted in \autoref{fig13} 
(besides the few objects observed so far in CO emission with ALMA).

The upper right panel of \autoref{fig13} shows that $\Sigma_{\rm sSFR}$ is positively correlated with the H \textsc{i} column density. The Spearman rank order correlation test  gives $r_{s}$ $=$ 0.42 and \emph{p} $=$ 8.30 $\times$ $10^{-3}$ between $N_{\rm H I}$ and $\rm \Sigma_{\rm sSFR}$ (although we note that this relation is not significant if only our sample galaxies are considered.) The correlation between $\rm \Sigma_{SFR}$ and $N_{\rm H I}$ (shown in the upper left panel of \autoref{fig13}) is not as significant even after including the literature galaxies, with $r_{s}$ $=$ 0.29 and \emph{p} $=$ 6.65 $\times$ $10^{-2}$. This finding seems to be at odds with the previous suggestions based on simulations \citep[e.g.][]{2004MNRAS.348..435N} that DLAs follow the KS law.

Indeed, most of the galaxies associated with the absorbers from our sample and the literature  lie substantially above the ``atomic K-S relation''. If taken at face value, this suggests that the absorption-selected galaxies (which have $z_{\rm med} \sim 0.8$) have more efficient star formation compared to the nearby galaxies. However, this  apparent discrepancy may also be caused in part by the fact that the local ``atomic K-S relation'' is based on observations of 21-cm emission (which is usually far less sensitive to low H \textsc{i} column densities than the Ly$\alpha$ absorption line technique used for the quasar absorber sample).

We also note that the galaxies with stellar mass greater than the median mass are located  farther away from %the dense region of the simulated galaxies and 
the ``atomic K-S'' relation in the upper left panel of \autoref{fig13}, as compared to lower-mass galaxies. This suggests that the more-massive absorption-selected galaxies have more efficient star formation, allowing them to reach comparable SFRs in lower H \textsc{i} column density regions.

The high values of $\Sigma_{\rm SFR}$ compared to the K-S relation for the galaxies in our sample and the literature seen in \autoref{fig13}  may appear surprising, given the findings of earlier studies based on indirect estimates of SFR surface density that the star formation efficiency in DLAs at $z \geq 1$ is 1-3\% of that predicted by the K-S relation \citep[e.g.][]{2006ApJ...652..981W,2011ApJ...736...48R,Rafelski_2016}. The latter 
measurements are also shown in the left panels of \autoref{fig13} for comparison. 
A substantial fraction of our galaxies 
have 18 $<$ log $N_{\rm H I} < 20$.  But even focusing on just the absorbers with log $N_{\rm H I} \geq 20.3$ (i.e. the DLAs) in our sample, the star formation efficiency of the absorber-associated galaxies (with a median redshift of $\sim 1.01$) still appears to be comparable to or (for some galaxies) substantially higher than predicted by the KS relation. 

The large difference between our values of $\Sigma_{\rm SFR}$ and 
those in the past DLA studies may result from the fact that, unlike our 
the study, the past  studies were based not on direct SFR measurements for  galaxies associated with DLAs, but 
on measurements in the outskirts of isolated star-forming galaxies and on the assumption that the latter are related to DLAs. It is also noteworthy in this context that the surface brightness of the galaxies associated with DLAs 
detected in our 
sample and those from the literature are in fact several magnitudes brighter than the upper limit of 29 mag arcsec$^{-2}$ estimated in 
prior studies of star formation efficiency in DLAs at $z \sim 3$ \citep[]{2006ApJ...652..981W,2011ApJ...736...48R}. We note, however, that we cannot make a precise statement about the agreement between those DLAs and the simulation data due to the limited number of simulated galaxies with 
$N_{\rm H I}> 10^{21}$ cm$^{-2}$.  
The ability of IFS studies to reveal star-forming galaxies 
associated with DLAs (and lower H \textsc{i} column density absorbers) demonstrated by our MUSE-ALMA-Halos sample 
and other IFS studies in the literature indeed mark a 
huge improvement in detecting star formation in galaxies associated with gas-rich quasar absorbers compared to past searches. We also note, in passing, that the previous studies of the SFR surface density in DLAs mentioned above \citep[e.g.,][]{2004MNRAS.348..435N, Rafelski_2016} adopted the original KS relation for $\rm \Sigma_{SFR}$ vs $\rm \Sigma_{gas}$ instead of the ``atomic'' version of this relation $\rm \Sigma_{SFR}$ vs $\rm \Sigma_{H I}$.

To examine whether the star formation efficiency of absorber-selected gas-rich galaxies may have increased dramatically at $z < 1$, we compare the $\Sigma_{\rm SFR}$ vs. log $N_{\rm H I}$ trends for galaxies with $z<1$ and galaxies at $z \geq 1$ in the lower panels of  \autoref{fig13}. The lower redshift galaxies appear to have comparable SFR surface densities for lower H \textsc{i} column densities. This suggests a higher star formation efficiency of absorber-selected galaxies at $z < 1$ compared to those at $z \geq 1$. Also shown for comparison in \autoref{fig13} are 
%The cyan dots are the 
simulated galaxies at $z$ = 0.5, based on an analysis of the data obtained from the Illustris TNG simulations \citep[][]{2018MNRAS.475..624N,2018MNRAS.475..648P,2018MNRAS.475..676S,2018MNRAS.477.1206N,2018MNRAS.480.5113M}. These state-of-the-art cosmological simulations incorporate essential physical processes pertinent to the formation and evolution of galaxies, such as gravity, hydrodynamics, gas cooling, star formation, stellar feedback, black hole feedback, and more. In this study, we employ the highest-resolution version of the TNG100 simulation, executed within a $\sim 100 \, \rm Mpc$ box. Haloes are initially identified by employing the Friends of Friends (FOF) algorithm \citep{1985ApJ...292..371D}. Galaxies, regarded as substructures, are recognized as gravitationally-bound assemblies of particles within these FOF haloes performing the \textsc{SUBFIND} algorithm \citep{2001MNRAS.328..726S}. Each FOF halo encompasses a central galaxy, generally the most massive galaxy within that halo, and all additional galaxies within the halo are labeled as satellites.

Properties of galaxies, including stellar masses and star formation rates (SFRs), are taken from the principal TNG galaxy catalogs. These are measured within the stellar half-mass radius by utilizing the particle data of the simulation. Furthermore, the neutral hydrogen (H \textsc{i}) content of TNG galaxies has been extracted from catalogs provided by \citet{2018ApJS..238...33D}, based on the analytic model of \citet{2014ApJ...790...10S}, where they use an optimized post-processing framework for estimating the abundance of atomic and molecular hydrogen. This method uses the surface density of neutral hydrogen and the ultraviolet (UV) flux within the Lyman-Werner band, with all computations being performed through face-on projections within a two-dimensional model. The UV radiation emitted from young stars is modeled by assuming a constant escape fraction and optically thin propagation across the galaxy. The simulation demonstrates a relatively satisfactory agreement with the measurements of the SFR surface density based on emission-line observations for galaxies from both our sample and the literature shown in \autoref{fig13} (and, like these data, also lie substantially above the upper limits for the DLAs from past studies). 

The agreement between the simulated and observed data 
appears to be better for the sSFR surface density than for the SFR surface density, as seen in the right panels of \autoref{fig13}. This difference may result from the differences in the stellar mass distributions for the simulated and observed galaxies. The stellar mass distribution of the observed galaxies (from our sample and the literature) peaks around log $M_{*}=9.67$ and shows fewer low-$M_{*}$ galaxies compared to the distribution for the simulated galaxies. 
The higher $M_{*}$ galaxies have lower $\Sigma_{\rm sSFR}$ (as expected from the negative correlation between sSFR and $M_{*}$, see \autoref{table5}), giving better consistency between the $\Sigma_{\rm sSFR}$ vs. $M{*}$ trends for the simulated and observed galaxies, compared to the median $\Sigma_{\rm SFR}$ vs. $M{*}$ trends.

To summarize, we find interdependence between the stellar properties and the absorption properties. In particular, the H \textsc{i} column density and the absorption metallicity show correlations with \emph{M*}, ssfr, $\Sigma_{\rm sSFR}$, but not with the SFR and $\Sigma_{\rm SFR}$.

\section{Summary and Conclusions}

We have analyzed the morphological and stellar properties of 66 galaxies detected within $\pm$ 500 km s$^{-1}$ of the redshifts of strong intervening quasar absorbers at $0.2 \lesssim z \lesssim 1.4$ with $N_{\rm HI}$ > $10^{18}$ $\rm cm^{-2}$ (that also have MUSE and/or ALMA data). The structural parameters of these absorption-selected galaxies were determined using \textsc{galfit}. The galaxies were found to have S\'ersic indices ranging from 0.3 to 2.3 and effective radii ranging from 0.7 to 7.6 kpc. The $\rm k$-corrected absolute magnitudes of these galaxies range from -15.8 to -23.7 mag. Our main findings are as follows:

\begin{enumerate}
    \item The absolute (rest-frame) surface brightness shows a strong positive correlation with the galaxy luminosity. The trend appears flatter at lower luminosities for those galaxies that have high H \textsc{i} column densities. This suggests dwarf galaxies are associated with high H \textsc{i} column densities.

    \item The star formation rate correlates well with the stellar mass. Most galaxies associated with intervening quasar absorbers are consistent with the star formation main sequence. 
    
    \item Larger galaxies are found to be more massive compared to smaller galaxies. Furthermore, massive galaxies are more centrally concentrated, as observed for nearby galaxies. Overall the absorption-selected galaxies follow similar trends as those shown by the general galaxy population.

    \item For most ($\sim$85\%) of the galaxies in our sample, the impact parameters are smaller than the virial radius. Most of the sight lines with high $N_{\rm H I}$ in our sample probe the CGM of the associated galaxies at impact parameters less than half the virial radius, and only a small fraction have impact parameters larger than the virial radius.
    
    \item The rest-frame equivalent widths of Mg II $\lambda$ 2796  show a negative correlation with stellar mass. Such trend suggests that lower $N_{\rm H I}$ absorbers
    are associated with more massive galaxies that have undergone high past star formation and gas consumption activity.
    %Similarly, the gas-rich galaxies trace a similar trend as previously determined mass-metallicity relation. While we apply an impact parameter correction to assume metallicities at the galaxy's center, such corrections seem severe for the galaxies located at larger impact parameters. In the same way, the galaxies' metallicity is lower with the increase in sSFR.}

    \item The absorption metallicity and emission metallicity show a positive correlation with the stellar mass for many of the absorption-selected galaxies and are consistent with the mass-metallicity relation for star-forming galaxies. While the metallicity shows no correlation with SFR, the metallicity is negatively correlated with the specific SFR, suggesting that the low-metallicity absorbers are associated with galaxies with vigorous current 
    star formation but low past star formation activity.
    
    \item Metallicity appears to be positively correlated with the impact parameter and normalized impact parameter. This suggests that metal-poor galaxies are found in smaller haloes and are, therefore, detectable in galaxies at smaller impact parameters.

    \item The H \textsc{i} column density is negatively correlated with the normalized impact parameter and the effective radius of the galaxies. This shows that galaxies associated with higher $N_{\rm H I}$ absorbers are smaller in size and, therefore, likely to show strong absorption at small impact parameters.

    \item The sSFR surface density is positively correlated with the H \textsc{i} column density, but no correlation is seen 
    between SFR surface density and H \textsc{i} column density. Furthermore, the $\rm \Sigma_{SFR}$ for the absorber-associated galaxies is substantially higher than predicted from the atomic K-S relation for nearby galaxies, suggesting higher star formation efficiency in the absorber-selected galaxies. The SFR surface density is also substantially higher than the upper limits on $\rm \Sigma_{SFR}$ for DLAs estimated in past studies. Moreover, the star formation efficiency for absorber-associated galaxies at $z<1$ appears to be higher than 
    for those at $z \geq 1$.

\end{enumerate}

The overall conclusions from our results are: the stellar and morphological properties of absorption-selected galaxies are consistent with the star formation main sequence of galaxies and show the mass-metallicity relation. Furthermore, the higher H \textsc{i} column density absorbers are associated with smaller galaxies in smaller halos that have generally not experienced much star formation in the past and have thus remained metal-poor. However, these galaxies associated with the higher $N_{\rm H I}$ absorbers have more active current star formation and exhibit higher surface densities of star formation and gas consumption. Our study also reveals that a substantial fraction of gas-rich quasar absorbers arises in groups of galaxies. 

Our study of structural and stellar properties of 66 associated galaxies associated with 
gas-rich quasar absorbers has thus allowed us to search for correlations  between a variety of morphological, stellar, and gas properties. 
%While the current sample size is small,
However, our sample is still relatively small. Increasing the number of absorption-selected galaxies with measurements of the various properties with future MUSE, ALMA, and \emph{HST} observations is essential to more robustly establish the trends suggested by our study and to fully interpret their implications for the evolution of galaxies and their CGM.

\section*{Acknowledgements}

We thank an anonymous referee for constructive comments that have helped to improve this manuscript. AK and VPK acknowledge support from a grant from the Space Telescope Science Institute for GO program 15939 (PI: P\'eroux). AK and VPK also acknowledge additional partial support from US National Science Foundation grant AST/2007538 and NASA grants  NNX17AJ26G and 80NSSC20K0887 (PI: Kulkarni). GK and SW acknowledge the financial support of the Australian Research Council through grant CE170100013 (ASTRO3D). MA acknowledges funding from the Deutsche Forschungsgemeinschaft (DFG) through an Emmy Noether Research Group (grant number NE 2441/1-1). We also thank the International Space Science Institute (ISSI) (\url{https://www.issibern.ch/}) for financial support. This research made use of following python packages: \textsc{astropy}, a community-developed core Python package for Astronomy \citep[e.g.][]{2013A&A...558A..33A, 2018zndo...1461536C}, \textsc{matplotlib} \citep{4160265}, \textsc{pandas} \citep{mckinney2010data}, \textsc{photutils} \citep{larry_bradley_2022_6825092} and \textsc{numpy} \citep{Harris_2020}. 

%%%%%%%%%%%%%%%%%%%%%%%%%%%%%%%%%%%%%%%%%%%%%%%%%%
\section*{Data Availability}

Data directly related to this publication and its figures can be requested from the authors. The raw data can be downloaded from the public
archives. The TNG simulation \citep{2019ComAC...6....2N} is publicly available at \url{https://www.tng-project.org/}.



\bibliographystyle{mnras}
\bibliography{References} % if your bibtex file is called example.bib

\begin{thebibliography}{}
\makeatletter
\relax
\def\mn@urlcharsother{\let\do\@makeother \do\$\do\&\do\#\do\^\do\_\do\%\do\~}
\def\mn@doi{\begingroup\mn@urlcharsother \@ifnextchar [ {\mn@doi@}
  {\mn@doi@[]}}
\def\mn@doi@[#1]#2{\def\@tempa{#1}\ifx\@tempa\@empty \href
  {http://dx.doi.org/#2} {doi:#2}\else \href {http://dx.doi.org/#2} {#1}\fi
  \endgroup}
\def\mn@eprint#1#2{\mn@eprint@#1:#2::\@nil}
\def\mn@eprint@arXiv#1{\href {http://arxiv.org/abs/#1} {{\tt arXiv:#1}}}
\def\mn@eprint@dblp#1{\href {http://dblp.uni-trier.de/rec/bibtex/#1.xml}
  {dblp:#1}}
\def\mn@eprint@#1:#2:#3:#4\@nil{\def\@tempa {#1}\def\@tempb {#2}\def\@tempc
  {#3}\ifx \@tempc \@empty \let \@tempc \@tempb \let \@tempb \@tempa \fi \ifx
  \@tempb \@empty \def\@tempb {arXiv}\fi \@ifundefined
  {mn@eprint@\@tempb}{\@tempb:\@tempc}{\expandafter \expandafter \csname
  mn@eprint@\@tempb\endcsname \expandafter{\@tempc}}}

\bibitem[\protect\citeauthoryear{{Arnouts}, {Cristiani}, {Moscardini},
  {Matarrese}, {Lucchin}, {Fontana}  \& {Giallongo}}{{Arnouts}
  et~al.}{1999}]{1999MNRAS.310..Arnouts}
{Arnouts} S.,  {Cristiani} S.,  {Moscardini} L.,  {Matarrese} S.,  {Lucchin}
  F.,  {Fontana} A.,   {Giallongo} E.,  1999, \mn@doi [\mnras]
  {10.1046/j.1365-8711.1999.02978.x}, \href
  {https://ui.adsabs.harvard.edu/abs/1999MNRAS.310..540A} {310, 540}

\bibitem[\protect\citeauthoryear{{Astropy Collaboration} et~al.,}{{Astropy
  Collaboration} et~al.}{2013}]{2013A&A...558A..33A}
{Astropy Collaboration} et~al., 2013, \mn@doi [\aap]
  {10.1051/0004-6361/201322068}, \href
  {https://ui.adsabs.harvard.edu/abs/2013A&A...558A..33A} {558, A33}

\bibitem[\protect\citeauthoryear{{Augustin} et~al.,}{{Augustin}
  et~al.}{2018}]{2018MNRAS.478.3120A}
{Augustin} R.,  et~al., 2018, \mn@doi [\mnras] {10.1093/mnras/sty1287}, \href
  {https://ui.adsabs.harvard.edu/abs/2018MNRAS.478.3120A} {478, 3120}

\bibitem[\protect\citeauthoryear{{Bacon} et~al.,}{{Bacon}
  et~al.}{2010}]{2010SPIE.7735E..08B}
{Bacon} R.,  et~al., 2010, in {McLean} I.~S.,  {Ramsay} S.~K.,   {Takami} H.,
  eds,  Society of Photo-Optical Instrumentation Engineers (SPIE) Conference
  Series Vol. 7735, Ground-based and Airborne Instrumentation for Astronomy
  III. p. 773508 (\mn@eprint {arXiv} {2211.16795}), \mn@doi{10.1117/12.856027}

\bibitem[\protect\citeauthoryear{{Bashir}, {Zafar}, {Khan}  \&
  {Chishtie}}{{Bashir} et~al.}{2019}]{2019NewA...66....9B}
{Bashir} W.,  {Zafar} T.,  {Khan} F.~M.,   {Chishtie} F.,  2019, \mn@doi [\na]
  {10.1016/j.newast.2018.07.001}, \href
  {https://ui.adsabs.harvard.edu/abs/2019NewA...66....9B} {66, 9}

\bibitem[\protect\citeauthoryear{{Berg} et~al.,}{{Berg}
  et~al.}{2016}]{2016MNRAS.463.3021B}
{Berg} T.~A.~M.,  et~al., 2016, \mn@doi [\mnras] {10.1093/mnras/stw2232}, \href
  {https://ui.adsabs.harvard.edu/abs/2016MNRAS.463.3021B} {463, 3021}

\bibitem[\protect\citeauthoryear{{Berg} et~al.,}{{Berg}
  et~al.}{2017}]{2017MNRAS.464L..56B}
{Berg} T.~A.~M.,  et~al., 2017, \mn@doi [\mnras] {10.1093/mnrasl/slw185}, \href
  {https://ui.adsabs.harvard.edu/abs/2017MNRAS.464L..56B} {464, L56}

\bibitem[\protect\citeauthoryear{{Berg} et~al.,}{{Berg}
  et~al.}{2022}]{2022arXiv220413229B}
{Berg} M.~A.,  et~al., 2022, \mn@doi [arXiv e-prints]
  {10.48550/arXiv.2204.13229}, \href
  {https://ui.adsabs.harvard.edu/abs/2022arXiv220413229B} {p. arXiv:2204.13229}

\bibitem[\protect\citeauthoryear{{Bergeron}}{{Bergeron}}{1986}]{1986A&A...155L...8B}
{Bergeron} J.,  1986, \aap, \href
  {https://ui.adsabs.harvard.edu/abs/1986A&A...155L...8B} {155, L8}

\bibitem[\protect\citeauthoryear{{Bergeron} \& {Boiss{\'e}}}{{Bergeron} \&
  {Boiss{\'e}}}{1991}]{1991A&A...243..344B}
{Bergeron} J.,  {Boiss{\'e}} P.,  1991, \aap, \href
  {https://ui.adsabs.harvard.edu/abs/1991A&A...243..344B} {243, 344}

\bibitem[\protect\citeauthoryear{{Bertin} \& {Arnouts}}{{Bertin} \&
  {Arnouts}}{1996}]{1996A&AS..117..393B}
{Bertin} E.,  {Arnouts} S.,  1996, \mn@doi [\aaps] {10.1051/aas:1996164}, \href
  {https://ui.adsabs.harvard.edu/abs/1996A&AS..117..393B} {117, 393}

\bibitem[\protect\citeauthoryear{{Blanton} \& {Roweis}}{{Blanton} \&
  {Roweis}}{2007}]{2007AJ..Blanton}
{Blanton} M.~R.,  {Roweis} S.,  2007, \mn@doi [\aj] {10.1086/510127}, \href
  {https://ui.adsabs.harvard.edu/abs/2007AJ....133..734B} {133, 734}

\bibitem[\protect\citeauthoryear{{Boisse}, {Le Brun}, {Bergeron}  \&
  {Deharveng}}{{Boisse} et~al.}{1998}]{1998A&A...333..841B}
{Boisse} P.,  {Le Brun} V.,  {Bergeron} J.,   {Deharveng} J.-M.,  1998, \aap,
  \href {https://ui.adsabs.harvard.edu/abs/1998A&A...333..841B} {333, 841}

\bibitem[\protect\citeauthoryear{{Boogaard} et~al.,}{{Boogaard}
  et~al.}{2018}]{2018A&A...619A..27B}
{Boogaard} L.~A.,  et~al., 2018, \mn@doi [\aap] {10.1051/0004-6361/201833136},
  \href {https://ui.adsabs.harvard.edu/abs/2018A&A...619A..27B} {619, A27}

\bibitem[\protect\citeauthoryear{{Bordoloi} et~al.,}{{Bordoloi}
  et~al.}{2011}]{2011ApJ...743...10B}
{Bordoloi} R.,  et~al., 2011, \mn@doi [\apj] {10.1088/0004-637X/743/1/10},
  \href {https://ui.adsabs.harvard.edu/abs/2011ApJ...743...10B} {743, 10}

\bibitem[\protect\citeauthoryear{{Bouch{\'e}} et~al.,}{{Bouch{\'e}}
  et~al.}{2012}]{2012MNRAS.419....2B}
{Bouch{\'e}} N.,  et~al., 2012, \mn@doi [\mnras]
  {10.1111/j.1365-2966.2011.19500.x}, \href
  {https://ui.adsabs.harvard.edu/abs/2012MNRAS.419....2B} {419, 2}

\bibitem[\protect\citeauthoryear{{Bouch{\'e}}, {Murphy}, {Kacprzak},
  {P{\'e}roux}, {Contini}, {Martin}  \& {Dessauges-Zavadsky}}{{Bouch{\'e}}
  et~al.}{2013}]{2013Sci..Bouche}
{Bouch{\'e}} N.,  {Murphy} M.~T.,  {Kacprzak} G.~G.,  {P{\'e}roux} C.,
  {Contini} T.,  {Martin} C.~L.,   {Dessauges-Zavadsky} M.,  2013, \mn@doi
  [Science] {10.1126/science.1234209}, \href
  {https://ui.adsabs.harvard.edu/abs/2013Sci...341...50B} {341, 50}

\bibitem[\protect\citeauthoryear{Bradley et~al.,}{Bradley
  et~al.}{2022}]{larry_bradley_2022_6825092}
Bradley L.,  et~al., 2022, astropy/photutils: 1.5.0,
  \mn@doi{10.5281/zenodo.6825092}, \url
  {https://doi.org/10.5281/zenodo.6825092}

\bibitem[\protect\citeauthoryear{{Buta}}{{Buta}}{2011}]{2011arXiv1102.0550B}
{Buta} R.~J.,  2011, arXiv e-prints, \href
  {https://ui.adsabs.harvard.edu/abs/2011arXiv1102.0550B} {p. arXiv:1102.0550}

\bibitem[\protect\citeauthoryear{Carollo et~al.,}{Carollo
  et~al.}{2013}]{Carollo_2013}
Carollo C.~M.,  et~al., 2013, \mn@doi [\apj] {10.1088/0004-637x/773/2/112},
  773, 112

\bibitem[\protect\citeauthoryear{{Chen} et~al.,}{{Chen}
  et~al.}{2020}]{2020MNRAS.497..498C}
{Chen} H.-W.,  et~al., 2020, \mn@doi [\mnras] {10.1093/mnras/staa1773}, \href
  {https://ui.adsabs.harvard.edu/abs/2020MNRAS.497..498C} {497, 498}

\bibitem[\protect\citeauthoryear{{Christensen}, {M{\o}ller}, {Fynbo}  \&
  {Zafar}}{{Christensen} et~al.}{2014}]{2014MNRAS.445..225C}
{Christensen} L.,  {M{\o}ller} P.,  {Fynbo} J.~P.~U.,   {Zafar} T.,  2014,
  \mn@doi [\mnras] {10.1093/mnras/stu1726}, \href
  {https://ui.adsabs.harvard.edu/abs/2014MNRAS.445..225C} {445, 225}

\bibitem[\protect\citeauthoryear{{Chun}, {Kulkarni}, {Gharanfoli}  \&
  {Takamiya}}{{Chun} et~al.}{2010}]{2010AJ....139..296C}
{Chun} M.~R.,  {Kulkarni} V.~P.,  {Gharanfoli} S.,   {Takamiya} M.,  2010,
  \mn@doi [\aj] {10.1088/0004-6256/139/1/296}, \href
  {https://ui.adsabs.harvard.edu/abs/2010AJ....139..296C} {139, 296}

\bibitem[\protect\citeauthoryear{{Churchill}}{{Churchill}}{2001}]{2001ApJ...560...92C}
{Churchill} C.~W.,  2001, \mn@doi [\apj] {10.1086/322512}, \href
  {https://ui.adsabs.harvard.edu/abs/2001ApJ...560...92C} {560, 92}

\bibitem[\protect\citeauthoryear{{Collaboration}}{{Collaboration}}{2018}]{2018zndo...1461536C}
{Collaboration} T.~A.,  2018, {astropy v3.1: a core python package for
  astronomy}, Zenodo, \mn@doi{10.5281/zenodo.1461536}

\bibitem[\protect\citeauthoryear{{Curti}, {Cresci}, {Mannucci}, {Marconi},
  {Maiolino}  \& {Esposito}}{{Curti} et~al.}{2017}]{2017MNRAS.465.1384C}
{Curti} M.,  {Cresci} G.,  {Mannucci} F.,  {Marconi} A.,  {Maiolino} R.,
  {Esposito} S.,  2017, \mn@doi [\mnras] {10.1093/mnras/stw2766}, \href
  {https://ui.adsabs.harvard.edu/abs/2017MNRAS.465.1384C} {465, 1384}

\bibitem[\protect\citeauthoryear{{Davis}, {Efstathiou}, {Frenk}  \&
  {White}}{{Davis} et~al.}{1985}]{1985ApJ...292..371D}
{Davis} M.,  {Efstathiou} G.,  {Frenk} C.~S.,   {White} S.~D.~M.,  1985,
  \mn@doi [\apj] {10.1086/163168}, \href
  {https://ui.adsabs.harvard.edu/abs/1985ApJ...292..371D} {292, 371}

\bibitem[\protect\citeauthoryear{{Diemer} et~al.,}{{Diemer}
  et~al.}{2018}]{2018ApJS..238...33D}
{Diemer} B.,  et~al., 2018, \mn@doi [\apjs] {10.3847/1538-4365/aae387}, \href
  {https://ui.adsabs.harvard.edu/abs/2018ApJS..238...33D} {238, 33}

\bibitem[\protect\citeauthoryear{{Dutta} et~al.,}{{Dutta}
  et~al.}{2020}]{2020MNRAS.499.5022D}
{Dutta} R.,  et~al., 2020, \mn@doi [\mnras] {10.1093/mnras/staa3147}, \href
  {https://ui.adsabs.harvard.edu/abs/2020MNRAS.499.5022D} {499, 5022}

\bibitem[\protect\citeauthoryear{{Dutta} et~al.,}{{Dutta}
  et~al.}{2021}]{2021MNRAS.508.4573D}
{Dutta} R.,  et~al., 2021, \mn@doi [\mnras] {10.1093/mnras/stab2752}, \href
  {https://ui.adsabs.harvard.edu/abs/2021MNRAS.508.4573D} {508, 4573}

\bibitem[\protect\citeauthoryear{{Fossati} et~al.,}{{Fossati}
  et~al.}{2019}]{2019MNRAS.490.1451F}
{Fossati} M.,  et~al., 2019, \mn@doi [\mnras] {10.1093/mnras/stz2693}, \href
  {https://ui.adsabs.harvard.edu/abs/2019MNRAS.490.1451F} {490, 1451}

\bibitem[\protect\citeauthoryear{{Fumagalli}, {O'Meara}  \&
  {Prochaska}}{{Fumagalli} et~al.}{2016}]{2016MNRAS.455.4100F}
{Fumagalli} M.,  {O'Meara} J.~M.,   {Prochaska} J.~X.,  2016, \mn@doi [\mnras]
  {10.1093/mnras/stv2616}, \href
  {https://ui.adsabs.harvard.edu/abs/2016MNRAS.455.4100F} {455, 4100}

\bibitem[\protect\citeauthoryear{{Fynbo}, {Prochaska}, {Sommer-Larsen},
  {Dessauges-Zavadsky}  \& {M{\o}ller}}{{Fynbo}
  et~al.}{2008}]{2008ApJ...683..321F}
{Fynbo} J. P.~U.,  {Prochaska} J.~X.,  {Sommer-Larsen} J.,
  {Dessauges-Zavadsky} M.,   {M{\o}ller} P.,  2008, \mn@doi [\apj]
  {10.1086/589555}, \href
  {https://ui.adsabs.harvard.edu/abs/2008ApJ...683..321F} {683, 321}

\bibitem[\protect\citeauthoryear{{Fynbo} et~al.,}{{Fynbo}
  et~al.}{2011}]{2011MNRAS.413.2481F}
{Fynbo} J.~P.~U.,  et~al., 2011, \mn@doi [\mnras]
  {10.1111/j.1365-2966.2011.18318.x}, \href
  {https://ui.adsabs.harvard.edu/abs/2011MNRAS.413.2481F} {413, 2481}

\bibitem[\protect\citeauthoryear{{Graham} \& {Driver}}{{Graham} \&
  {Driver}}{2005}]{2005PASA...22..118G}
{Graham} A.~W.,  {Driver} S.~P.,  2005, \mn@doi [\pasa] {10.1071/AS05001},
  \href {https://ui.adsabs.harvard.edu/abs/2005PASA...22..118G} {22, 118}

\bibitem[\protect\citeauthoryear{{Hamanowicz} et~al.,}{{Hamanowicz}
  et~al.}{2020}]{2020MNRAS.492.2347H}
{Hamanowicz} A.,  et~al., 2020, \mn@doi [\mnras] {10.1093/mnras/stz3590}, \href
  {https://ui.adsabs.harvard.edu/abs/2020MNRAS.492.2347H} {492, 2347}

\bibitem[\protect\citeauthoryear{Harris et~al.,}{Harris
  et~al.}{2020}]{Harris_2020}
Harris C.~R.,  et~al., 2020, \mn@doi [Nature] {10.1038/s41586-020-2649-2}, 585,
  357

\bibitem[\protect\citeauthoryear{{Hartoog}, {Fynbo}, {Kaper}, {De Cia}  \&
  {Bagdonaite}}{{Hartoog} et~al.}{2015}]{2015MNRAS.447.2738H}
{Hartoog} O.~E.,  {Fynbo} J.~P.~U.,  {Kaper} L.,  {De Cia} A.,   {Bagdonaite}
  J.,  2015, \mn@doi [\mnras] {10.1093/mnras/stu2578}, \href
  {https://ui.adsabs.harvard.edu/abs/2015MNRAS.447.2738H} {447, 2738}

\bibitem[\protect\citeauthoryear{{H{\"a}u{\ss}ler}, {Barden}, {Bamford}  \&
  {Rojas}}{{H{\"a}u{\ss}ler} et~al.}{2011}]{2011ASPC..442..155H}
{H{\"a}u{\ss}ler} B.,  {Barden} M.,  {Bamford} S.~P.,   {Rojas} A.,  2011, in
  {Evans} I.~N.,  {Accomazzi} A.,  {Mink} D.~J.,   {Rots} A.~H.,  eds,
  Astronomical Society of the Pacific Conference Series Vol. 442, Astronomical
  Data Analysis Software and Systems XX. p.~155

\bibitem[\protect\citeauthoryear{Hilz, Naab  \& Ostriker}{Hilz
  et~al.}{2013}]{10.1093..Hilz}
Hilz M.,  Naab T.,   Ostriker J.~P.,  2013, \mn@doi [\mnras]
  {10.1093/mnras/sts501}, 429, 2924

\bibitem[\protect\citeauthoryear{Hunter}{Hunter}{2007}]{4160265}
Hunter J.~D.,  2007, \mn@doi [Computing in Science & Engineering]
  {10.1109/MCSE.2007.55}, 9, 90

\bibitem[\protect\citeauthoryear{{Ichikawa}, {Kajisawa}  \&
  {Akhlaghi}}{{Ichikawa} et~al.}{2012}]{2012MNRAS.422.1014I}
{Ichikawa} T.,  {Kajisawa} M.,   {Akhlaghi} M.,  2012, \mn@doi [\mnras]
  {10.1111/j.1365-2966.2012.20674.x}, \href
  {https://ui.adsabs.harvard.edu/abs/2012MNRAS.422.1014I} {422, 1014}

\bibitem[\protect\citeauthoryear{{Ilbert} et~al.,}{{Ilbert}
  et~al.}{2006}]{2006A&A...457..Ilbert}
{Ilbert} O.,  et~al., 2006, \mn@doi [\aap] {10.1051/0004-6361:20065138}, \href
  {https://ui.adsabs.harvard.edu/abs/2006A&A...457..841I} {457, 841}

\bibitem[\protect\citeauthoryear{Jenkins}{Jenkins}{2009}]{Jenkins_2009}
Jenkins E.~B.,  2009, \mn@doi [\apj] {10.1088/0004-637X/700/2/1299}, 700, 1299

\bibitem[\protect\citeauthoryear{{Kacprzak}, {Churchill}, {Evans}, {Murphy}  \&
  {Steidel}}{{Kacprzak} et~al.}{2011}]{2011MNRAS.416.3118K}
{Kacprzak} G.~G.,  {Churchill} C.~W.,  {Evans} J.~L.,  {Murphy} M.~T.,
  {Steidel} C.~C.,  2011, \mn@doi [\mnras] {10.1111/j.1365-2966.2011.19261.x},
  \href {https://ui.adsabs.harvard.edu/abs/2011MNRAS.416.3118K} {416, 3118}

\bibitem[\protect\citeauthoryear{Karachentsev, Makarov  \&
  Kaisina}{Karachentsev et~al.}{2013}]{Karachentsev_2013}
Karachentsev I.~D.,  Makarov D.~I.,   Kaisina E.~I.,  2013, \mn@doi [\aj]
  {10.1088/0004-6256/145/4/101}, 145, 101

\bibitem[\protect\citeauthoryear{{Kashikawa}, {Misawa}, {Minowa}, {Okoshi},
  {Hattori}, {Toshikawa}, {Ishikawa}  \& {Onoue}}{{Kashikawa}
  et~al.}{2014}]{2014ApJ...780..116K}
{Kashikawa} N.,  {Misawa} T.,  {Minowa} Y.,  {Okoshi} K.,  {Hattori} T.,
  {Toshikawa} J.,  {Ishikawa} S.,   {Onoue} M.,  2014, \mn@doi [\apj]
  {10.1088/0004-637X/780/2/116}, \href
  {https://ui.adsabs.harvard.edu/abs/2014ApJ...780..116K} {780, 116}

\bibitem[\protect\citeauthoryear{{Kauffmann} et~al.,}{{Kauffmann}
  et~al.}{2003}]{2003MNRAS.Kauffmann}
{Kauffmann} G.,  et~al., 2003, \mn@doi [\mnras]
  {10.1046/j.1365-8711.2003.06292.x}, \href
  {https://ui.adsabs.harvard.edu/abs/2003MNRAS.341...54K} {341, 54}

\bibitem[\protect\citeauthoryear{Kelvin et~al.,}{Kelvin
  et~al.}{2012}]{10.1111/j.1365-2966.2012.20355.x}
Kelvin L.~S.,  et~al., 2012, Monthly Notices of the Royal Astronomical Society,
  421, 1007

\bibitem[\protect\citeauthoryear{{Kennicutt}}{{Kennicutt}}{1998}]{1998ApJ...498..541K}
{Kennicutt} Robert~C. J.,  1998, \mn@doi [\apj] {10.1086/305588}, \href
  {https://ui.adsabs.harvard.edu/abs/1998ApJ...498..541K} {498, 541}

\bibitem[\protect\citeauthoryear{{Kravtsov}}{{Kravtsov}}{2013}]{2013ApJ...764L..31K}
{Kravtsov} A.~V.,  2013, \mn@doi [\apjl] {10.1088/2041-8205/764/2/L31}, \href
  {https://ui.adsabs.harvard.edu/abs/2013ApJ...764L..31K} {764, L31}

\bibitem[\protect\citeauthoryear{{Krogager}, {Fynbo}, {M{\o}ller}, {Ledoux},
  {Noterdaeme}, {Christensen}, {Milvang-Jensen}  \& {Sparre}}{{Krogager}
  et~al.}{2012}]{2012MNRAS.424L...1K}
{Krogager} J.~K.,  {Fynbo} J.~P.~U.,  {M{\o}ller} P.,  {Ledoux} C.,
  {Noterdaeme} P.,  {Christensen} L.,  {Milvang-Jensen} B.,   {Sparre} M.,
  2012, \mn@doi [\mnras] {10.1111/j.1745-3933.2012.01272.x}, \href
  {https://ui.adsabs.harvard.edu/abs/2012MNRAS.424L...1K} {424, L1}

\bibitem[\protect\citeauthoryear{{Krogager}, {M{\o}ller}, {Fynbo}  \&
  {Noterdaeme}}{{Krogager} et~al.}{2017}]{2017MNRAS.469.2959K}
{Krogager} J.~K.,  {M{\o}ller} P.,  {Fynbo} J.~P.~U.,   {Noterdaeme} P.,  2017,
  \mn@doi [\mnras] {10.1093/mnras/stx1011}, \href
  {https://ui.adsabs.harvard.edu/abs/2017MNRAS.469.2959K} {469, 2959}

\bibitem[\protect\citeauthoryear{{Kulkarni}, {Hill}, {Schneider}, {Weymann},
  {Storrie-Lombardi}, {Rieke}, {Thompson}  \& {Jannuzi}}{{Kulkarni}
  et~al.}{2000}]{2000ApJ...536...36K}
{Kulkarni} V.~P.,  {Hill} J.~M.,  {Schneider} G.,  {Weymann} R.~J.,
  {Storrie-Lombardi} L.~J.,  {Rieke} M.~J.,  {Thompson} R.~I.,   {Jannuzi}
  B.~T.,  2000, \mn@doi [\apj] {10.1086/308904}, \href
  {https://ui.adsabs.harvard.edu/abs/2000ApJ...536...36K} {536, 36}

\bibitem[\protect\citeauthoryear{{Kulkarni}, {Hill}, {Schneider}, {Weymann},
  {Storrie-Lombardi}, {Rieke}, {Thompson}  \& {Jannuzi}}{{Kulkarni}
  et~al.}{2001}]{2001ApJ...551...37K}
{Kulkarni} V.~P.,  {Hill} J.~M.,  {Schneider} G.,  {Weymann} R.~J.,
  {Storrie-Lombardi} L.~J.,  {Rieke} M.~J.,  {Thompson} R.~I.,   {Jannuzi}
  B.~T.,  2001, \mn@doi [\apj] {10.1086/320079}, \href
  {https://ui.adsabs.harvard.edu/abs/2001ApJ...551...37K} {551, 37}

\bibitem[\protect\citeauthoryear{{Kulkarni}, {Fall}, {Lauroesch}, {York},
  {Welty}, {Khare}  \& {Truran}}{{Kulkarni} et~al.}{2005}]{2005ApJ...618...68K}
{Kulkarni} V.~P.,  {Fall} S.~M.,  {Lauroesch} J.~T.,  {York} D.~G.,  {Welty}
  D.~E.,  {Khare} P.,   {Truran} J.~W.,  2005, \mn@doi [\apj] {10.1086/425956},
  \href {https://ui.adsabs.harvard.edu/abs/2005ApJ...618...68K} {618, 68}

\bibitem[\protect\citeauthoryear{{Kulkarni}, {Khare}, {Som}, {Meiring}, {York},
  {P{\'e}roux}  \& {Lauroesch}}{{Kulkarni} et~al.}{2010}]{2010NewA...15..735K}
{Kulkarni} V.~P.,  {Khare} P.,  {Som} D.,  {Meiring} J.,  {York} D.~G.,
  {P{\'e}roux} C.,   {Lauroesch} J.~T.,  2010, \mn@doi [\na]
  {10.1016/j.newast.2010.05.006}, \href
  {https://ui.adsabs.harvard.edu/abs/2010NewA...15..735K} {15, 735}

\bibitem[\protect\citeauthoryear{{Kulkarni}, {Bowen}, {Straka}, {York},
  {Gupta}, {Noterdaeme}  \& {Srianand}}{{Kulkarni}
  et~al.}{2022}]{2022ApJ...929..150K}
{Kulkarni} V.~P.,  {Bowen} D.~V.,  {Straka} L.~A.,  {York} D.~G.,  {Gupta} N.,
  {Noterdaeme} P.,   {Srianand} R.,  2022, \mn@doi [\apj]
  {10.3847/1538-4357/ac5fab}, \href
  {https://ui.adsabs.harvard.edu/abs/2022ApJ...929..150K} {929, 150}

\bibitem[\protect\citeauthoryear{{Lan}, {M{\'e}nard}  \& {Zhu}}{{Lan}
  et~al.}{2014}]{2014ApJ...795...31L}
{Lan} T.-W.,  {M{\'e}nard} B.,   {Zhu} G.,  2014, \mn@doi [\apj]
  {10.1088/0004-637X/795/1/31}, \href
  {https://ui.adsabs.harvard.edu/abs/2014ApJ...795...31L} {795, 31}

\bibitem[\protect\citeauthoryear{{Ledoux}, {Petitjean}, {Fynbo}, {M{\o}ller}
  \& {Srianand}}{{Ledoux} et~al.}{2006}]{2006A&A...457...71L}
{Ledoux} C.,  {Petitjean} P.,  {Fynbo} J.~P.~U.,  {M{\o}ller} P.,   {Srianand}
  R.,  2006, \mn@doi [\aap] {10.1051/0004-6361:20054242}, \href
  {https://ui.adsabs.harvard.edu/abs/2006A&A...457...71L} {457, 71}

\bibitem[\protect\citeauthoryear{{Lima-Dias} et~al.,}{{Lima-Dias}
  et~al.}{2021}]{2021MNRAS..Lima}
{Lima-Dias} C.,  et~al., 2021, \mn@doi [\mnras] {10.1093/mnras/staa3326}, \href
  {https://ui.adsabs.harvard.edu/abs/2021MNRAS.500.1323L} {500, 1323}

\bibitem[\protect\citeauthoryear{{Lofthouse} et~al.,}{{Lofthouse}
  et~al.}{2020}]{2020MNRAS.491.2057L}
{Lofthouse} E.~K.,  et~al., 2020, \mn@doi [\mnras] {10.1093/mnras/stz3066},
  \href {https://ui.adsabs.harvard.edu/abs/2020MNRAS.491.2057L} {491, 2057}

\bibitem[\protect\citeauthoryear{{Lofthouse} et~al.,}{{Lofthouse}
  et~al.}{2023}]{2023MNRAS.518..305L}
{Lofthouse} E.~K.,  et~al., 2023, \mn@doi [\mnras] {10.1093/mnras/stac3089},
  \href {https://ui.adsabs.harvard.edu/abs/2023MNRAS.518..305L} {518, 305}

\bibitem[\protect\citeauthoryear{{Lundgren} et~al.,}{{Lundgren}
  et~al.}{2012}]{2012ApJ...760...49L}
{Lundgren} B.~F.,  et~al., 2012, \mn@doi [\apj] {10.1088/0004-637X/760/1/49},
  \href {https://ui.adsabs.harvard.edu/abs/2012ApJ...760...49L} {760, 49}

\bibitem[\protect\citeauthoryear{{Ly}, {Malkan}, {Rigby}  \& {Nagao}}{{Ly}
  et~al.}{2016}]{2016ApJ...828...67L}
{Ly} C.,  {Malkan} M.~A.,  {Rigby} J.~R.,   {Nagao} T.,  2016, \mn@doi [\apj]
  {10.3847/0004-637X/828/2/67}, \href
  {https://ui.adsabs.harvard.edu/abs/2016ApJ...828...67L} {828, 67}

\bibitem[\protect\citeauthoryear{{Ma}, {Brammer}, {Ge}, {Prochaska}  \&
  {Lundgren}}{{Ma} et~al.}{2018}]{2018ApJ...857L..12M}
{Ma} J.,  {Brammer} G.,  {Ge} J.,  {Prochaska} J.~X.,   {Lundgren} B.,  2018,
  \mn@doi [\apjl] {10.3847/2041-8213/aabc51}, \href
  {https://ui.adsabs.harvard.edu/abs/2018ApJ...857L..12M} {857, L12}

\bibitem[\protect\citeauthoryear{{Marinacci} et~al.,}{{Marinacci}
  et~al.}{2018}]{2018MNRAS.480.5113M}
{Marinacci} F.,  et~al., 2018, \mn@doi [\mnras] {10.1093/mnras/sty2206}, \href
  {https://ui.adsabs.harvard.edu/abs/2018MNRAS.480.5113M} {480, 5113}

\bibitem[\protect\citeauthoryear{McConnachie}{McConnachie}{2012}]{McConnachie_2012}
McConnachie A.~W.,  2012, \mn@doi [\aj] {10.1088/0004-6256/144/1/4}, 144, 4

\bibitem[\protect\citeauthoryear{McKinney et~al.}{McKinney
  et~al.}{2010}]{mckinney2010data}
McKinney W.,  et~al., 2010, in Proceedings of the 9th Python in Science
  Conference. pp 51--56

\bibitem[\protect\citeauthoryear{{Meiring}, {Lauroesch}, {Kulkarni},
  {P{\'e}roux}, {Khare}, {York}  \& {Crotts}}{{Meiring}
  et~al.}{2007}]{2007MNRAS.376..557M}
{Meiring} J.~D.,  {Lauroesch} J.~T.,  {Kulkarni} V.~P.,  {P{\'e}roux} C.,
  {Khare} P.,  {York} D.~G.,   {Crotts} A. P.~S.,  2007, \mn@doi [\mnras]
  {10.1111/j.1365-2966.2007.11521.x}, \href
  {https://ui.adsabs.harvard.edu/abs/2007MNRAS.376..557M} {376, 557}

\bibitem[\protect\citeauthoryear{{Meiring}, {Lauroesch}, {Kulkarni},
  {P{\'e}roux}, {Khare}  \& {York}}{{Meiring}
  et~al.}{2009}]{2009MNRAS.397.2037M}
{Meiring} J.~D.,  {Lauroesch} J.~T.,  {Kulkarni} V.~P.,  {P{\'e}roux} C.,
  {Khare} P.,   {York} D.~G.,  2009, \mn@doi [\mnras]
  {10.1111/j.1365-2966.2009.15064.x}, \href
  {https://ui.adsabs.harvard.edu/abs/2009MNRAS.397.2037M} {397, 2037}

\bibitem[\protect\citeauthoryear{{M{\'e}nard} \& {Chelouche}}{{M{\'e}nard} \&
  {Chelouche}}{2009}]{2009MNRAS.393..808M}
{M{\'e}nard} B.,  {Chelouche} D.,  2009, \mn@doi [\mnras]
  {10.1111/j.1365-2966.2008.14225.x}, \href
  {https://ui.adsabs.harvard.edu/abs/2009MNRAS.393..808M} {393, 808}

\bibitem[\protect\citeauthoryear{Mowla et~al.,}{Mowla
  et~al.}{2019}]{Mowla_2019}
Mowla L.~A.,  et~al., 2019, \mn@doi [\apj] {10.3847/1538-4357/ab290a}, 880, 57

\bibitem[\protect\citeauthoryear{{Muzahid}, {Kacprzak}, {Charlton}  \&
  {Churchill}}{{Muzahid} et~al.}{2016}]{2016ApJ...823...66M}
{Muzahid} S.,  {Kacprzak} G.~G.,  {Charlton} J.~C.,   {Churchill} C.~W.,  2016,
  \mn@doi [\apj] {10.3847/0004-637X/823/1/66}, \href
  {https://ui.adsabs.harvard.edu/abs/2016ApJ...823...66M} {823, 66}

\bibitem[\protect\citeauthoryear{{Muzahid} et~al.,}{{Muzahid}
  et~al.}{2020}]{2020MNRAS.496.1013M}
{Muzahid} S.,  et~al., 2020, \mn@doi [\mnras] {10.1093/mnras/staa1347}, \href
  {https://ui.adsabs.harvard.edu/abs/2020MNRAS.496.1013M} {496, 1013}

\bibitem[\protect\citeauthoryear{{Nagamine}, {Springel}  \&
  {Hernquist}}{{Nagamine} et~al.}{2004}]{2004MNRAS.348..435N}
{Nagamine} K.,  {Springel} V.,   {Hernquist} L.,  2004, \mn@doi [\mnras]
  {10.1111/j.1365-2966.2004.07180.x}, \href
  {https://ui.adsabs.harvard.edu/abs/2004MNRAS.348..435N} {348, 435}

\bibitem[\protect\citeauthoryear{{Naiman} et~al.,}{{Naiman}
  et~al.}{2018}]{2018MNRAS.477.1206N}
{Naiman} J.~P.,  et~al., 2018, \mn@doi [\mnras] {10.1093/mnras/sty618}, \href
  {https://ui.adsabs.harvard.edu/abs/2018MNRAS.477.1206N} {477, 1206}

\bibitem[\protect\citeauthoryear{{Nelson} et~al.,}{{Nelson}
  et~al.}{2018}]{2018MNRAS.475..624N}
{Nelson} D.,  et~al., 2018, \mn@doi [\mnras] {10.1093/mnras/stx3040}, \href
  {https://ui.adsabs.harvard.edu/abs/2018MNRAS.475..624N} {475, 624}

\bibitem[\protect\citeauthoryear{{Nelson} et~al.,}{{Nelson}
  et~al.}{2019}]{2019ComAC...6....2N}
{Nelson} D.,  et~al., 2019, \mn@doi [Computational Astrophysics and Cosmology]
  {10.1186/s40668-019-0028-x}, \href
  {https://ui.adsabs.harvard.edu/abs/2019ComAC...6....2N} {6, 2}

\bibitem[\protect\citeauthoryear{{Nielsen}, {Churchill}  \&
  {Kacprzak}}{{Nielsen} et~al.}{2013}]{2013ApJ...776..115N}
{Nielsen} N.~M.,  {Churchill} C.~W.,   {Kacprzak} G.~G.,  2013, \mn@doi [\apj]
  {10.1088/0004-637X/776/2/115}, \href
  {https://ui.adsabs.harvard.edu/abs/2013ApJ...776..115N} {776, 115}

\bibitem[\protect\citeauthoryear{{Noterdaeme} et~al.,}{{Noterdaeme}
  et~al.}{2012}]{2012A&A...547L...1N}
{Noterdaeme} P.,  et~al., 2012, \mn@doi [\aap] {10.1051/0004-6361/201220259},
  \href {https://ui.adsabs.harvard.edu/abs/2012A&A...547L...1N} {547, L1}

\bibitem[\protect\citeauthoryear{{Noterdaeme}, {Petitjean}, {P{\^a}ris}, {Cai},
  {Finley}, {Ge}, {Pieri}  \& {York}}{{Noterdaeme}
  et~al.}{2014}]{2014A&A...566A..24N}
{Noterdaeme} P.,  {Petitjean} P.,  {P{\^a}ris} I.,  {Cai} Z.,  {Finley} H.,
  {Ge} J.,  {Pieri} M.~M.,   {York} D.~G.,  2014, \mn@doi [\aap]
  {10.1051/0004-6361/201322809}, \href
  {https://ui.adsabs.harvard.edu/abs/2014A&A...566A..24N} {566, A24}

\bibitem[\protect\citeauthoryear{{Peng}, {Ho}, {Impey}  \& {Rix}}{{Peng}
  et~al.}{2002}]{2002AJ....124..266P}
{Peng} C.~Y.,  {Ho} L.~C.,  {Impey} C.~D.,   {Rix} H.-W.,  2002, \mn@doi [\aj]
  {10.1086/340952}, \href
  {https://ui.adsabs.harvard.edu/abs/2002AJ....124..266P} {124, 266}

\bibitem[\protect\citeauthoryear{{P{\'e}roux} \& {Howk}}{{P{\'e}roux} \&
  {Howk}}{2020}]{2020ARA&A..58..363P}
{P{\'e}roux} C.,  {Howk} J.~C.,  2020, \mn@doi [\araa]
  {10.1146/annurev-astro-021820-120014}, \href
  {https://ui.adsabs.harvard.edu/abs/2020ARA&A..58..363P} {58, 363}

\bibitem[\protect\citeauthoryear{{P{\'e}roux}, {Dessauges-Zavadsky},
  {D'Odorico}, {Kim}  \& {McMahon}}{{P{\'e}roux}
  et~al.}{2003}]{2003MNRAS.345..480P}
{P{\'e}roux} C.,  {Dessauges-Zavadsky} M.,  {D'Odorico} S.,  {Kim} T.-S.,
  {McMahon} R.~G.,  2003, \mn@doi [\mnras] {10.1046/j.1365-8711.2003.06952.x},
  \href {https://ui.adsabs.harvard.edu/abs/2003MNRAS.345..480P} {345, 480}

\bibitem[\protect\citeauthoryear{{P{\'e}roux}, {Meiring}, {Kulkarni}, {Khare},
  {Lauroesch}, {Vladilo}  \& {York}}{{P{\'e}roux}
  et~al.}{2008}]{2008MNRAS.386.2209P}
{P{\'e}roux} C.,  {Meiring} J.~D.,  {Kulkarni} V.~P.,  {Khare} P.,  {Lauroesch}
  J.~T.,  {Vladilo} G.,   {York} D.~G.,  2008, \mn@doi [\mnras]
  {10.1111/j.1365-2966.2008.13186.x}, \href
  {https://ui.adsabs.harvard.edu/abs/2008MNRAS.386.2209P} {386, 2209}

\bibitem[\protect\citeauthoryear{{P{\'e}roux}, {Bouch{\'e}}, {Kulkarni}, {York}
   \& {Vladilo}}{{P{\'e}roux} et~al.}{2011}]{2011MNRAS.410.2237P}
{P{\'e}roux} C.,  {Bouch{\'e}} N.,  {Kulkarni} V.~P.,  {York} D.~G.,
  {Vladilo} G.,  2011, \mn@doi [\mnras] {10.1111/j.1365-2966.2010.17598.x},
  \href {https://ui.adsabs.harvard.edu/abs/2011MNRAS.410.2237P} {410, 2237}

\bibitem[\protect\citeauthoryear{{P{\'e}roux}, {Bouch{\'e}}, {Kulkarni}, {York}
   \& {Vladilo}}{{P{\'e}roux} et~al.}{2012}]{2012MNRAS..Celine}
{P{\'e}roux} C.,  {Bouch{\'e}} N.,  {Kulkarni} V.~P.,  {York} D.~G.,
  {Vladilo} G.,  2012, \mn@doi [\mnras] {10.1111/j.1365-2966.2011.19947.x},
  \href {https://ui.adsabs.harvard.edu/abs/2012MNRAS.419.3060P} {419, 3060}

\bibitem[\protect\citeauthoryear{{P{\'e}roux} et~al.,}{{P{\'e}roux}
  et~al.}{2016}]{2016MNRAS.457..903P}
{P{\'e}roux} C.,  et~al., 2016, \mn@doi [\mnras] {10.1093/mnras/stw016}, \href
  {https://ui.adsabs.harvard.edu/abs/2016MNRAS.457..903P} {457, 903}

\bibitem[\protect\citeauthoryear{{P{\'e}roux} et~al.,}{{P{\'e}roux}
  et~al.}{2019}]{2019MNRAS.485.1595P}
{P{\'e}roux} C.,  et~al., 2019, \mn@doi [\mnras] {10.1093/mnras/stz202}, \href
  {https://ui.adsabs.harvard.edu/abs/2019MNRAS.485.1595P} {485, 1595}

\bibitem[\protect\citeauthoryear{{P{\'e}roux} et~al.,}{{P{\'e}roux}
  et~al.}{2022}]{2022MNRAS.516.5618P}
{P{\'e}roux} C.,  et~al., 2022, \mn@doi [\mnras] {10.1093/mnras/stac2546},
  \href {https://ui.adsabs.harvard.edu/abs/2022MNRAS.516.5618P} {516, 5618}

\bibitem[\protect\citeauthoryear{{Pillepich} et~al.,}{{Pillepich}
  et~al.}{2018}]{2018MNRAS.475..648P}
{Pillepich} A.,  et~al., 2018, \mn@doi [\mnras] {10.1093/mnras/stx3112}, \href
  {https://ui.adsabs.harvard.edu/abs/2018MNRAS.475..648P} {475, 648}

\bibitem[\protect\citeauthoryear{{Prochaska} \& {Wolfe}}{{Prochaska} \&
  {Wolfe}}{2009}]{2009ApJ...696.1543P}
{Prochaska} J.~X.,  {Wolfe} A.~M.,  2009, \mn@doi [\apj]
  {10.1088/0004-637X/696/2/1543}, \href
  {https://ui.adsabs.harvard.edu/abs/2009ApJ...696.1543P} {696, 1543}

\bibitem[\protect\citeauthoryear{{Prochaska}, {Wolfe}, {Howk}, {Gawiser},
  {Burles}  \& {Cooke}}{{Prochaska} et~al.}{2007}]{2007ApJS..171...29P}
{Prochaska} J.~X.,  {Wolfe} A.~M.,  {Howk} J.~C.,  {Gawiser} E.,  {Burles}
  S.~M.,   {Cooke} J.,  2007, \mn@doi [\apjs] {10.1086/513714}, \href
  {https://ui.adsabs.harvard.edu/abs/2007ApJS..171...29P} {171, 29}

\bibitem[\protect\citeauthoryear{{Rafelski}, {Wolfe}  \& {Chen}}{{Rafelski}
  et~al.}{2011}]{2011ApJ...736...48R}
{Rafelski} M.,  {Wolfe} A.~M.,   {Chen} H.-W.,  2011, \mn@doi [\apj]
  {10.1088/0004-637X/736/1/48}, \href
  {https://ui.adsabs.harvard.edu/abs/2011ApJ...736...48R} {736, 48}

\bibitem[\protect\citeauthoryear{{Rafelski}, {Wolfe}, {Prochaska}, {Neeleman}
  \& {Mendez}}{{Rafelski} et~al.}{2012}]{2012ApJ...755...89R}
{Rafelski} M.,  {Wolfe} A.~M.,  {Prochaska} J.~X.,  {Neeleman} M.,   {Mendez}
  A.~J.,  2012, \mn@doi [\apj] {10.1088/0004-637X/755/2/89}, \href
  {https://ui.adsabs.harvard.edu/abs/2012ApJ...755...89R} {755, 89}

\bibitem[\protect\citeauthoryear{Rafelski, Gardner, Fumagalli, Neeleman,
  Teplitz, Grogin, Koekemoer  \& Scarlata}{Rafelski
  et~al.}{2016}]{Rafelski_2016}
Rafelski M.,  Gardner J.~P.,  Fumagalli M.,  Neeleman M.,  Teplitz H.~I.,
  Grogin N.,  Koekemoer A.~M.,   Scarlata C.,  2016, \mn@doi [\apj]
  {10.3847/0004-637X/825/2/87}, 825, 87

\bibitem[\protect\citeauthoryear{{Rahmani} et~al.,}{{Rahmani}
  et~al.}{2016}]{2016MNRAS.463..980R}
{Rahmani} H.,  et~al., 2016, \mn@doi [\mnras] {10.1093/mnras/stw1965}, \href
  {https://ui.adsabs.harvard.edu/abs/2016MNRAS.463..980R} {463, 980}

\bibitem[\protect\citeauthoryear{{Rahmani} et~al.,}{{Rahmani}
  et~al.}{2018}]{2018MNRAS.480.5046R}
{Rahmani} H.,  et~al., 2018, \mn@doi [\mnras] {10.1093/mnras/sty2216}, \href
  {https://ui.adsabs.harvard.edu/abs/2018MNRAS.480.5046R} {480, 5046}

\bibitem[\protect\citeauthoryear{{Rao}, {Turnshek}  \& {Nestor}}{{Rao}
  et~al.}{2006}]{2006ApJ...636..610R}
{Rao} S.~M.,  {Turnshek} D.~A.,   {Nestor} D.~B.,  2006, \mn@doi [\apj]
  {10.1086/498132}, \href
  {https://ui.adsabs.harvard.edu/abs/2006ApJ...636..610R} {636, 610}

\bibitem[\protect\citeauthoryear{{Rhodin}, {Christensen}, {M{\o}ller}, {Zafar}
  \& {Fynbo}}{{Rhodin} et~al.}{2018}]{2018A&A...618A.129R}
{Rhodin} N.~H.~P.,  {Christensen} L.,  {M{\o}ller} P.,  {Zafar} T.,   {Fynbo}
  J.~P.~U.,  2018, \mn@doi [\aap] {10.1051/0004-6361/201832992}, \href
  {https://ui.adsabs.harvard.edu/abs/2018A&A...618A.129R} {618, A129}

\bibitem[\protect\citeauthoryear{{Rhodin}, {Krogager}, {Christensen},
  {Valentino}, {Heintz}, {M{\o}ller}, {Zafar}  \& {Fynbo}}{{Rhodin}
  et~al.}{2021}]{2021MNRAS.506..546R}
{Rhodin} N.~H.~P.,  {Krogager} J.~K.,  {Christensen} L.,  {Valentino} F.,
  {Heintz} K.~E.,  {M{\o}ller} P.,  {Zafar} T.,   {Fynbo} J.~P.~U.,  2021,
  \mn@doi [\mnras] {10.1093/mnras/stab1691}, \href
  {https://ui.adsabs.harvard.edu/abs/2021MNRAS.506..546R} {506, 546}

\bibitem[\protect\citeauthoryear{{Rubin}, {Diamond-Stanic}, {Coil}, {Crighton}
  \& {Moustakas}}{{Rubin} et~al.}{2018}]{2018ApJ...853...95R}
{Rubin} K. H.~R.,  {Diamond-Stanic} A.~M.,  {Coil} A.~L.,  {Crighton} N. H.~M.,
    {Moustakas} J.,  2018, \mn@doi [\apj] {10.3847/1538-4357/aa9792}, \href
  {https://ui.adsabs.harvard.edu/abs/2018ApJ...853...95R} {853, 95}

\bibitem[\protect\citeauthoryear{{Schroetter} et~al.,}{{Schroetter}
  et~al.}{2016}]{2016ApJ...833...39S}
{Schroetter} I.,  et~al., 2016, \mn@doi [\apj] {10.3847/1538-4357/833/1/39},
  \href {https://ui.adsabs.harvard.edu/abs/2016ApJ...833...39S} {833, 39}

\bibitem[\protect\citeauthoryear{{Schroetter} et~al.,}{{Schroetter}
  et~al.}{2019}]{2019MNRAS.490.4368S}
{Schroetter} I.,  et~al., 2019, \mn@doi [\mnras] {10.1093/mnras/stz2822}, \href
  {https://ui.adsabs.harvard.edu/abs/2019MNRAS.490.4368S} {490, 4368}

\bibitem[\protect\citeauthoryear{Seo \& Ann}{Seo \& Ann}{2022}]{10.1093...Seo}
Seo M.,  Ann H.~B.,  2022, \mn@doi [\mnras] {10.1093/mnras/stac1719}, 514, 5853

\bibitem[\protect\citeauthoryear{{Shen}, {Mo}, {White}, {Blanton}, {Kauffmann},
  {Voges}, {Brinkmann}  \& {Csabai}}{{Shen} et~al.}{2003}]{2003MNRAS..Shen}
{Shen} S.,  {Mo} H.~J.,  {White} S. D.~M.,  {Blanton} M.~R.,  {Kauffmann} G.,
  {Voges} W.,  {Brinkmann} J.,   {Csabai} I.,  2003, \mn@doi [\mnras]
  {10.1046/j.1365-8711.2003.06740.x}, \href
  {https://ui.adsabs.harvard.edu/abs/2003MNRAS.343..978S} {343, 978}

\bibitem[\protect\citeauthoryear{{Som}, {Kulkarni}, {Meiring}, {York},
  {P{\'e}roux}, {Lauroesch}, {Aller}  \& {Khare}}{{Som}
  et~al.}{2015}]{2015ApJ...806...25S}
{Som} D.,  {Kulkarni} V.~P.,  {Meiring} J.,  {York} D.~G.,  {P{\'e}roux} C.,
  {Lauroesch} J.~T.,  {Aller} M.~C.,   {Khare} P.,  2015, \mn@doi [\apj]
  {10.1088/0004-637X/806/1/25}, \href
  {https://ui.adsabs.harvard.edu/abs/2015ApJ...806...25S} {806, 25}

\bibitem[\protect\citeauthoryear{{Springel}, {White}, {Tormen}  \&
  {Kauffmann}}{{Springel} et~al.}{2001}]{2001MNRAS.328..726S}
{Springel} V.,  {White} S. D.~M.,  {Tormen} G.,   {Kauffmann} G.,  2001,
  \mn@doi [\mnras] {10.1046/j.1365-8711.2001.04912.x}, \href
  {https://ui.adsabs.harvard.edu/abs/2001MNRAS.328..726S} {328, 726}

\bibitem[\protect\citeauthoryear{{Springel} et~al.,}{{Springel}
  et~al.}{2018}]{2018MNRAS.475..676S}
{Springel} V.,  et~al., 2018, \mn@doi [\mnras] {10.1093/mnras/stx3304}, \href
  {https://ui.adsabs.harvard.edu/abs/2018MNRAS.475..676S} {475, 676}

\bibitem[\protect\citeauthoryear{{Sternberg}, {Le Petit}, {Roueff}  \& {Le
  Bourlot}}{{Sternberg} et~al.}{2014}]{2014ApJ...790...10S}
{Sternberg} A.,  {Le Petit} F.,  {Roueff} E.,   {Le Bourlot} J.,  2014, \mn@doi
  [\apj] {10.1088/0004-637X/790/1/10}, \href
  {https://ui.adsabs.harvard.edu/abs/2014ApJ...790...10S} {790, 10}

\bibitem[\protect\citeauthoryear{{Straka}, {Kulkarni}  \& {York}}{{Straka}
  et~al.}{2011}]{2011AJ....141..206S}
{Straka} L.~A.,  {Kulkarni} V.~P.,   {York} D.~G.,  2011, \mn@doi [\aj]
  {10.1088/0004-6256/141/6/206}, \href
  {https://ui.adsabs.harvard.edu/abs/2011AJ....141..206S} {141, 206}

\bibitem[\protect\citeauthoryear{{Tody}}{{Tody}}{1986}]{1986SPIE..627..733T}
{Tody} D.,  1986, in {Crawford} D.~L.,  ed.,  Society of Photo-Optical
  Instrumentation Engineers (SPIE) Conference Series Vol. 627, Instrumentation
  in astronomy VI. p.~733, \mn@doi{10.1117/12.968154}

\bibitem[\protect\citeauthoryear{{Tumlinson}, {Peeples}  \& {Werk}}{{Tumlinson}
  et~al.}{2017}]{2017ARA&A..55..389T}
{Tumlinson} J.,  {Peeples} M.~S.,   {Werk} J.~K.,  2017, \mn@doi [\araa]
  {10.1146/annurev-astro-091916-055240}, \href
  {https://ui.adsabs.harvard.edu/abs/2017ARA&A..55..389T} {55, 389}

\bibitem[\protect\citeauthoryear{{Vladilo}, {Abate}, {Yin}, {Cescutti}  \&
  {Matteucci}}{{Vladilo} et~al.}{2011}]{2011A&A...530A..33V}
{Vladilo} G.,  {Abate} C.,  {Yin} J.,  {Cescutti} G.,   {Matteucci} F.,  2011,
  \mn@doi [\aap] {10.1051/0004-6361/201016330}, \href
  {https://ui.adsabs.harvard.edu/abs/2011A&A...530A..33V} {530, A33}

\bibitem[\protect\citeauthoryear{{Weng} et~al.,}{{Weng}
  et~al.}{2023}]{2023MNRAS.519..931W}
{Weng} S.,  et~al., 2023, \mn@doi [\mnras] {10.1093/mnras/stac3497}, \href
  {https://ui.adsabs.harvard.edu/abs/2023MNRAS.519..931W} {519, 931}

\bibitem[\protect\citeauthoryear{{Wolfe} \& {Chen}}{{Wolfe} \&
  {Chen}}{2006}]{2006ApJ...652..981W}
{Wolfe} A.~M.,  {Chen} H.-W.,  2006, \mn@doi [\apj] {10.1086/507574}, \href
  {https://ui.adsabs.harvard.edu/abs/2006ApJ...652..981W} {652, 981}

\bibitem[\protect\citeauthoryear{{Wolfe} \& {Wills}}{{Wolfe} \&
  {Wills}}{1977}]{1977ApJ...218...39W}
{Wolfe} A.~M.,  {Wills} B.~J.,  1977, \mn@doi [\apj] {10.1086/155655}, \href
  {https://ui.adsabs.harvard.edu/abs/1977ApJ...218...39W} {218, 39}

\bibitem[\protect\citeauthoryear{{Wolfe}, {Gawiser}  \& {Prochaska}}{{Wolfe}
  et~al.}{2005}]{2005ARA&A..43..861W}
{Wolfe} A.~M.,  {Gawiser} E.,   {Prochaska} J.~X.,  2005, \mn@doi [\araa]
  {10.1146/annurev.astro.42.053102.133950}, \href
  {https://ui.adsabs.harvard.edu/abs/2005ARA&A..43..861W} {43, 861}

\bibitem[\protect\citeauthoryear{{Wolfe}, {Prochaska}, {Jorgenson}  \&
  {Rafelski}}{{Wolfe} et~al.}{2008}]{2008ApJ...681..881W}
{Wolfe} A.~M.,  {Prochaska} J.~X.,  {Jorgenson} R.~A.,   {Rafelski} M.,  2008,
  \mn@doi [\apj] {10.1086/588090}, \href
  {https://ui.adsabs.harvard.edu/abs/2008ApJ...681..881W} {681, 881}

\bibitem[\protect\citeauthoryear{{Wootten} \& {Thompson}}{{Wootten} \&
  {Thompson}}{2009}]{2009IEEEP..97.1463W}
{Wootten} A.,  {Thompson} A.~R.,  2009, \mn@doi [IEEE Proceedings]
  {10.1109/JPROC.2009.2020572}, \href
  {https://ui.adsabs.harvard.edu/abs/2009IEEEP..97.1463W} {97, 1463}

\bibitem[\protect\citeauthoryear{{York}, {Dopita}, {Green}  \&
  {Bechtold}}{{York} et~al.}{1986}]{1986ApJ...311..610Y}
{York} D.~G.,  {Dopita} M.,  {Green} R.,   {Bechtold} J.,  1986, \mn@doi [\apj]
  {10.1086/164800}, \href
  {https://ui.adsabs.harvard.edu/abs/1986ApJ...311..610Y} {311, 610}

\bibitem[\protect\citeauthoryear{{Zabl} et~al.,}{{Zabl}
  et~al.}{2019}]{2019MNRAS.485.1961Z}
{Zabl} J.,  et~al., 2019, \mn@doi [\mnras] {10.1093/mnras/stz392}, \href
  {https://ui.adsabs.harvard.edu/abs/2019MNRAS.485.1961Z} {485, 1961}

\bibitem[\protect\citeauthoryear{{Zafar}, {M{\o}ller}, {P{\'e}roux}, {Quiret},
  {Fynbo}, {Ledoux}  \& {Deharveng}}{{Zafar}
  et~al.}{2017}]{2017MNRAS.465.1613Z}
{Zafar} T.,  {M{\o}ller} P.,  {P{\'e}roux} C.,  {Quiret} S.,  {Fynbo} J. P.~U.,
   {Ledoux} C.,   {Deharveng} J.-M.,  2017, \mn@doi [\mnras]
  {10.1093/mnras/stw2907}, \href
  {https://ui.adsabs.harvard.edu/abs/2017MNRAS.465.1613Z} {465, 1613}

\bibitem[\protect\citeauthoryear{{van der Wel} et~al.,}{{van der Wel}
  et~al.}{2014}]{2014ApJ...788...28V}
{van der Wel} A.,  et~al., 2014, \mn@doi [\apj] {10.1088/0004-637X/788/1/28},
  \href {https://ui.adsabs.harvard.edu/abs/2014ApJ...788...28V} {788, 28}

\makeatother
\end{thebibliography}

%\begin{thebibliography}{99}
%\end{thebibliography}

%%%%%%%%%%%%%%%%%%%%%%%%%%%%%%%%%%%%%%%%%%%%%%%%%%

%%%%%%%%%%%%%%%%% APPENDICES %%%%%%%%%%%%%%%%%%%%%

\appendix

\section{Images of Individual Fields}

We show below the broad-band \emph{HST} images in the reddest filter available for each of the remaining quasar fields, similar to the image of the field of Q0152+0023 shown in \autoref{fig1}.

\begin{figure*}
    \includegraphics[width=0.95\textwidth]{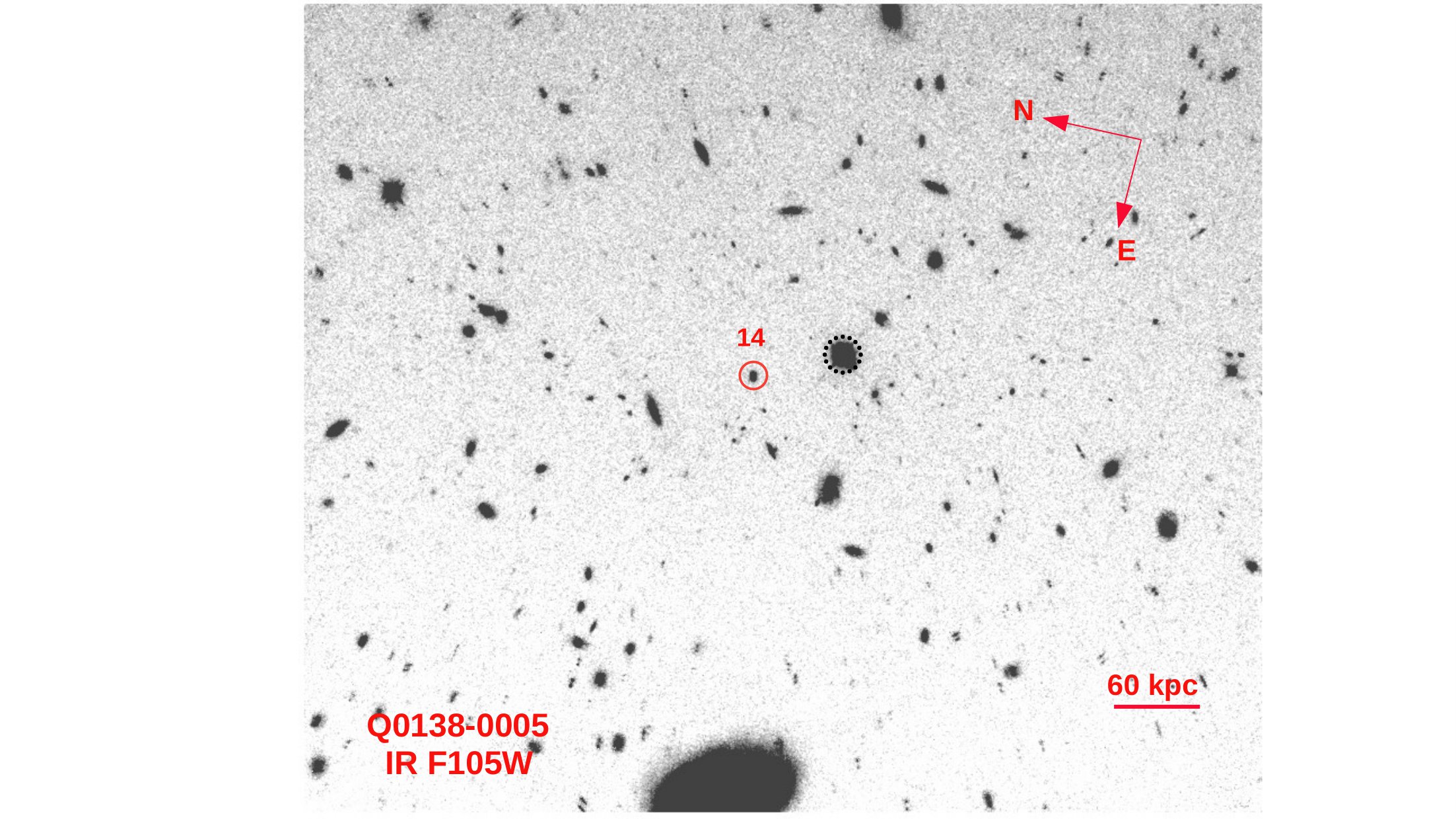}\\

    \vspace{1.0cm}
    \includegraphics[width=0.95\textwidth]{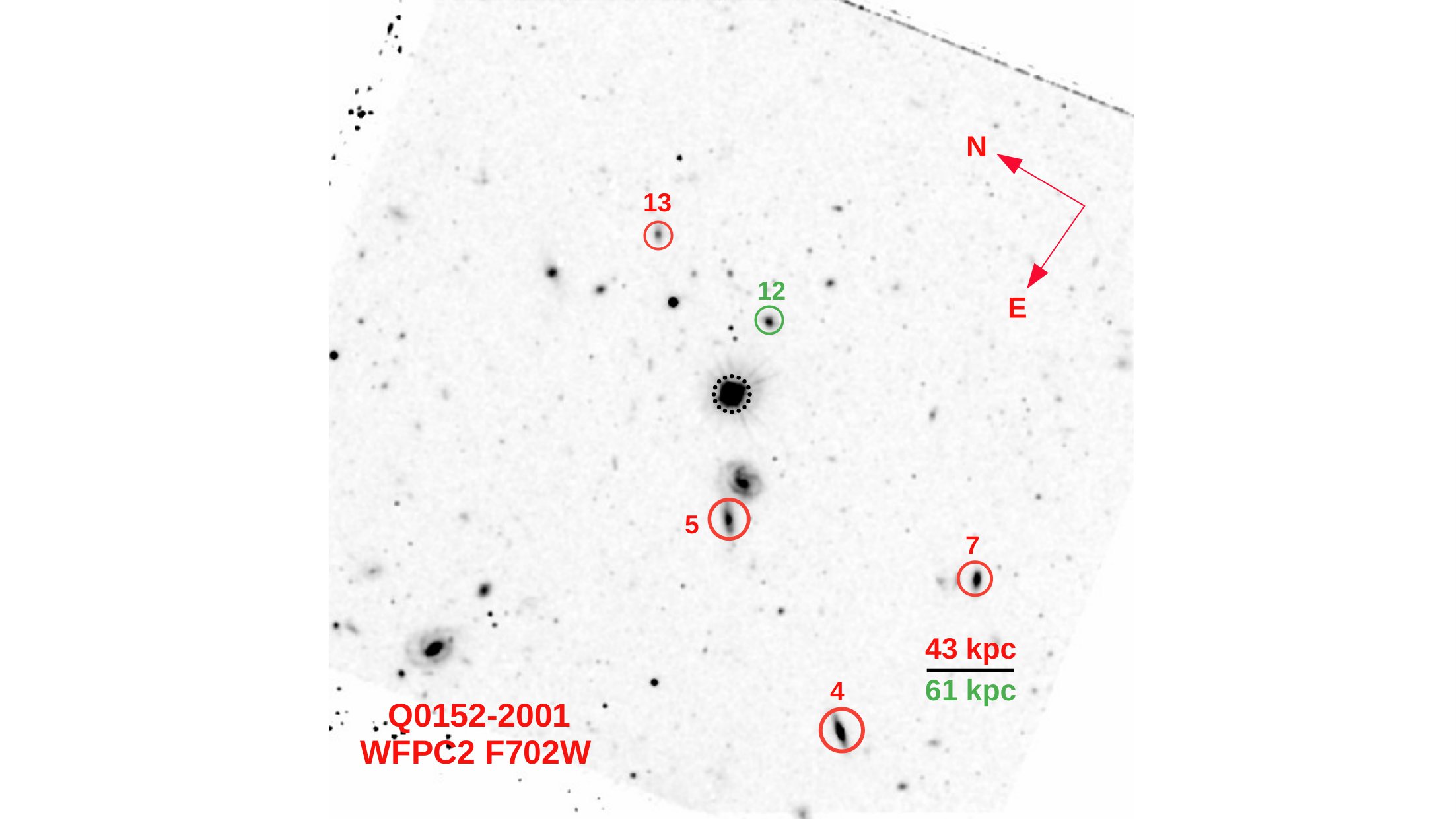}
    \vspace{1.0cm}
    
    \caption{Similar figures to \autoref{fig1}. Shown figures are the median stacked images of the quasar fields in the full-frame. The solid circles represent the position of the associated galaxies in the field of view, while a dotted black circle denotes the QSO's position. The associated galaxies are located within $\pm$ 500 $\rm km  s^{-1}$ of the absorber's redshift. The object identification number (ID) of these galaxies came from the MUSE-ALMA Haloes master table discussed in \citet{2022MNRAS.516.5618P}. While the scale corresponds to 60 kpc at $z_{\rm abs}$ = 0.7821 for the top figure, the scale corresponds to 43 kpc and 61 kpc at $z_{\rm abs}$ = 0.3830 and 0.7802 for the bottom figure, respectively. 
    }
    
    \label{fig14}

\end{figure*}

\begin{figure*}

    \includegraphics[width=0.9\textwidth]{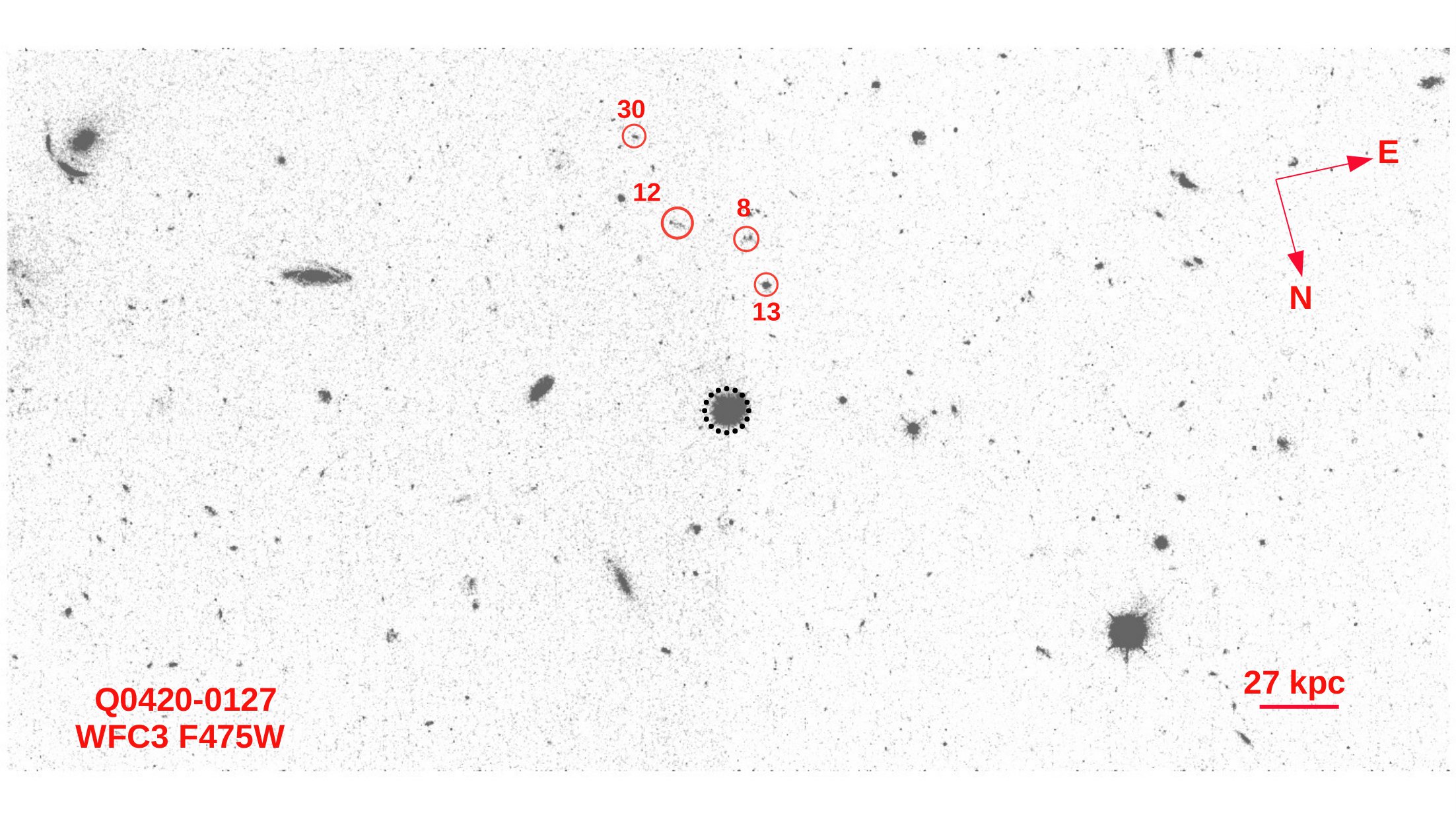}\\

    \vspace{1cm}

%    \label{fig20}

\end{figure*}

\begin{figure*}
  \centering
  \vspace{0.1cm}
  \begin{subfigure}{.4\linewidth}
    \centering
     \includegraphics[width = 1.0\linewidth]{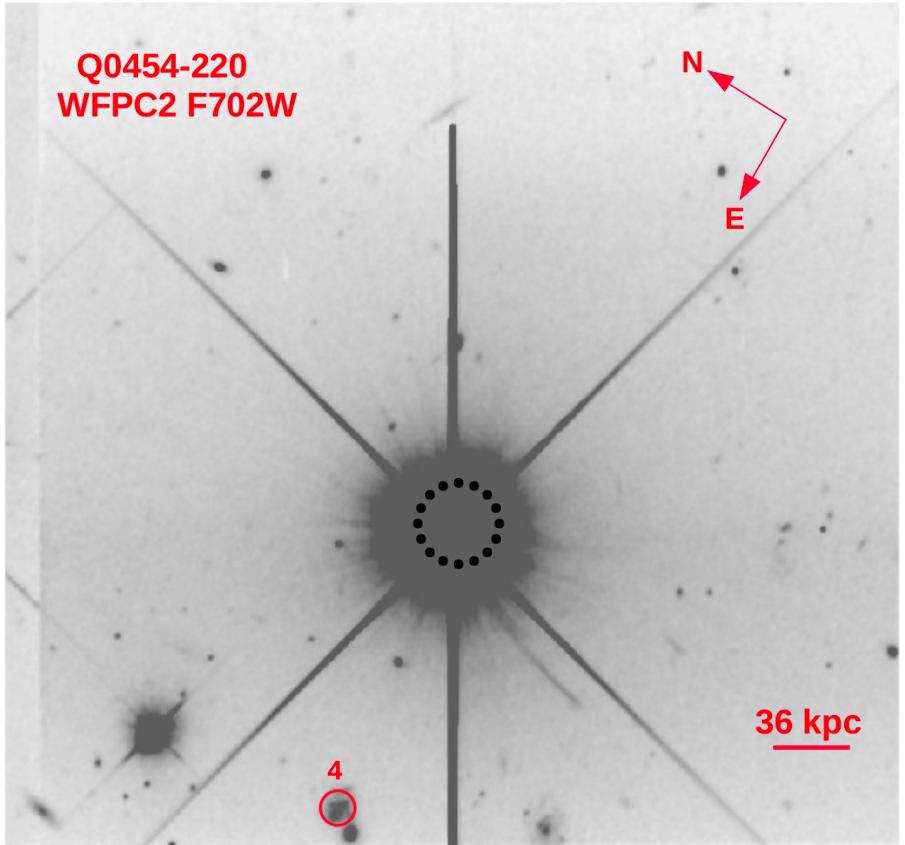}
  \end{subfigure}
  \hspace{1cm}
  %\vspace{0.1cm}
  \begin{subfigure}{.4\linewidth}
    \centering

      \includegraphics[width = 1.0\linewidth]{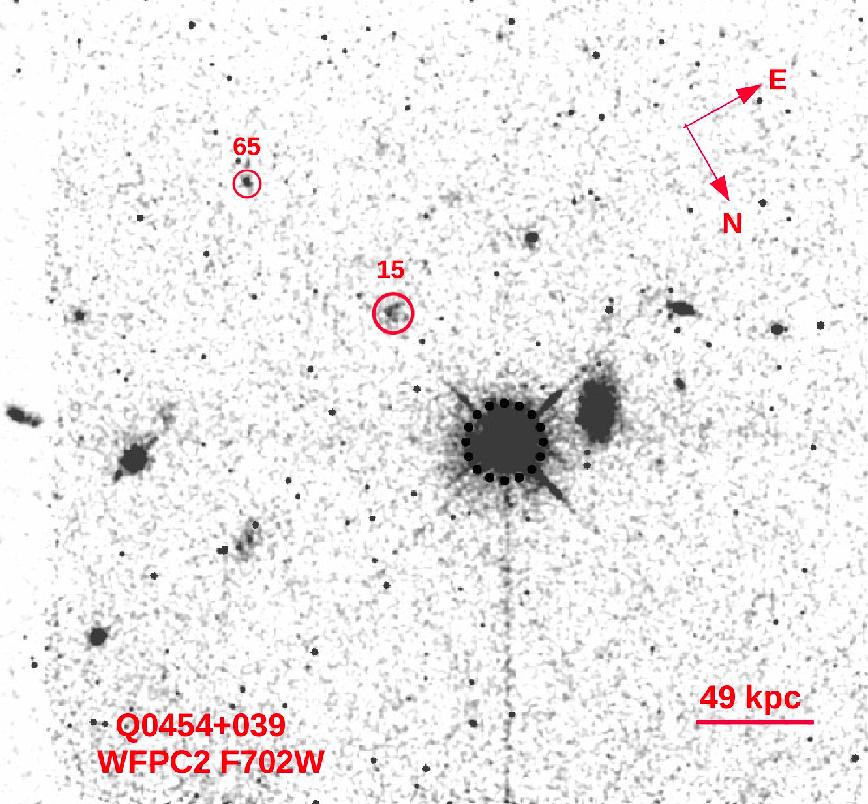}  
  \end{subfigure}

   \caption{See \autoref{fig14} caption. The scale corresponds to 27 kpc at $z_{\rm abs}$ = 0.6331 for the top figure, while in the bottom figures, the scale corresponds to 36 kpc and 49 kpc at $z_{\rm abs}$ = 0.4833 and 1.1532 for the left and right figure, respectively.}
 
\end{figure*}

\begin{figure*}

       \includegraphics[width=0.9\textwidth]{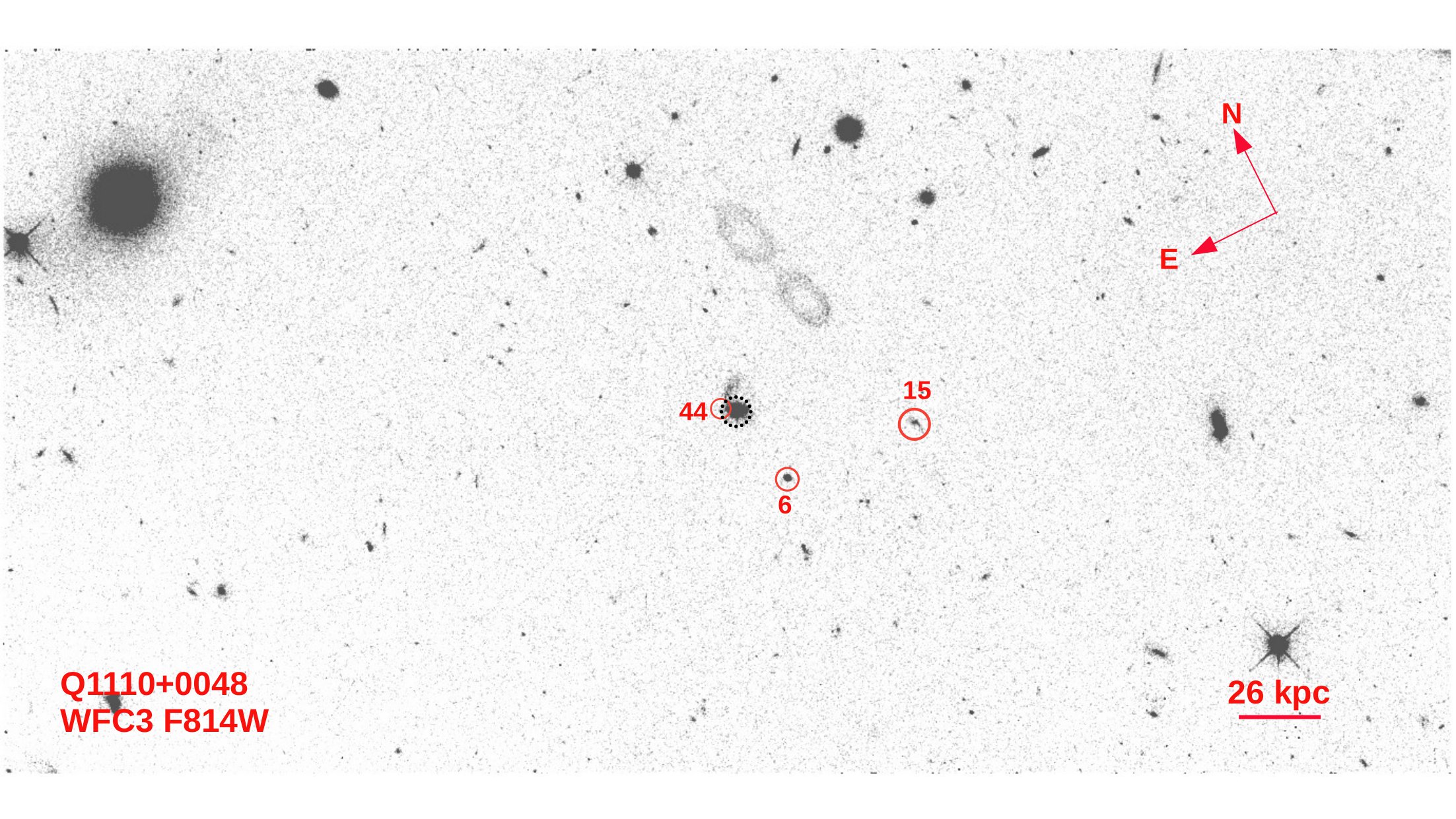}\\
 %   \caption{}
    \vspace{1cm}
   \includegraphics[width= 1.0\textwidth]{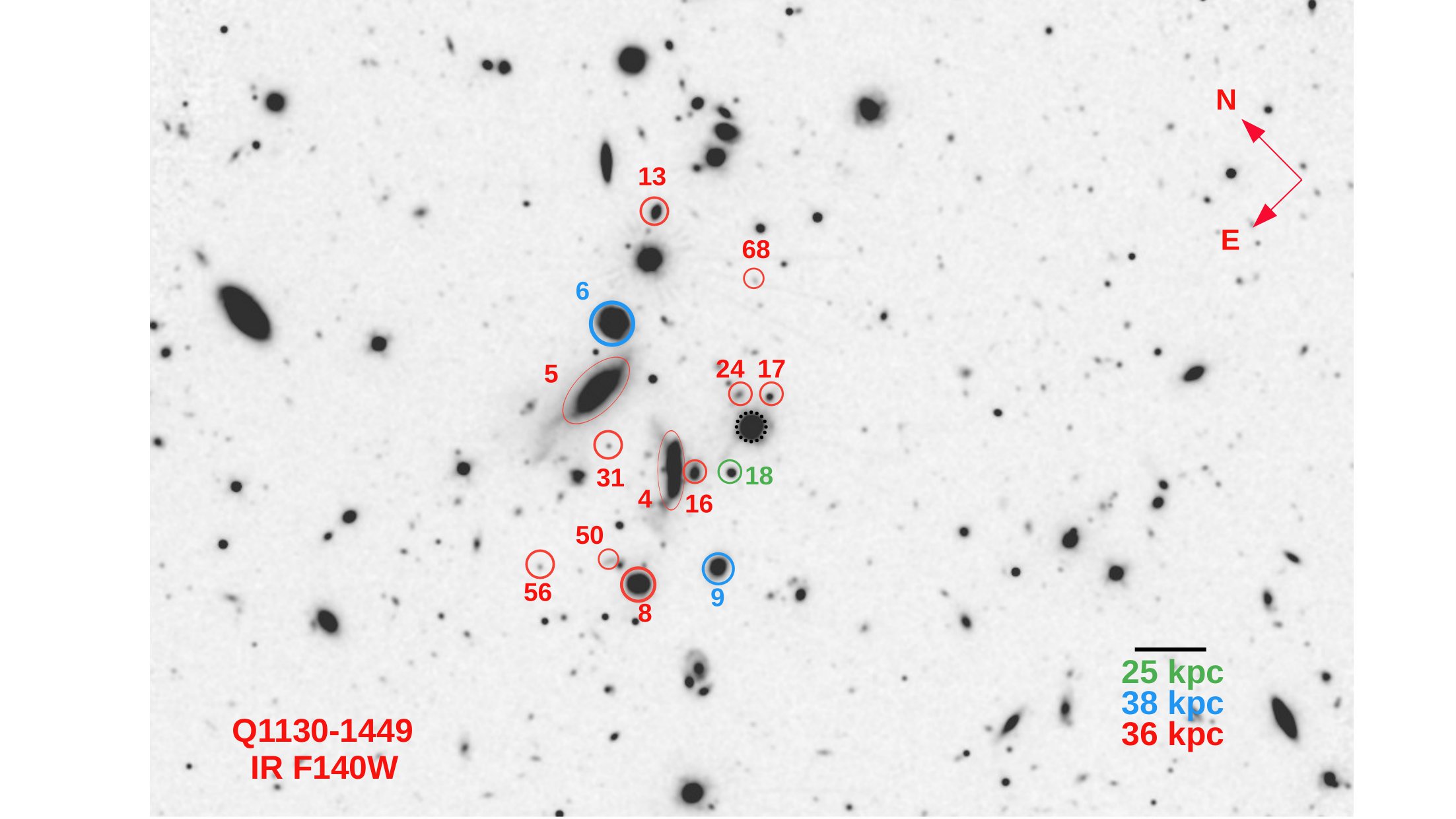}
 \caption{See \autoref{fig14} caption. The scale corresponds to 26 kpc at $z_{\rm abs}$ = 0.5604 for the top figure, while the scale corresponds to 25 kpc, 38 kpc and 36 kpc at $z_{\rm abs}$ = 0.1906, 0.3283, and 0.3130 for the bottom figure, respectively.}

    \label{fig21}
\end{figure*}

\begin{figure*}
     \includegraphics[width=0.9\textwidth]{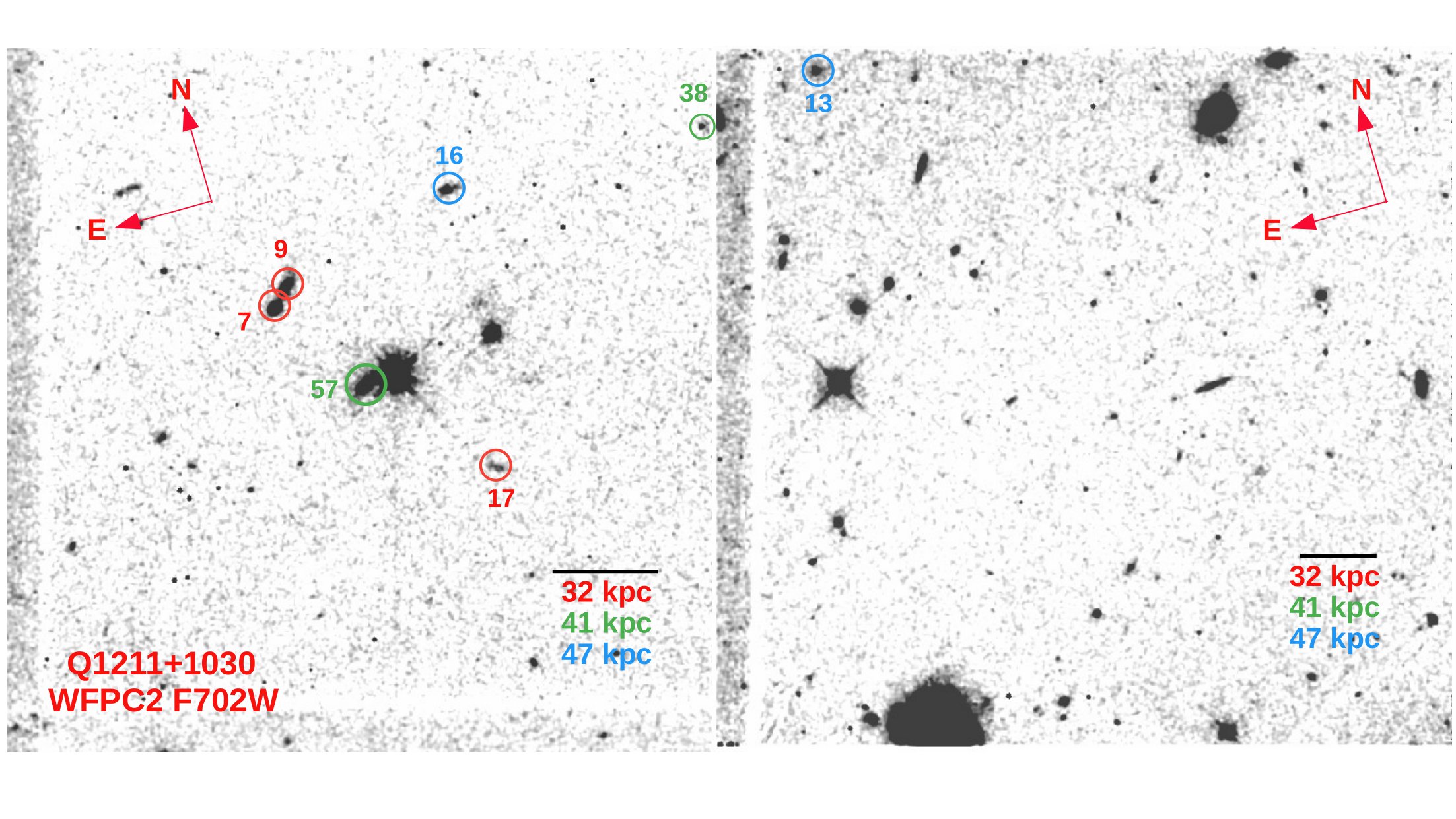}\\   
    \vspace{0.5cm}

    \includegraphics[width= 0.7\textwidth]{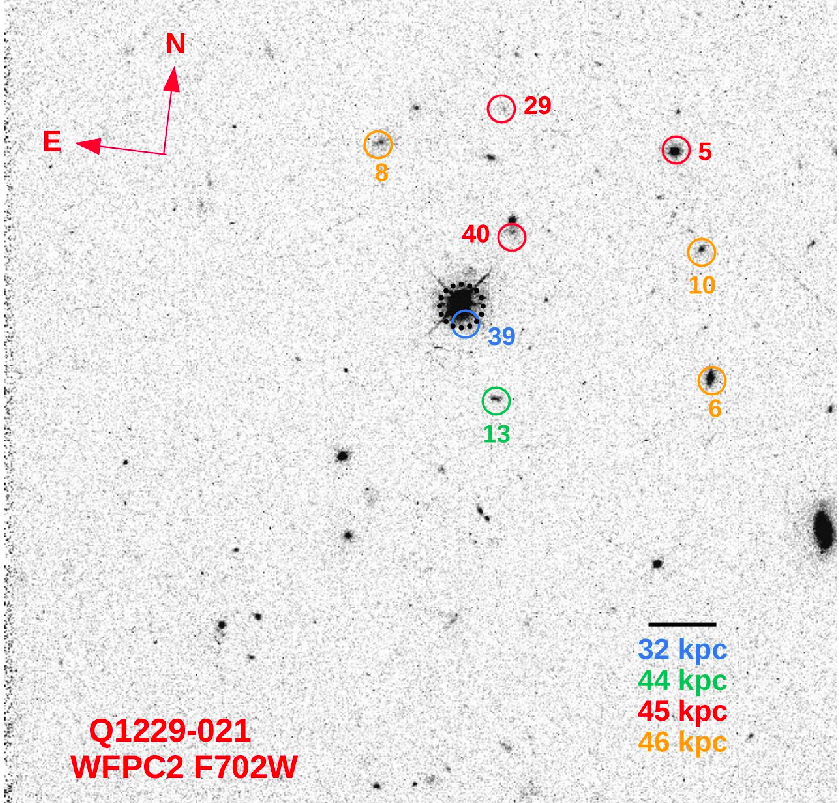}
    \label{fig22}

    \caption{See \autoref{fig14} caption. The scale corresponds to 32 kpc, 41 kpc, and 47 kpc at $z_{\rm abs}$ = 0.3929, 0.6296 and 0.8999 for the top figures, while the scale corresponds to 32 kpc, 44 kpc, 45 kpc and 46 kpc at $z_{\rm abs}$ = 0.3950, 0.7572, 0.7691 and 0.8311 for the bottom figure, respectively.}
    
\end{figure*}

\begin{figure*}

 \includegraphics[width=0.85\textwidth]{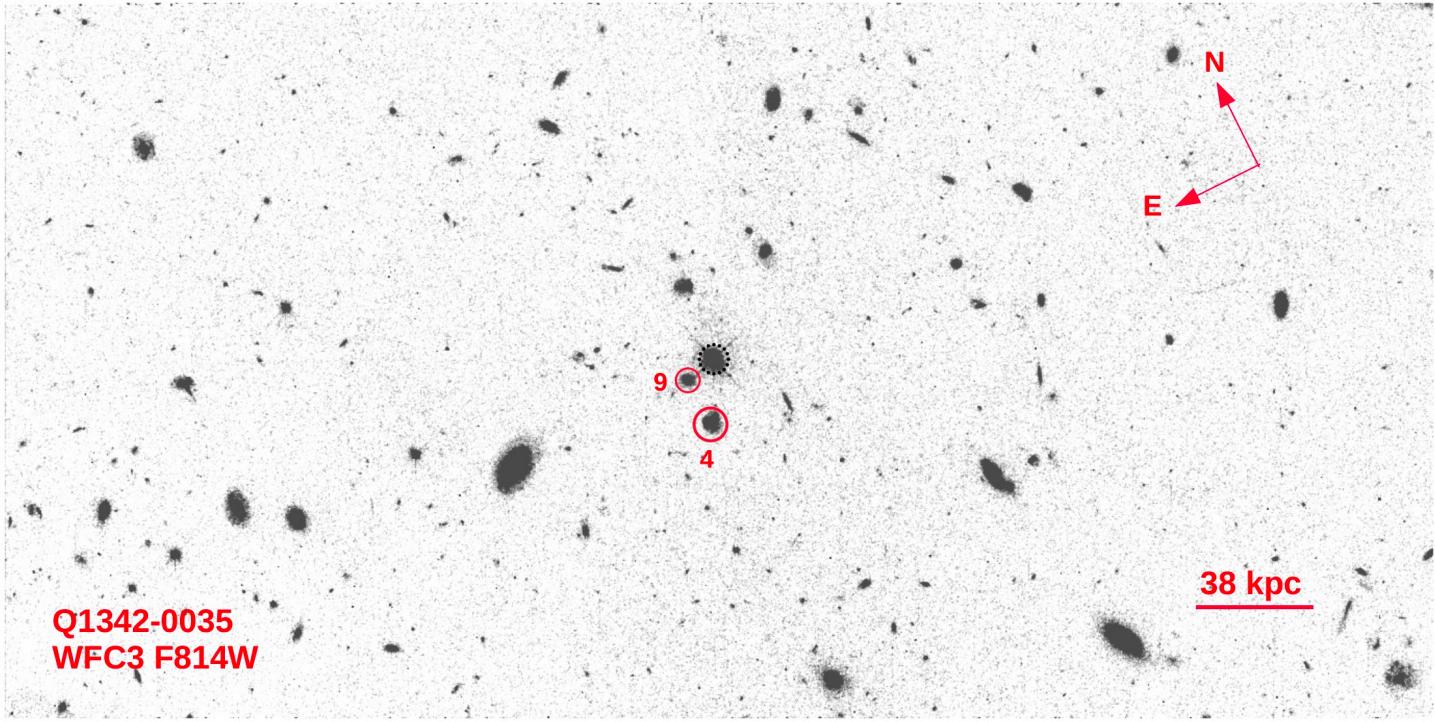}\\
 
    \vspace{0.05cm}

 \includegraphics[width=0.85\textwidth]{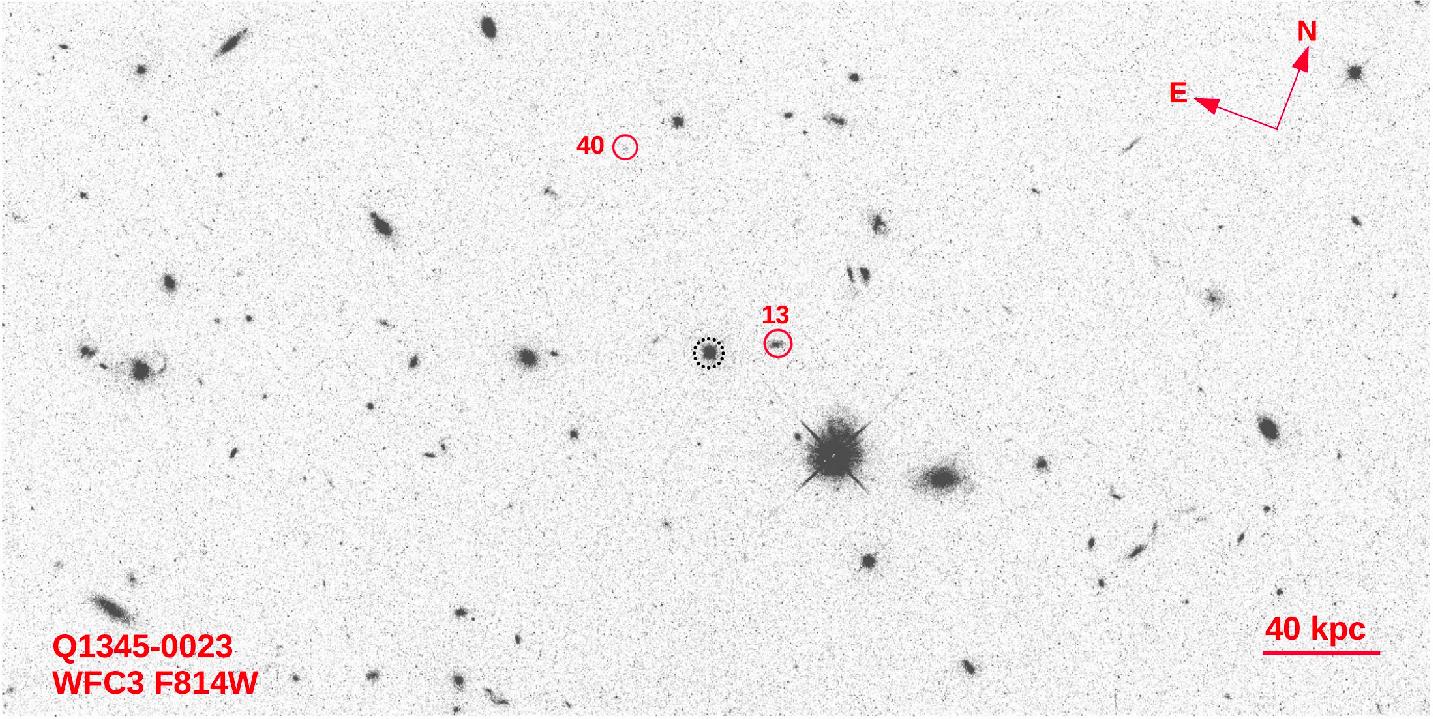}\\
    \vspace{0.05cm}

      \includegraphics[width=0.85\textwidth]{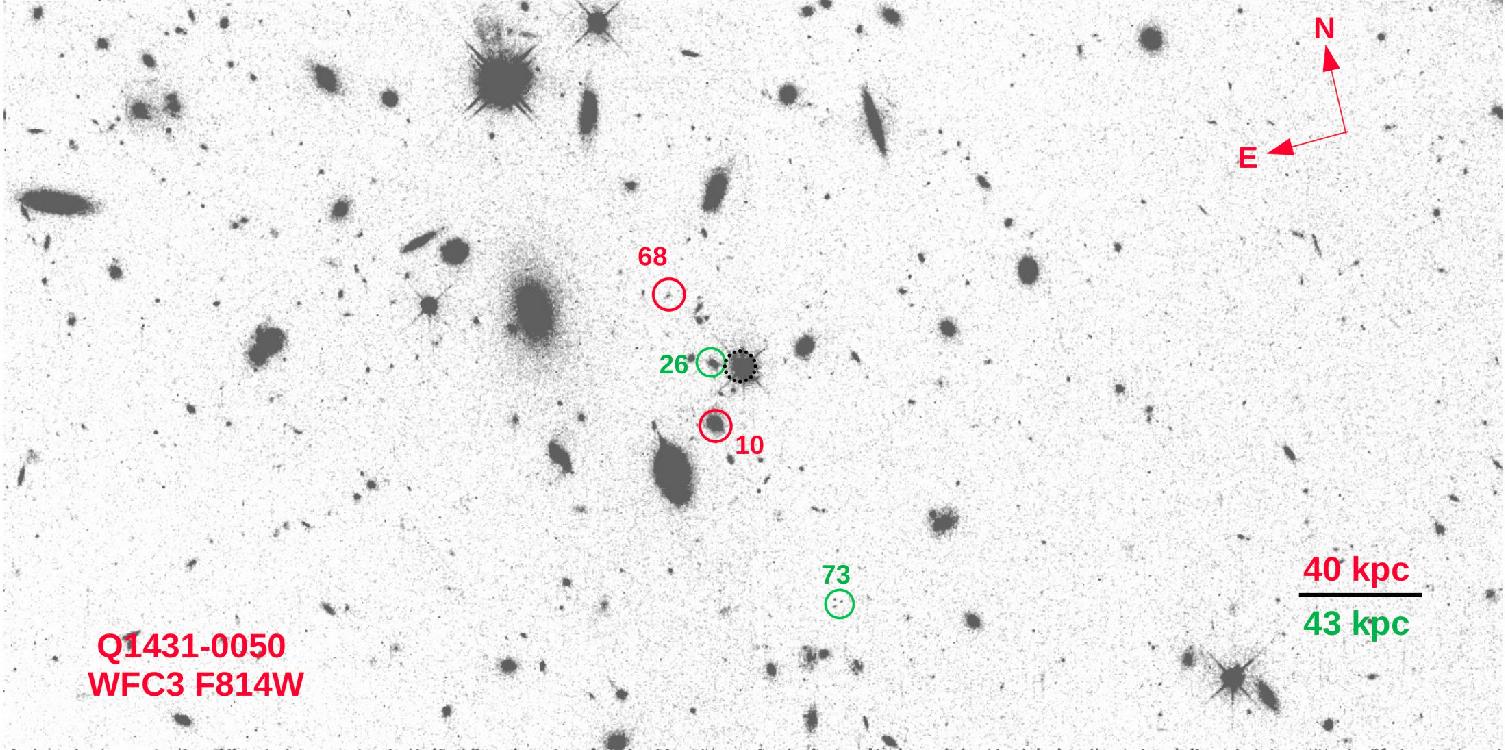} 

    \caption{See \autoref{fig14} caption. The scale corresponds to 38 kpc at $z_{\rm abs}$ = 0.5380 for the top figure, 40 kpc at $z_{\rm abs}$ = 0.6057 for the middle figure while the scale corresponds to 40 kpc and 43 kpc at $z_{\rm abs}$ = 0.6085 and 0.6868 for the bottom figure, respectively.}

    \label{fig23}
\end{figure*}

\begin{figure*}
    \includegraphics[width=0.9\textwidth]{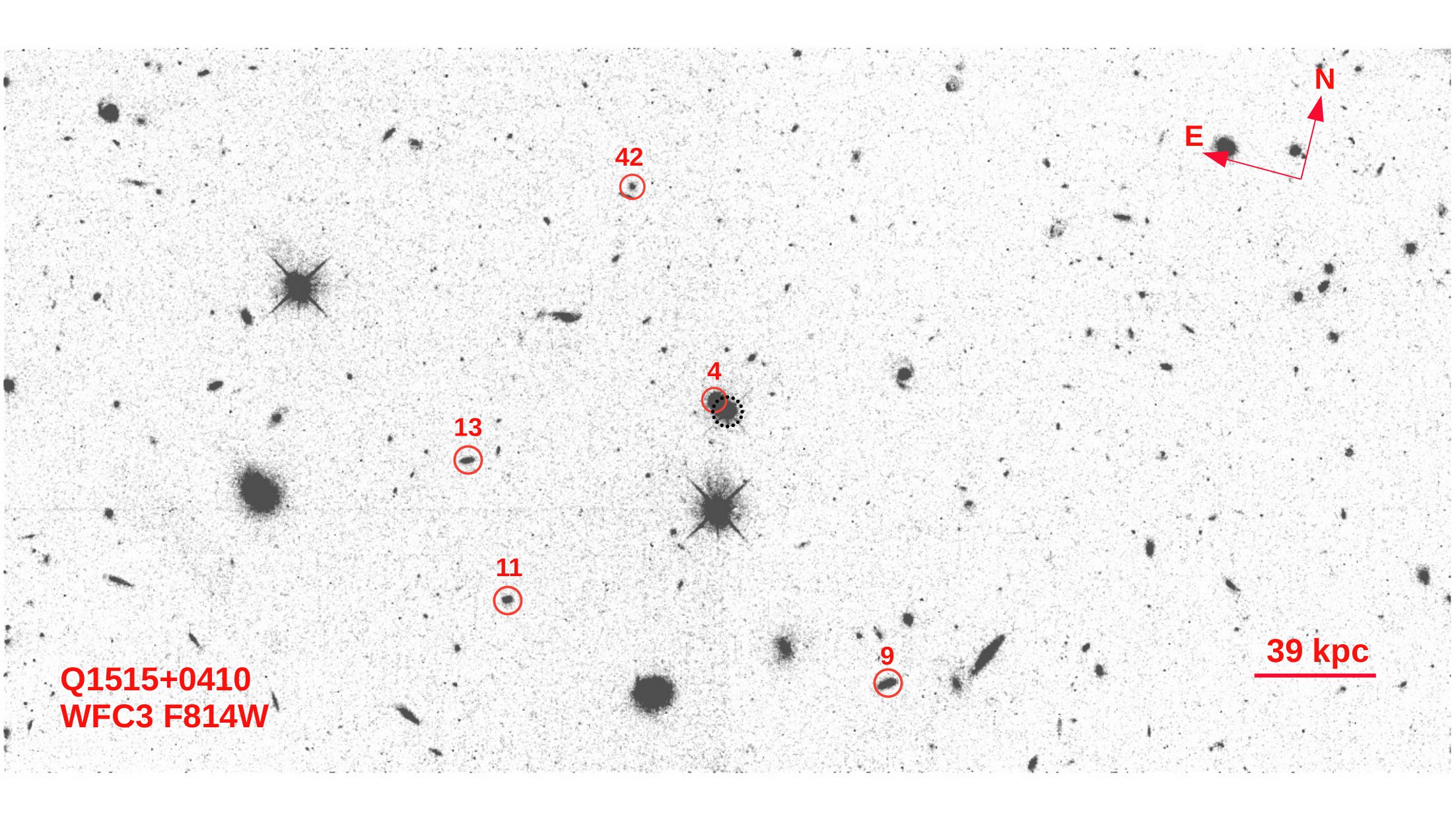}\\

  \vspace{0.1cm}

\end{figure*}

\begin{figure*}
  \centering
  
  \begin{subfigure}{.46\linewidth}
    \centering

    \includegraphics[width = \linewidth]{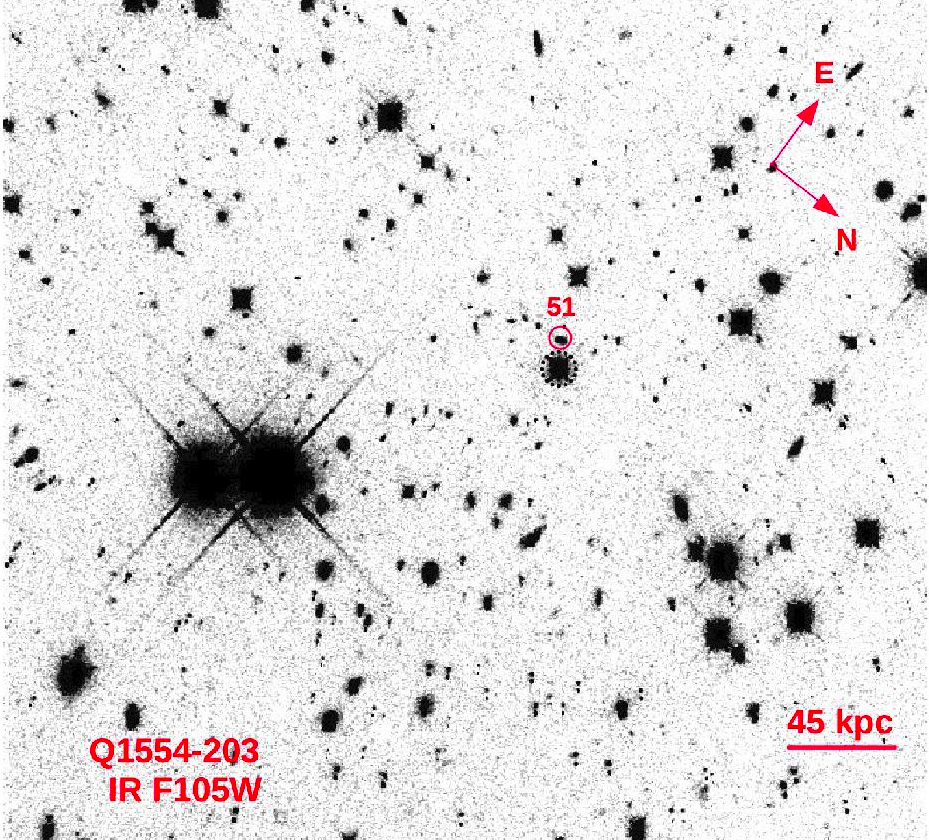}

    \vspace{0.1em}
    
   % \caption{Second image}
  \end{subfigure}%
  % Space between image B and C
  %\vspace{0.1cm}
  \begin{subfigure}{.45\linewidth}
    \centering
     \includegraphics[width = 0.95\linewidth]{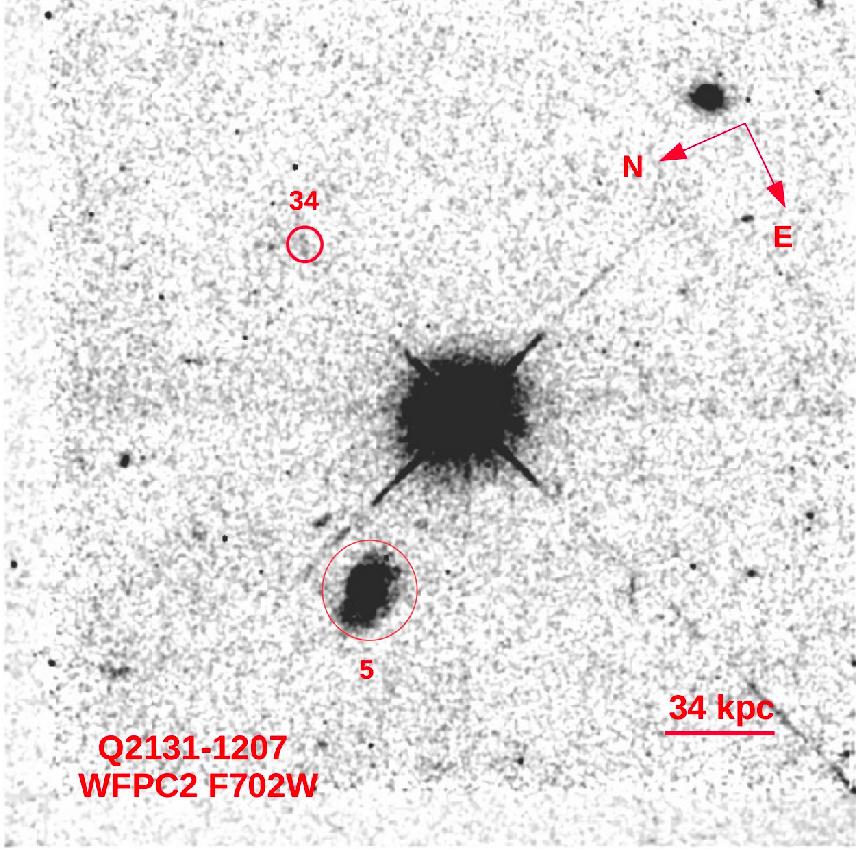}   
  \end{subfigure}

  \caption{See \autoref{fig14} caption. The scale corresponds to 39 kpc at $z_{\rm abs}$ = 0.5592 for the top figure, while in the bottom figures, the scale corresponds to 45 kpc and 34 kpc at $z_{\rm abs}$ = 0.7869 and 0.4298 for the left and right figure, respectively.}
 
\end{figure*}
%%%%%%%%%%%%%%%%%%%%%%%%%%%%%%%%%%%%%%%%%%%%%%%%%%

% Don't change these lines
\bsp	% typesetting comment
\label{lastpage}
\end{document}